\documentclass[aps,twocolumn,nofootinbib,floatfix,superscriptaddress]{revtex4}
\usepackage{amsfonts}
\usepackage{amssymb}
\usepackage{amsmath}
\usepackage{graphics}
\usepackage{graphicx}
\usepackage{soul}
\usepackage{natbib}
\usepackage{bm}
\usepackage{mathtools}
\usepackage{stmaryrd}
\usepackage{epstopdf}
\usepackage{color}
\usepackage{lipsum}
\usepackage{balance}
\usepackage{enumitem}
\usepackage[colorlinks=true,citecolor=black,linkcolor=black,urlcolor=black,filecolor=black]{hyperref}
\usepackage[dvipsnames]{xcolor}
\usepackage{bbold}
\usepackage{url}
\usepackage{tikz-feynman}
\usetikzlibrary{decorations.pathmorphing}

\usepackage{silence}
\tikzset{
  double wavy/.style={
    decorate,
    decoration={snake, amplitude=1pt, segment length=6pt},
    double,
    thick
  }
}

\defcitealias{MSR1973}{MSR}
\defcitealias{Flores+2025}{FF25}

\usepackage{acronym}
\newacro{VRR}{Vector Resonant Relaxation}
\newcommand{\VRR}{\ac{VRR}}
\newacro{DIA}{Direct Interaction Approximation}
\newcommand{\DIA}{\ac{DIA}}
\newacro{DF}{distribution function}
\newcommand{\DF}{\ac{DF}}
\newacro{MSR}{Martin--Siggia--Rose}
\newcommand{\MSR}{\ac{MSR}}
\newacro{FDT}{Fluctuation-Dissipation Theorem}
\newcommand{\FDT}{\ac{FDT}}
\newacro{BH}{black hole}
\newacroplural{BH}[BHs]{black holes}
\newcommand{\BH}{\ac{BH}}

\newcommand{\p}{\partial}
\newcommand{\rd}{\mathrm{d}}

\newcommand{\hbL}{\widehat{\mathbf{L}}}

\newcommand{\vphid}{\varphi_{\mathrm{d}}}
\newcommand{\vphi}{\varphi}
\newcommand{\deltaD}{\delta_{\mathrm{D}}}
\newcommand{\half}{\tfrac{1}{2}}
\newcommand{\Tc}{T_{\mathrm{c}}}

\newcommand{\mJ}{\mathcal{J}}
\newcommand{\ml}{\langle}
\newcommand{\mr}{\rangle}

\newcommand{\eps}{\epsilon}
\newcommand{\1}{\mathbf{1}}
\newcommand{\2}{\mathbf{2}}
\newcommand{\3}{\mathbf{3}}
\newcommand{\4}{\mathbf{4}}
\newcommand{\5}{\mathbf{5}}
\newcommand{\6}{\mathbf{6}}
\newcommand{\7}{\mathbf{7}}
\newcommand{\8}{\mathbf{8}}
\newcommand{\9}{\mathbf{9}}

\newcommand{\tmin}{\scalebox{0.6}{$-$}}
\newcommand{\tplus}{\scalebox{0.6}{$+$}}
\newcommand{\phh}{\scalebox{0.6}{1/2}}
\newcommand{\pth}{\scalebox{0.6}{3/2}}

\newcommand{\Min}{\mathrm{Min}}
\newcommand{\Max}{\mathrm{Max}}

\newcommand{\MBH}{M_\bullet}

\newcommand{\mO}{\mathcal{O}}

\newcommand{\rG}{\mathrm{G}}

\newcommand{\bLtot}{\mathbf{L}_{\mathrm{tot}}}

\newcommand{\NSEED}{\texttt{NSEED}}
\newcommand{\DT}{\texttt{DT}}
\newcommand{\NSTEPS}{\texttt{NSTEPS}}
\newcommand{\TCDT}{\texttt{TC/DT}}
\newcommand{\ITER}{\texttt{ITER}}
\newcommand{\LMAX}{\texttt{LMAX}}
\newcommand{\TMAXTC}{\texttt{TMAX/TC}}
\newcommand{\TMAX}{\texttt{TMAX}}
\newcommand{\LINT}{\texttt{LINT}}
\newcommand{\rT}{\mathrm{T}}
\newcommand{\br}{\mathbf{r}}
\newcommand{\bR}{\mathbf{R}}
\newcommand{\msfR}{\mathsf{R}}

\newcommand{\msfI}{\mathsf{I}}
\newcommand{\msfG}{\mathsf{G}}
\newcommand{\msfC}{\mathsf{C}}
\newcommand{\RePart}{\mathrm{Re}}
\newcommand{\ext}{\mathrm{ext}}
\newcommand{\even}{\mathrm{even}}
\newcommand{\odd}{\mathrm{odd}}

\usepackage{soul} \usepackage{xcolor}

\begin{document}

\title{Vector Resonant Relaxation and Statistical Closure Theory.
\\
II. One-loop Closure}

\author{Sofia Flores}
\affiliation{Institut d'Astrophysique de Paris, UMR 7095, 98 bis Boulevard Arago, F-75014 Paris, France}
\author{Jean-Baptiste Fouvry}
\affiliation{Institut d'Astrophysique de Paris, UMR 7095, 98 bis Boulevard Arago, F-75014 Paris, France}

\begin{abstract}
We use stellar dynamics as a testbed for statistical closure theory.
We focus on the process of ``Vector Resonant Relaxation,'' a long-range, non-linear,
and correlated relaxation mechanism that drives the reorientation of stellar orbital planes around a supermassive black hole.
This process provides a natural setting to evaluate the predictive power of generic statistical closure schemes for dynamical correlation functions,
in the fully non-linear and non-perturbative regime. 
We develop a numerical scheme that explicitly implements the seminal ``Martin--Siggia--Rose'' formalism at one-loop order via an iterative fixed-point approach, thereby improving upon the bare order from the ``Direct Interaction Approximation.'' 
Using this framework, we quantitatively validate the ability of the formalism to predict
(i) the two-point two-time correlation function;
(ii) the renormalised three-point interaction vertex;
(iii) the three-point three-time correlation function.
These predictions are compared to direct measurements from numerical simulations. 
We conclude by discussing the limitations of this approach and presenting possible future venues.
\end{abstract}
\maketitle

\section{Introduction}
\label{sec:Introduction}

A wide range of physical processes can generically be described
by a non-linear evolution equation of the form
\begin{equation}
\p_{t} \vphi_1 = \gamma_{12} \vphi_2 + \half \gamma_{123} \vphi_2 \vphi_3 + f^{\ext}_{1} ,
\label{eq:generic_equation} 
\end{equation}
where $1$ is the system's coordinate (that may include time)
and the summation over the repeated coordinates is implied.
In this equation, ${\vphi_1\!=\!\vphi(1)}$ is a field describing the state of the system,
$\gamma_{12}$ is a linear coupling coefficient in the evolution equation,
$\gamma_{123}$ a non-linear one,
and $f^{\ext}_{1}$ some stochastic (external) perturbation.
Equation~\eqref{eq:generic_equation} is particularly generic.
It applies to fluid dynamics
and captures the incompressible Navier--Stokes equation~\citep[see, e.g.\@,][]{Frisch1995, Berera+2013},
where $1$ denotes the position coordinate, $\vphi$ represents the velocity field, 
the linear term accounts for effects such as viscosity, 
and the non-linear term corresponds, for example, to convective acceleration.
Similarly, Eq.~\eqref{eq:generic_equation} also applies to electrostatic plasmas
and captures the Vlasov equation~\citep[see, e.g.\@,][]{Nicholson1992,Krommes2015}.
Therein, $1$ is the phase space coordinate, position and velocity,
$\vphi$ is the phase space density of, say, the electrons,
$\gamma_{12}$ describes the (linear) phase mixing,
while $\gamma_{123}$ captures the effect of the electrostatic fluctuations.
The same Vlasov equation is at the heart of the description
of self-gravitating systems such as galaxies~\cite[see, e.g.\@,][]{ Lynden1967,Binney+2008},
where, this time, the non-linearity stems from the gravitational Poisson equation.
As a last illustration, Eq.~\eqref{eq:generic_equation} is also relevant
to describe the formation of large-scale cosmological structures~\citep[see, e.g.\@,][]{Bernardeau+2002, Bernardeau+2012}.
In that case, the coordinate $1$ is the position coordinate,
$\vphi$ contains the density contrast,
$\gamma_{12}$ captures the effect of the cosmological expansion,
while $\gamma_{123}$ is associated with the non-linear contribution from self-gravity.
In the coming sections, we will focus on one other
realisation of Eq.~\eqref{eq:generic_equation},
in the context of gravitational dynamics---\VRR\@~\citep{Kocsis+2015}, that we will describe later on---viewing it as a particular regime of Eq.~\eqref{eq:generic_equation}
that conserves all its key properties.
Because it traverses so many fields of physics,
it is of prime importance to develop appropriate and efficient
theoretical tools to describe the statistical properties
of fields evolving according to Eq.~\eqref{eq:generic_equation}.

A first regime of Eq.~\eqref{eq:generic_equation}
is the so-called \textit{weak turbulence} regime~\citep{Krommes2002,Diamond+2010},
i.e.\ a regime in which there is a clear
separation between the linear timescale, ${\sim\!1/\gamma_{12}}$,
and the non-linear timescale, 
${\sim\!1/\gamma_{123}\vphi_3}$~\cite{Nazarenko2011,Schekochihin2025}.
In weak turbulence, non-linear effects are generically small
and can be treated perturbatively as small corrections to the linear dynamics~\citep[see, e.g.\@,][]{Micha+2004,Nazarenko2011}.

In the \textit{strongly turbulent} regime, both timescales are comparable, 
and non-linearities compete with linear effects~\cite{Yokoyama+2014,Schekochihin2025}.
This causes standard perturbative approaches to fail. 
This situation commonly arises, for example, 
when modelling stationary turbulence in fluid dynamics~\cite{Tarpin+2019,Zhou2021}.

In the regime of \VRR\@,
as will be detailed later,
the importance of non-linearities
is made all the clearer,
since one has
${ \gamma_{12} \!=\! 0 }$ (no linear dynamics)
and ${ f^{\ext}_{1} \!=\! 0 }$ (no external excitation).
This is an extreme limit of strong turbulence,
in which the evolution is driven entirely by the non-linear coupling, $\gamma_{123}$.
As such, in the present work, Eq.~\eqref{eq:generic_equation}
reduces to the (apparently) simpler form
\begin{equation}
\p_{t} \vphi_{1} = \half \gamma_{123} \vphi_{2} \vphi_{3} ,
\label{eq:nonlinear}
\end{equation}
of a purely quadratically non-linear evolution equation.
Interestingly, the same equation applies
to the Maier--Saupe model of liquid crystals~\citep[see, e.g.\@,][]{Maier+1958,Plischke+2006,Roupas+2017}.

When describing the statistical properties of a system evolving according to Eq.~\eqref{eq:nonlinear},
a first quantity of key interest is the system's
two-point cumulant (in space and time).
It is generically defined as
\begin{equation}
C_{1 2} = \ml \vphi_1 \vphi_2 \mr - \ml \vphi_1 \mr \ml \vphi_2 \mr ,
\label{eq:def_C_generic}
\end{equation}
where ${ \ml ... \mr }$ stands for the ensemble average
over independent realisations. 
Here, stochasticity is said to be extrinsic,
since $\gamma_{123}$ in Eq.~\eqref{eq:nonlinear}
is non-stochastic; there is no external excitation;
and randomness only enters through the system's initial conditions.
Characterising in detail the two-point correlation function is particularly important
since it provides a wealth of statistical information
about the temporal and spatial coherence that may emerge spontaneously
in the system, e.g.\@, like vortices in fluid systems~\citep[see, e.g.\@,][]{Kraichnan1958}, 
eddies in electrostatic plasmas~\citep[see, e.g.\@,][]{Dupree1972,Boutros+1981},
or large-scale structures in cosmology~\citep[see, e.g.\@,][]{Valageas2004,Bernardeau+2012Corr}.

A related quantity of interest is the generalised skewness 
\begin{equation}
S_{123} = \frac{ \ml \vphi_{1} \vphi_{2} \vphi_{3} \mr}{ \sqrt{\ml \vphi^2_{1} \mr \ml \vphi^2_{2} \mr \ml \vphi^2_{3} \mr}},
\label{eq:intro_skew}
\end{equation}
where ${ \ml \vphi_{1} \vphi_{2} \vphi_{3} \mr}$ is the three-point correlation function.
For simplicity, we assumed here that the mean field vanishes,
i.e.\ ${ \ml \vphi \mr \!=\! 0 }$. This is correct for \VRR\ in the isotropic limit.
The dimensionless skewness, $S$, quantifies the degree of non-Gaussianity in the system.
Indeed, for Gaussian statistics, one has ${S\!=\!0}$,
while ${S \!\sim\! 1}$ indicates strong non-Gaussianities.
In fluid turbulence, the three-point function can be related to energy dissipation via Kolmogorov ${4/5}$-law~\citep[see, e.g.\@,][]{Kolmogorov1941c,Frisch1995}; while in cosmology, its scale-dependent counterpart, the bispectrum, probes non-Gaussianities beyond the power spectrum~\citep[see, e.g.\@,][]{Baldauf+2011}.

From Eq.~\eqref{eq:nonlinear}, and assuming ${ \ml \vphi \mr \!=\! 0 }$,
we readily find that $C_{12}$ evolves according to
\begin{equation}
\p_{t} C_{12} = \tfrac{1}{2} \gamma_{134} \ml \vphi_{2} \vphi_{3} \vphi_{4} \mr .
\label{eq:Evol_C_approx}
\end{equation}
Equation~\eqref{eq:Evol_C_approx} clearly emphasises the well known closure problem: cumulants follow a hierarchy that is not closed~\cite{Kraichnan1959,Orszag1970,Krommes2015}.
This is a direct consequence of the quadratic non-linearity in Eq.~\eqref{eq:nonlinear}.
Assuming Gaussian initial conditions, a non-zero skewness appears at infinitesimal times, and higher-order cumulants follow.
Phrased differently, as evolution proceeds,
the PDF of $\vphi$ must necessarily deviate from the initial Gaussian form~\cite{McComb1990,McComb2014}.

A system evolving according to Eq.~\eqref{eq:Evol_C_approx}
offers no obvious path for perturbative expansion, as there is a non-perturbative relationship between correlations.
Indeed, without a linear timescale, non-Gaussianities grow on the intrinsic non-linear timescale set by the non-linear coupling coefficient and field amplitudes~\cite{Frisch1995,Nazarenko2011}.
Consequently, traditional quasi-normal closures applicable in weak turbulence~\cite{Orszag1970,Krommes2002} are bound to fail.
As a result, self-consistent closures,
such as renormalisation methods, are mandatory~\cite[see, e.g.\@,][]{McComb2014,Canet2025}.

In this work, we tackle this intrinsic difficulty by explicitly implementing and benchmarking a general statistical closure scheme---the \MSR\ formalism~\cite{MSR1973}---using the \VRR\ system as a controlled toy model. The simplifying assumptions of time stationarity, isotropy, 
equilibrium dynamics, and initial Gaussian statistics, make \VRR\  an ideal, tractable framework to test and validate such closure methods in the fully non-linear regime.

This paper is organised as follows.
In Section~\ref{sec:VRR_dyn}, we introduce the equations governing the \VRR\ dynamics.
In Section~\ref{sec:MSR_general}, we briefly recast the general \MSR\ equations and the associated hierarchy of closures.
In Section~\ref{sec:Numerical}, we describe the numerical scheme developed to solve the \MSR\ closure equations.
In Section~\ref{sec:Application}, we present the bare and one-loop predictions for the two- and three-point correlation functions, as well as for the renormalised vertex, in the isotropic \VRR\ system, and compare these results with measurements from $N$-body simulations.
In Section~\ref{sec:Discussion}, we explore the invariance of the statistical properties with respect to $N$ and discuss the underlying difficulty 
of extending the scheme to two-loop order.
Finally, we conclude in Section~\ref{sec:Conclusion}.

\section{ \VRR: A playground for closure theory}
\label{sec:VRR_dyn}

\begin{figure}[htbp!]
\begin{center}
\includegraphics[width=0.3\linewidth]{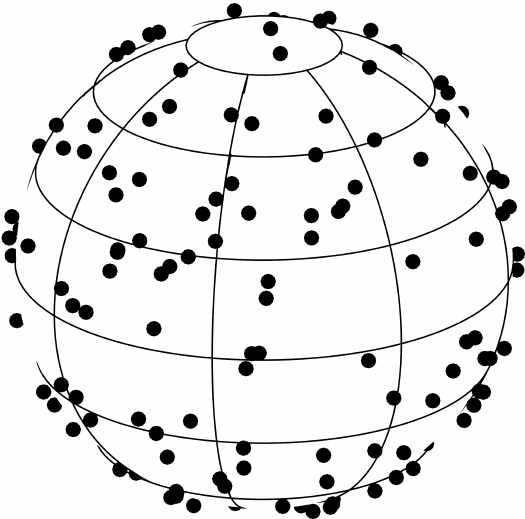}
\includegraphics[width=0.3\linewidth]{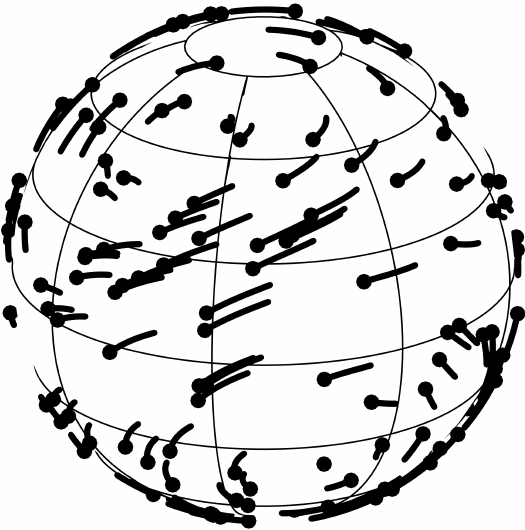}
\includegraphics[width=0.3\linewidth]{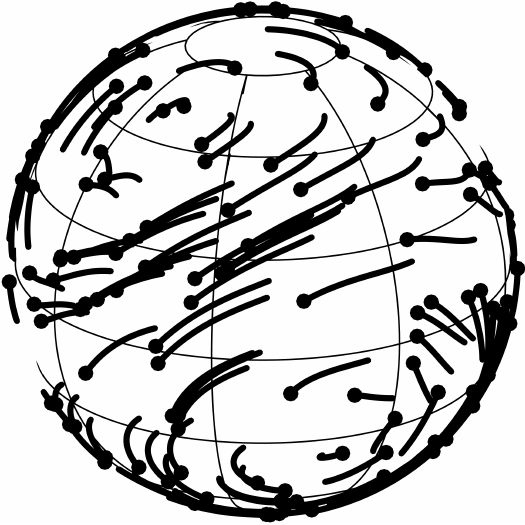}
\caption{Illustration of the process of \VRR\@.
Each point on the unit sphere depicts the instantaneous orbital orientation of a star around a supermassive \BH, 
with the tail showing its past trajectory. 
Left: snapshot at ${t\!=\!0}$. Middle: ${t\!=\!\Tc/8}$. Right: ${t\!=\!\Tc/4}$, 
where $\Tc$ is the coherence time defined in Eq.~\eqref{eq:def_Tc}. 
Characterising correlation functions in this system
offers detailed information on the statistical properties
of the stars' rearrangement on the unit sphere. 
}
\label{fig:Sphere}
\end{center}
\end{figure}

We study the same physical setup as in~\citep{Flores+2025}, which we refer to as \citetalias{Flores+2025}.
We now briefly recall the main elements of this dynamical model.
We consider a system of ${ N \!\gg\! 1 }$ stars orbiting a supermassive \BH\@,
just like SgrA* in the centre of the Milky Way~\citep{Gillessen+2017}.
At leading order, the central \BH\ dominates the system's mean potential,
so that stars follow quasi-Keplerian orbits.
Yet, as a result of the gravitational perturbations
that they impose onto one another,
these orbits slowly diffuse in orbital orientation,
eccentricity and semi-major axis~\cite{Alexander2017}.
Considering timescales of the order of ${1\,\mathrm{Myr}}$,
one can average the stars' dynamics over their Keplerian motion
and their in-plane precession~\citep{Rauch+1996}, thus assuming that the associated eccentricities and semi-major axes
remain fixed.
On those timescales, stars are formally replaced with massive annuli,
with their orbital orientation being the only dynamical quantity evolving with time.
This is the process of \VRR\@~\citep{Kocsis+2015},
i.e.\ the efficient process of gravitational relaxation
associated with the long-term evolution
of each star's normalised angular momentum vector, $\hbL$.
Each unit vector can be tracked via a \textit{point particle}
evolving on the unit sphere---which is phase space for \VRR---through long-range, correlated motions. 
This is illustrated in Fig.~\ref{fig:Sphere}.
The process of \VRR\ leads to rich phenomenology.
For example, it plays an important role
in warping the stellar disc around SgrA*~\cite{Kocsis+2011}
and impacting the rate of binary mergers therein~\citep{Hamers+2018}.
It can also drive some spontaneous phase transitions~\citep[see, e.g.\@,][]{Roupas+2017,Szolgyen+2018,Touma+2019,Tremaine2020a,Tremaine2020b,Gruzinov+2020,Magnan+2022,Mathe+2023},
drive an efficient dynamical friction on heavy objects~\citep{Ginat+2023,Levin+2024},
and cause the dilution of stellar discs~\citep{Panamarev+2022,Panamarev+2025}.
Finally, it may also be used to constrain the stellar,
and even dark matter, content in the vicinity of SgrA*~\citep[see, e.g.\@,][]{Fouvry+2023,Ginat+2025Axion}.
In the present work, we are mainly interested
in using \VRR\ as an explicit, physically-informed, playground
for fully non-linear dynamics, with which we test methods from stastical closure theory.

In order to simplify the description of \VRR\@,
we consider a single-population system,
i.e.\ we assume that all the underlying orbits
have the same semi-major axis and eccentricity.
Although it makes the calculation more cumbersome, the case of a multi-population system
can readily be included in the present formalism~\citepalias{Flores+2025}.
We may then introduce the empirical \DF\ of the system via
\begin{equation}
\vphid (\hbL , t) = \sum_{i = 1}^{N} \deltaD \big[ \hbL \!-\! \hbL_{i} (t) \big] \, ,
\label{eq:def_Fd}
\end{equation}
with $\deltaD$ the Dirac delta,
and ${ \hbL_{i} (t) }$ the orbital orientation at time $t$ of particle $i$.
Expanding the \DF\ over the (real) spherical harmonics,
we can write it as\footnote{We use
the same normalisation as in~\citetalias{Flores+2025},
namely ${ \! \int \! \rd \hbL \, Y_{a} Y_{b} \!=\! \delta_{a}^{b} }$.}
\begin{equation}
\vphid (\hbL , t) = \sum_{a} \vphi_{a} (t) \, Y_{a} (\hbL) ,
\label{eq:expansion_vphid}
\end{equation}
with ${ a \!=\! (\ell,m) }$.
From the Hamiltonian of \VRR\ and the continuity equation
in phase space~\cite{Fouvry+2019},
the dynamics of \VRR\ is driven by
\begin{equation}
\p_{t} \vphi_1 = \half \, \gamma_{1 2 3} \, \vphi_2 \, \vphi_3 ,
\label{eq:Master_Equation_short}
\end{equation}
where ${1 \!\!=\!\!(a,t)}$, and ${\gamma_{1 2 3}}$ is the \textit{bare} coupling coefficient,
whose expression is reproduced in Appendix~\ref{app:Bare_gamma}.
Importantly, ${\gamma\!=\!\gamma_{1 2 3}}$ is local in time,
non-stochastic, and it contains all the information
on the gravitational interaction between the massive annuli of \VRR\@.
To further simplify the model, we restrict \VRR\ to a ${\ell \!=\! 2}$ coupling.\footnote{This retains the coupling at large scales,
similar to the Batchelor regime in plasma physics~\citep[see, e.g.\@][and references therein]{Nastac+2025}.}
This choice is physically motivated, since the \VRR\ interaction is primarily dominated by the large-scale quadrupolar component~\cite[see appendix~{B} in][]{Kocsis+2015}.
The case of a multipole interaction spectrum
is considered in Appendix~\ref{app:InteractionSpectrum}.
As previously emphasised, Eq.~\eqref{eq:Master_Equation_short} is fully non-linear: there is no linear term associated with kinetic energy.

In \VRR, fluctuations are generated by the finite number of particles $N$.
In practice, for the rest of this work,
we will focus on \VRR\ in the isotropic limit, 
i.e.\ assuming an initial statistically uniform distribution on the unit sphere,
as illustrated in Fig.~\ref{fig:Sphere}.
In particular, for any ${ \ell_{a} \!>\! 0 }$,
this imposes ${ \ml \vphi_{a} \mr \!=\! 0 }$.
In that limit, the evolution equation for the two-point correlation function $C_{12}$, as defined in Eq.~\eqref{eq:def_C_generic},
is exactly Eq.~\eqref{eq:Evol_C_approx}.

\section{The \MSR\ formalism}
\label{sec:MSR_general}

The aim of statistical closure theory is to close
the infinite hierarchy of evolution equations
for the cumulants in non-linear systems, such as \VRR.
One possible approach for closure
is the seminal Martin--Siggia--Rose formalism~\citep{MSR1973}.
In~\citetalias{Flores+2025}, we focused on
the leading (i.e.\ bare) order approximation of this formalism,
namely the so-called \DIA\@~\citep{Kraichnan1959}.
In the present work, our goal is to go beyond this low-order approximation
and extend it, self-consistently, to next order.

\subsection{The general \MSR\ equations}
\label{sec:MSR_eqsl}

In this section, we briefly review the key steps leading to the \MSR\ closure formula~\citep{MSR1973}.
Firstly, \MSR\ notes that in order to have a full statistical description of a non-linear system, one has to consider not only its correlation functions but also its response functions to infinitesimal perturbations.
This is crucial because while correlations tell us about internal fluctuations,
the response functions characterise causal relationships and susceptibility,
namely how a small perturbation at any point in space and time propagates 
and influences the system's future evolution. Additionally, under certain conditions (such as thermal equilibrium), correlation and response are related by the fluctuation-dissipation theorem~\citep{Kraichnan1959a}, which states that they are directly proportional to one another.

To this end, we introduce the response function by adding a non-stochastic infinitesimal perturbation $\eta$ to the equation of motion of the field $\vphi$.
As such, starting from Eq.~\eqref{eq:Master_Equation_short},
we consider
\begin{equation}
\p_{t} \vphi_1 = \half \, \gamma_{1 2 3} \, \vphi_2 \, \vphi_3 + \eta_1 .
\label{eq:Eq_perturb}
\end{equation}
The mean response function is then defined as
\begin{equation}
R_{1 2} = \ml \delta \vphi_1 / \delta \eta_2 \mr |_{\eta=0} . 
\label{eq:def_R}
\end{equation}
It quantifies the average response of the field at point $1$ to an infinitesimal perturbation applied at point $2$.
Importantly, for causality to hold, the response function is non-zero
only for ${ t_1 \!\geq\! t_2 }$.

To simultaneously track the time evolution of correlation and response, a key step in~\citep{MSR1973} 
is the introduction of the \textit{two-point propagator} $G_{\1 \2}$, with ${\1 \!=\! (\eps,1)}$ 
and ${\eps \!=\! \pm}$, encapsulating both quantities within a single object.
More precisely, it reads
\begin{align}
G_{12}^{\tplus\tplus} {} & = C_{12} ; \quad \;\;\; G_{12}^{\tplus\tmin} = R_{12} ;
\nonumber
\\
G_{12}^{\tmin\tplus} {} & = R_{21} ; \quad \;\;\; G_{12}^{\tmin\tmin} = 0 .
\label{eq:components_G}
\end{align}
We note that the component ${G_{12}^{\tplus\tplus} }$ contains the information about the initial conditions of the system,
as detailed in Eq.~\eqref{eq:def_C0}.
Similarly, one may introduce a \textit{three-point propagator} $G_{\1\2\3}$, whose ${(+\!+\!+)}$ component exactly reproduces the three-point correlation. 
See Appendix~\ref{app:Prediction_3_point} for more detail.

From the evolution equations of $C_{12}$ and $R_{12}$~\citepalias[see, e.g.\@, appendix~B in][]{Flores+2025},
one can derive an evolution equation for $G_{\1\2}$ in terms of the three-point propagator $G_{\1\2\3}$.
It reads
\begin{equation}
\p_{t_1} G_{\1 \2} = \half \varsigma_{\1\3} \gamma_{\3\4\5} G_{\4\5\2} + \varsigma_{\1\2},
\label{eq:Evol_G_3}
\end{equation}
where the Pauli-like delta matrix $\varsigma_{\1\2}$ and the \textit{bare interaction vertex} $\gamma_{\1\2\3}$ are given in Appendix~\ref{app:gamma_def}.
Importantly, we recall that for isotropic \VRR\@, the mean field vanishes. Consequently,  Eq.~\eqref{eq:Evol_G_3} contains no \textit{one-point propagator} term~\citepalias[see, e.g.\@, eq.~B17 in][]{Flores+2025}.
Thus, by considering the ${ (+ +) }$ component of Eq.~\eqref{eq:Evol_G_3},
one recovers Eq.~\eqref{eq:Evol_C_approx} exactly.

Crucially, introducing the additional degree of freedom ${ \eps \!=\! \pm }$ plays a key role in fully symmetrizing Eq.~\eqref{eq:Evol_G_3}. 
In that equation, the coupling $\gamma_{\1\2\3}$ is now fully symmetric in all its coordinates, 
whereas $\gamma_{123}$ in Eq.~\eqref{eq:Master_Equation_short} is symmetric only in its last two coordinates.
This greatly simplifies the treatment of the equations, 
as detailed later on,
along with their diagrammatic representation.

Of course, Eq.~\eqref{eq:Evol_G_3} is still not closed, as the two-point propagator still depends on the three-point one.
The next important step in the \MSR\ derivation is to rewrite Eq.~\eqref{eq:Evol_G_3}
in a form suitable for systematic closure. This is achieved by recasting it as a Dyson equation.
In that view, one introduces $\Gamma_{\1\2\3}$, the \textit{renormalised interaction vertex}, defined in terms of the three-point propagator ${G_{\1\2\3}}$.
It reads
\begin{equation}
\Gamma_{\1\2\3} = G^{-1}_{\1 \1'} G^{-1}_{\2 \2'} G^{-1}_{\3\3'}G_{\1'\2'\3'} .
\label{eq:def_Gamma_3_point}
\end{equation}
Just like the bare vertex, $\Gamma_{\1\2\3}$ is fully symmetric.
From this definition, the renormalised vertex can be interpreted as a proxy for the skewness (Eq.~\ref{eq:intro_skew}),
therefore encapsulating the complexity of all the non-linear interactions that go beyond Gaussian approximations.
With the rewriting from Eq.~\eqref{eq:def_Gamma_3_point}, no information about the three-point correlation has been lost.

Injecting Eq.~\eqref{eq:def_Gamma_3_point} into Eq.~\eqref{eq:Evol_G_3}, we find
\begin{equation}
\p_{t_1} G_{\1 \2} = \varsigma_{\1\3} \Sigma_{\3\4} G_{\4\2} + \varsigma_{\1\2},
\end{equation}
where we introduced $\Sigma$, the \textit{self-energy}, as
\begin{equation}
\label{eq:Sigma}
\Sigma_{\1\2} = \half \, \gamma_{\1\3\4} G_{\3\3'} G_{\4\4'} \Gamma_{\2\3'\4'} .
\end{equation}
We now define the (inverse) two-point \textit{bare propagator}, ${g^{-1}_{\1\2}}$, which captures the time derivative as\footnote{The definition of the bare propagator in Eq.~\eqref{eq:def_g}
corrects a mistake in eq.~{(17)} of~\citetalias{Flores+2025}.}
\begin{equation}
g^{-1}_{\1\2} = - \p_{t_1} \varsigma_{\1\2}.
\label{eq:def_g}
\end{equation}
With this definition, we can rewrite the evolution equation for the two-point propagator as
\begin{equation}
G^{-1}_{\1\2} = g^{-1}_{\1\2} - \Sigma_{\1\2} ,
\label{eq:Dyson_inv}
\end{equation}
which takes the form of a Dyson equation.

At this stage, Eq.~\eqref{eq:Dyson_inv} is only
a convenient rewriting of Eq.~\eqref{eq:Evol_G_3}:
no information has been lost. 
In practice, this rewriting allows one to shift the focus of the statistical closure
from the three-point propagator, $G_{\1\2\3}$,
to the three-point renormalised vertex, $\Gamma_{\1\2\3}$.
By analogy with quantum field theory, ${\Gamma\!=\!\Gamma_{\1\2\3}}$ corresponds to the \textit{charge} of the system. 
More generically, the \MSR\ equations share deep connections with the Schwinger--Dyson equations~\cite[see, e.g.\@,][]{Bellon+2021}. 

The last stage of the \MSR\ approach is to rewrite the renormalised vertex $\Gamma$
(Eq.~\ref{eq:def_Gamma_3_point}) in a way that enables systematic closure. 
As shown in~\cite{MSR1973}, $\Gamma$ satisfies the functional relation
\begin{equation}
\Gamma_{\1\2\3} = \gamma_{\1\2\3} + \frac{\delta \Sigma_{\1\2}}{\delta G_{\4\5}} \, G_{\4\4'} G_{\5\5'} \Gamma_{\3\4'\5'} .
\label{eq:func_Gamma}
\end{equation}
This is an intricate calculation that requires careful treatment of the functional derivatives of the Dyson equation~\eqref{eq:Dyson_inv}.
We refer to~\cite{MSR1973} and appendix~{B} of~\citetalias{Flores+2025} for the explicit derivation
of Eq.~\eqref{eq:func_Gamma}.
Together, Eqs.~{\eqref{eq:Dyson_inv}-\eqref{eq:func_Gamma}}
form a self-consistent system of (functional) equations for
the two-point propagator, ${G\!=\!G_{\1\2}}$, and the renormalised vertex, $\Gamma$.
The second term on the right-hand side of Eq.~\eqref{eq:func_Gamma}
generates corrections to $\Gamma$, 
i.e.\ the hierarchy of closures.

\subsection{Diagrammatic representation}
\label{sec:Diags}

Because they can easily become quite cumbersome,
it is very convenient to represent the \MSR\ equations using diagrams.
In particular, this helps with the identification of the system's symmetries and
the design of efficient numerical implementations.
We choose the following diagrammatic representation for the bare/dressed two-point propagator
and three-point vertex
\begin{equation}
\begin{aligned}
g_{\1\2} &=
\begin{tikzpicture}
  \draw[double wavy] (0,0) -- (1,0)
    node[pos=0, above=-0.5pt] {\scalebox{0.8}{$\1$}}
    node[pos=1, above=-0.5pt] {\scalebox{0.8}{$\2$}};
\end{tikzpicture}
\, ;
\quad
& G_{\1\2} &= 
\begin{tikzpicture}
\begin{feynman}
\draw[double, thick] (0,0) -- (1,0)
node[pos=0, above=-0.025] {$\scalebox{0.8}{$\1$}$}
node[pos=1, above=-0.025] {$\scalebox{0.8}{$\2$}$};
\end{feynman}
\end{tikzpicture} \, ;
\\
\gamma_{\1\2\3} &= \;
\begin{tikzpicture}[baseline=(current bounding box.center)]
  \begin{feynman}
    \vertex [dot, minimum size=8pt, line width=1pt] (v) at (2.3,0) {};
    \vertex (a) at (1.75,0);
    \vertex (b) at (2.75,0.5);
    \vertex (c) at (2.75,-0.5);
    \draw[plain, thick] (a) -- (v) node[pos=0.3, above=-0.05]  {$\scalebox{0.8}{$\1$}$};
    \draw[plain, thick] (v) -- (b) node[pos=0.45, above] {$\scalebox{0.8}{$\2$}$};
    \draw[plain, thick] (v) -- (c) node[pos=0.35, below=-0.025] {$\scalebox{0.8}{$\3$}$};
  \end{feynman}
\end{tikzpicture} \, ;
\quad
& \Gamma_{\1\2\3} &= \;
\begin{tikzpicture}[baseline=(current bounding box.center)]
\begin{feynman}
\vertex [blob, minimum size=12pt, line width=1pt] (v) at (2.3,0) {};
\vertex (a) at (1.75,0);
\vertex (b) at (2.75,0.5);
\vertex (c) at (2.75,-0.5);
\draw[plain, thick] (a) -- (v) node[pos=0.3, above=-0.05]  {$\scalebox{0.8}{$\1$}$};
\draw[plain, thick] (v) -- (b) node[pos=0.45, above] {$\scalebox{0.8}{$\2$}$};
\draw[plain, thick] (v) -- (c) node[pos=0.35, below=-0.025] {$\scalebox{0.8}{$\3$}$};
\end{feynman}
\end{tikzpicture} \, ,
\end{aligned}
\label{diag:full}
\end{equation}
where the external legs are associated with coordinates, i.e.\@, ${ \1 \!=\! (\eps , 1) \!=\! (\eps , t , a) \!=\! (\eps , t , \ell , m) }$.

With this convention, the self-energy from Eq.~\eqref{eq:Sigma} can be represented as
\begin{equation}
\begin{tikzpicture}[baseline=(current bounding box.center)]
\node at (-1,0) {$\Sigma_{\bm{1}\bm{2}}$};
\node at (-0.5,0) {$=$};
\node at (0.0,0) {$\frac{1}{2}$};
\begin{feynman}
\vertex (a) at (0.25,0);
\vertex (b) at (1,0);
\vertex (d) at (3,0);
\vertex [dot, minimum size=8pt, line width=1pt] (v1) at (0.75,0) {};
\vertex [blob, minimum size=12pt, line width=1pt] (v2) at (2.5,0) {};
\draw[plain, thick] (a) -- (v1) node[pos=0.3, above=-0.05]  {$\scalebox{0.8}{$\1$}$} 
node[pos=2, above=0.2]  {$\scalebox{0.8}{$\3$}$} node[pos=2, above=-0.65]  {$\scalebox{0.8}{$\4$}$};
\draw[plain, thick] (d) -- (v2) node[pos=0.3, above=-0.05]  {$\scalebox{0.8}{$\2$}$}
node[pos=2.6, above=0.2]  {$\scalebox{0.8}{$\3'$}$} node[pos=2.6, above=-0.65]  {$\scalebox{0.8}{$\4'$}$};
\diagram* {
(v1) -- [double, thick, quarter left, looseness=1]  (v2) ,
(v1) -- [double, thick, quarter right, looseness=1] (v2)
};
\end{feynman}
\end{tikzpicture} \, .
\label{diag:Sigma}
\end{equation}
Here, we explicitly wrote the internal coordinates associated with the internal sums in Eq.~\eqref{eq:Sigma}.
From now on, these coordinates will be omitted.

The three-point propagator from Eq.~\eqref{eq:def_Gamma_3_point} becomes
\begin{equation}
G_{\1\2\3} = \;
\begin{tikzpicture}[baseline=(current bounding box.center)]
\begin{feynman}
\vertex [blob, minimum size=12pt, line width=1pt] (v) at (2.3,0) {};
\vertex (a) at (1.5,0);
\vertex (b) at (3,0.65);
\vertex (c) at (3,-0.65);
\draw[plain, thick] (a) -- (v) node[pos=0.3, above=-0.05]  {$\scalebox{0.8}{$\1$}$};
\draw[plain, thick] (v) -- (b) node[pos=0.5, above] {$\scalebox{0.8}{$\2$}$};
\draw[plain, thick] (v) -- (c) node[pos=0.5, below] {$\scalebox{0.8}{$\3$}$};
\diagram* {
(a) -- [double, thick] (v) -- [double, thick] (b),
(v) -- [double, thick] (c)
};
\end{feynman}
\end{tikzpicture} \, .
\label{diag:G3}
\end{equation}
Similarly, we can write the Dyson equation (Eq.~\ref{eq:Dyson_inv}) as
\begin{equation}
\begin{tikzpicture}[baseline=(current bounding box.center)]
\draw[double, thick] (-2,0) -- (-1,0);
\node at (-1,0.25) {$\scalebox{0.7}{$-1$}$};
\node at (-0.5,0) {$=$};
\draw[double wavy] (0,0) -- (1,0);
\node at (1,0.25) {$\scalebox{0.7}{$-1$}$};
\node at (1.5,0) {$-$};
\node at (2.0,0) {$\frac{1}{2}$};
\begin{feynman}
\vertex (a) at (2.25,0);
\vertex (b) at (3,0);
\vertex (d) at (5,0);
\vertex [dot, minimum size=8pt, line width=1pt] (v1) at (2.75,0) {};
\vertex [blob, minimum size=12pt, line width=1pt] (v2) at (4.5,0) {};
\diagram* {
(a) -- [plain, thick] (v1),
(d) -- [plain, thick] (v2),
(v1) -- [double, thick, quarter left, looseness=1] (v2) ,
(v1) -- [double, thick, quarter right, looseness=1] (v2)
};
\end{feynman}
\end{tikzpicture} \, ,
\label{diag:Dyson}
\end{equation}
where, here, we also omitted the writing of the external coordinates.
Thus, conveniently, the diagrammatic representation 
allows one to drop the explicit writing of all coordinates,
while retaining the whole complexity of the involved equations.

\subsection{Closures}
\label{sec:Closures}

Determining the renormalised vertex $\Gamma$ effectively closes
the Dyson equation~\eqref{diag:Dyson}.
In turn, this allows for the determination of the two-point propagator, $G$,
and therefore the two-point correlation, $C$.
Thus, in the \MSR\ formalism,
$\Gamma$ becomes the central quantity one should self-consistently determine
to characterise the statistics of the system's fluctuations.
Phrased differently, in the \MSR\ scheme,
the approximation---hence, closure---is performed directly on $\Gamma$.

\subsubsection{Bare order}
\label{sec:ZeroLoop}

At lowest order in the closure scheme,
following Eq.~\eqref{eq:func_Gamma},
the renormalised vertex $\Gamma$ is simply set to the bare vertex $\gamma$.
It reads
\begin{equation}
\begin{tikzpicture}[baseline=(current bounding box.center)]
\begin{feynman}
\vertex [blob, minimum size=12pt, line width=1pt] (v) at (0,0) {};
\vertex (a) at (-0.55,0);
\vertex (b) at (0.45,0.5);
\vertex (c) at (0.45,-0.5);
\diagram* {
(a) -- [plain, thick] (v) -- [plain, thick] (b),
(v) -- [plain, thick] (c)
};
\end{feynman}
\node at (0.8,0) {$=$};
\begin{feynman}
\vertex [dot, minimum size=8pt, line width=1pt] (v) at (1.8,0) {};
\vertex (a) at (1.25,0);
\vertex (b) at (2.25,0.5);
\vertex (c) at (2.25,-0.5);
\diagram* {
(a) -- [plain, thick] (v) -- [plain, thick] (b),
(v) -- [plain, thick] (c)
};
\end{feynman}
\end{tikzpicture} \, .
\label{diag:bare}
\end{equation}
This approximation yields the bare prediction
for the two-point correlation and response functions.
It coincides exactly with the \DIA\ equations~\citep{MSR1973}.
This bare prediction is the one that was investigated in detail
for \VRR\@ in~\citetalias{Flores+2025}.
Later in this paper, we will use these previous results
to validate our general implementation of the \MSR\ scheme.

At this stage, it is crucial to stress that, for \VRR\@,
the bare prediction is finite,
i.e.\ the predicted two-point correlation function obtained through Eq.~\eqref{diag:bare}
is bounded and well-behaved~\citepalias{Flores+2025}.
This is in sharp contrast with more traditional scenarios involving
charge renormalisation, where divergences
typically require further regularisation~\citep[see, e.g.\@][]{Peskin+1995,Lancaster+2014}.
As a result, in the case of isotropic \VRR\@,
going beyond bare order in the \MSR\ renormalisation scheme
is meant to provide controlled improvements
upon the already well-defined bare prediction.

\subsubsection{One-loop order}
\label{sec:One_loop}

The next-order closure in the \MSR\ scheme
accounts for the \textit{one-loop} contribution to the renormalised vertex.
One obtains from Eq.~\eqref{eq:func_Gamma}
the diagrammatic equation~\citep[see fig.~3 in][]{MSR1973} 
\begin{equation}
\begin{tikzpicture}[baseline=(current bounding box.center)]
\begin{feynman}
\vertex [blob, minimum size=12pt, line width=1pt] (v) at (0,0) {};
\vertex (a) at (-0.55,0);
\vertex (b) at (0.45,0.5);
\vertex (c) at (0.45,-0.5);
\diagram* {
(a) -- [plain, thick] (v) -- [plain, thick] (b),
(v) -- [plain, thick] (c)
};
\end{feynman}
\node at (0.8,0) {$=$};
\begin{feynman}
\vertex [dot, minimum size=8pt, line width=1pt] (v) at (1.8,0) {};
\vertex (a) at (1.25,0);
\vertex (b) at (2.25,0.5);
\vertex (c) at (2.25,-0.5);
\diagram* {
(a) -- [plain, thick] (v) -- [plain, thick] (b),
(v) -- [plain, thick] (c)
};
\end{feynman}
\node at (2.75,0) {$+$};
\begin{scope}[xshift=0.525\linewidth]
\begin{feynman}
\vertex [blob, minimum size=12pt, line width=1pt] (v1) at (-0.75,0) {};
\vertex [blob, minimum size=12pt, line width=1pt] (v2) at (0.3,0.6) {};
\vertex [blob, minimum size=12pt, line width=1pt] (v3) at (0.3,-0.6) {};
\vertex (in) at (-1.3,0);
\vertex (out1) at (0.8,0.9);
\vertex (out2) at (0.8,-0.9);
\diagram* {
(in) -- [plain, thick] (v1),
(v1) -- [double, thick] (v2),
(v1) -- [double, thick] (v3),
(v2) -- [double, thick] (v3),
(v2) -- [plain, thick] (out1),
(v3) -- [plain, thick] (out2)
};
\end{feynman}
\end{scope}
\end{tikzpicture} \, .
\label{diag:one_loop}
\end{equation}
In Appendix~\ref{app:Expansions}, we provide a heuristic derivation of Eq.~\eqref{diag:one_loop}.
We stress that Eq.~\eqref{diag:one_loop}
is an intricate self-consistent relation for $\Gamma$.
Phrased differently, the correction to $\gamma$ on the right-hand side
of Eq.~\eqref{diag:one_loop} involves \textit{powers} of $\Gamma$ itself~\cite{MSR1973}.
In Section~\ref{sec:Naive}, we will illustrate explicitly how the \textit{naive} approach
of expanding the right-hand side of Eq.~\eqref{diag:one_loop}
as a function of $\gamma$, rather than $\Gamma$, leads to ill-behaved predictions.

We also point out that Eq.~\eqref{diag:one_loop} involves
the two-point propagator $G$ as well.
Similarly, the Dyson equation~\eqref{diag:Dyson}
for $G$ involves $\Gamma$ and $G$ itself.
As such, at one-loop order,
one obtains from Eqs.~\eqref{diag:Dyson} and~\eqref{diag:one_loop}
a set of non-linear self-consistent equations involving both $G$ and $\Gamma$.
We postpone to Section~\ref{sec:FixedPoint} the description
of an effective approach to find solutions to this set of equations.
The present work is dedicated to solving the one-loop prediction from Eq.~\eqref{diag:one_loop}
for isotropic \VRR\@,
paying particular attention to the strengths and limitations
of the \MSR\ renormalisation scheme in that context.

\subsubsection{Two-loop order}
\label{sec:Two_loop}

Of course, one can build an even higher-order closure
from Eq.~\eqref{eq:func_Gamma}.
At two-loop order, the self-consistent equation satisfied
by $\Gamma$ becomes~\citep[see fig.~9 in][]{MSR1973}
\begin{equation}
\begin{tikzpicture}[baseline=(current bounding box.center)]
\begin{feynman}
\vertex [blob, minimum size=12pt, line width=1pt] (v) at (0,0) {};
\vertex (a) at (-0.55,0);
\vertex (b) at (0.45,0.5);
\vertex (c) at (0.45,-0.5);
\diagram* {
(a) -- [plain, thick] (v) -- [plain, thick] (b),
(v) -- [plain, thick] (c)
};
\end{feynman}
\node at (0.75,0) {$=$};
\begin{feynman}
\vertex [dot, minimum size=8pt, line width=1pt] (v) at (1.75,0) {};
\vertex (a) at (1.2,0);
\vertex (b) at (2.25,0.5);
\vertex (c) at (2.25,-0.5);
\diagram* {
(a) -- [plain, thick] (v) -- [plain, thick] (b),
(v) -- [plain, thick] (c)
};
\end{feynman}
\node at (2.4,0) {$+$};
\begin{scope}[xshift=0.375\linewidth]
\begin{feynman}
\vertex [blob, minimum size=12pt, line width=1pt] (v1) at (0,0) {};
\vertex [blob, minimum size=12pt, line width=1pt] (v2) at (1.05,0.6) {};
\vertex [blob, minimum size=12pt, line width=1pt] (v3) at (1.05,-0.6) {};
\vertex (in) at (-0.55,0);
\vertex (out1) at (1.55,0.9);
\vertex (out2) at (1.55,-0.9);
\diagram* {
(in) -- [plain, thick] (v1),
(v1) -- [double, thick] (v2),
(v1) -- [double, thick] (v3),
(v2) -- [double, thick] (v3),
(v2) -- [plain, thick] (out1),
(v3) -- [plain, thick] (out2)
};
\end{feynman}
\end{scope}
\node at (4.8,0) {$+ \, \half$};
\begin{scope}[xshift=0.315\linewidth]
\begin{feynman}
\vertex [blob, minimum size=12pt, line width=1pt] (v1) at (3,0) {};
\vertex [blob, minimum size=12pt, line width=1pt] (v2) at (3.75,0.6) {};
\vertex [blob, minimum size=12pt, line width=1pt] (v3) at (3.75,-0.6) {};
\vertex [blob, minimum size=12pt, line width=1pt] (v4) at (4.75,0.6) {};
\vertex [blob, minimum size=12pt, line width=1pt] (v5) at (4.75,-0.6) {};
\vertex (in) at (2.45,0);
\vertex (out1) at (5.25,0.6);
\vertex (out2) at (5.25,-0.6);
\diagram* {
(in) -- [plain, thick] (v1),
(v1) -- [double, thick] (v2),
(v1) -- [double, thick] (v3),
(v3) -- [double, thick] (v4),
(v2) -- [double, thick] (v5),
(v2) -- [double, thick] (v4),
(v3) -- [double, thick] (v5),
(v4) -- [plain, thick] (out1),
(v5) -- [plain, thick] (out2)
};
\end{feynman}
\end{scope}
\end{tikzpicture} .
\label{diag:two_loop}
\end{equation}
We refer to Appendix~\ref{app:Expansions}
for a heuristic derivation of Eq.~\eqref{diag:two_loop}.
Remarkably, Eq.~\eqref{eq:func_Gamma} only generates irreducible interaction diagrams.
Albeit more intricate than Eq.~\eqref{diag:one_loop},
formally, the two-loop closure from Eq.~\eqref{diag:two_loop},
when complemented with the Dyson equation (Eq.~\ref{diag:Dyson}),
also provides us with a set of non-linear and non-local equations
that must be satisfied jointly and self-consistently
by $G$ and $\Gamma$. Hence, the essence of the renormalisation procedure remains the same as the one at one-loop order.

\section{Numerical Scheme}
\label{sec:Numerical}

In this section, we detail our numerical scheme
to obtain the one-loop \MSR\ predictions
for the correlation functions of isotropic \VRR\@.
This section is meant to be concise:
more details are provided in Appendix~\ref{app:Numerical}.

\subsection{Fixed-point search}
\label{sec:FixedPoint}

At one-loop order, the two-point propagator, $G$,
and the three-point renormalised vertex, $\Gamma$,
are coupled through the Dyson equation~\eqref{diag:Dyson}
and the expansion from Eq.~\eqref{diag:one_loop}.
Together, these form a closed system of equations 
whose solution is required to predict the correlation functions of interest.
Both equations are highly non-local and non-linear,
e.g.\@, through the inverse propagator in Eq.~\eqref{diag:Dyson}
and the one-loop diagram in Eq.~\eqref{diag:one_loop}.
They are also high-dimensional, since we work in the extended coordinate space ${ \1 \!=\! (\eps , t , \ell , m) }$.
In the diagrams, each connected double line between two vertices
implies summation over a pair of these coordinates.
This complexity rules out analytical solutions, making numerical methods necessary.

\textit{Fixed-point}. Because the pair ${[G, \Gamma]}$ depends on itself through Eqs.~\eqref{diag:Dyson} and~\eqref{diag:one_loop}, 
we adopt a fixed-point approach to solve the system.
More precisely, given some state ${ [G^{(i)} , \Gamma^{(i)}] }$,
we may iterate once by evaluating Eqs.~\eqref{diag:Dyson} and~\eqref{diag:one_loop}
to obtain an updated state ${ [G^{(i+1)} , \Gamma^{(i+1)}] }$.
Schematically, this can be illustrated as 
\begin{samepage} \begin{subequations}
\begin{equation}
\begin{tikzpicture}[baseline=(current bounding box.center)]
\begin{feynman}
\vertex [blob, minimum size=12pt, line width=1pt, label=above:{$\scalebox{0.8}{$(i\!+\!1)$}$}] (v) at (0,0) {};
\vertex (a) at (-0.55,0);
\vertex (b) at (0.45,0.5);
\vertex (c) at (0.45,-0.5);
\diagram* {
(a) -- [plain, thick] (v) -- [plain, thick] (b),
(v) -- [plain, thick] (c)
};
\end{feynman}
\node at (0.8,0) {$=$};
\begin{feynman}
\vertex [dot, minimum size=8pt, line width=1pt] (v) at (1.8,0) {};
\vertex (a) at (1.25,0);
\vertex (b) at (2.25,0.5);
\vertex (c) at (2.25,-0.5);
\diagram* {
(a) -- [plain, thick] (v) -- [plain, thick] (b),
(v) -- [plain, thick] (c)
};
\end{feynman}
\node at (2.75,0) {$+$};
\begin{scope}[xshift=0.525\linewidth]
\begin{feynman}
\vertex [blob, minimum size=12pt, line width=1pt, label=above:{$\scalebox{0.8}{$(i)$}$}] (v1) at (-0.75,0) {};
\vertex [blob, minimum size=12pt, line width=1pt, label=above:{$\scalebox{0.8}{$(i)$}$}] (v2) at (0.3,0.6) {};
\vertex [blob, minimum size=12pt, line width=1pt, label=below:{$\scalebox{0.8}{$(i)$}$}] (v3) at (0.3,-0.6) {};
\vertex (in) at (-1.3,0);
\vertex (out1) at (0.8,0.9);
\vertex (out2) at (0.8,-0.9);
\diagram* {
(in) -- [plain, thick] (v1),
(v2) -- [plain, thick] (out1),
(v3) -- [plain, thick] (out2)
};
\draw[double, thick] (v1) to node[midway, above] {\scalebox{0.8}{$(i)$}} (v2);
\draw[double, thick] (v1) to node[midway, below] {\scalebox{0.8}{$(i)$}} (v3);
\draw[double, thick] (v2) to node[midway, right] {\scalebox{0.8}{$(i)$}} (v3);
\end{feynman}
\end{scope}
\end{tikzpicture}\, ,
\label{diag:one_loop_iter}
\end{equation}
\begin{equation}
\begin{tikzpicture}[baseline=(current bounding box.center)]
\draw[double, thick] (-2,0) to node[midway, below] {\scalebox{0.8}{$(i\!+\!1)$}} (-1,0);
\node at (-1,0.25) {$\scalebox{0.7}{$-1$}$};
\node at (-0.5,0) {$=$};
\draw[double wavy] (0,0) -- (1,0);
\node at (1,0.25) {$\scalebox{0.7}{$-1$}$};
\node at (1.5,0) {$-$};
\node at (2.0,0) {$\frac{1}{2}$};
\begin{feynman}
\vertex (a) at (2.25,0);
\vertex (b) at (3,0);
\vertex (d) at (5,0);
\vertex [dot, minimum size=8pt, line width=1pt] (v1) at (2.75,0) {};
\vertex [blob, minimum size=12pt, line width=1pt, label=below:{$\scalebox{0.8}{$(i)$}$}] (v2) at (4.5,0) {};
\diagram* {
(a) -- [plain, thick] (v1),
(d) -- [plain, thick] (v2)
};
\draw[double, thick, quarter left, looseness=1] (v1) to node[midway, above] {\scalebox{0.8}{$(i)$}} (v2);
\draw[double, thick, quarter right, looseness=1] (v1) to node[midway, below] {\scalebox{0.8}{$(i)$}} (v2);
\end{feynman}
\end{tikzpicture}\, .
\label{diag:Dyson_iter}
\end{equation}
\label{diag:Iter}\end{subequations}
\end{samepage}The scheme is iterated until convergence,
i.e.\ until reaching some iteration $i$ where ${ [G^{(i)} , \Gamma^{(i)}] }$ and ${ [G^{(i+1)} , \Gamma^{(i+1)}] }$
are sufficiently close, with a stopping criterion illustrated in Fig.~\ref{fig:CV_iter}.
At this stage, a unique, well-behaved fixed-point solution for ${[G,\Gamma]}$ in the \MSR\ scheme is not guaranteed. 
Iterations may fail to converge or yield unphysical solutions.

\textit{Initial conditions}. We must now specify the initial conditions of the scheme, ${ [G^{(0)} , \Gamma^{(0)}] }$.
In order to accelerate convergence, it is useful
to consider initial conditions that are in the vicinity
of the final fixed-point. 
A natural choice is to impose ${ [G^{(0)} , \Gamma^{(0)}] \!=\! [g , \gamma] }$,
i.e.\ initiate the fixed-point search with the bare vertex and propagator
(see Appendix~\ref{app:Bare}).
In practice, convergence is significantly hastened
by initialising the two-point propagator with ${G^{(0)} \!=\! G_{\rG}}$,
where $G_{\rG}$ is a propagator with a Gaussian time-dependence
inferred from the system's coherence time $\Tc$ (see Appendix~\ref{app:IC_TC}).
For the vertex, we always initialise it to ${ \Gamma^{(0)} \!=\! \gamma }$,
the bare vertex, i.e.\ it is initialised to a value determined entirely by the equations of motion.
For the bare order closure, we checked explicitly
that both choices of initial conditions yield the same final prediction.
In contrast, for the one-loop closure,
starting with ${ G^{(0)} \!=\! g }$ leads to a slow and marginally stable convergence,
while choosing ${ G^{(0)} \!=\! G_{\rG} }$ performs much better,
offering a fast and stable convergence.
By the end of the fixed-point search, both $\Gamma$ and $G$ deviate from their
initial conditions, i.e.\ they become \textit{dressed}.
Let us now make a few more remarks
on the present scheme.

\textit{Generic approach}. Although we focused on the one-loop closure
of Eq.~\eqref{diag:one_loop},
we note that the fixed-point approach from Eq.~\eqref{diag:Iter}
is generic to other truncation orders.
At bare order (Eq.~\ref{diag:bare}),
since one has ${ \Gamma \!=\! \gamma }$,
one may limit the search to iterating upon $G$ only.
At two-loop order (Eq.~\ref{diag:two_loop}),
one would use the exact same scheme as in Eq.~\eqref{diag:Iter},
while considering a more intricate self-consistent relation
for $\Gamma$.

\textit{Convergence}. The fixed-point algorithm from Eq.~\eqref{diag:Iter}
provides a simple yet robust approach
for achieving convergence in ${[G , \Gamma]}$.
It is a direct implementation that exhibits good stability, 
whereas more sophisticated methods may fail. 
To further accelerate convergence, one could consider complementing it
with information from the gradients with respect to ${[G , \Gamma]}$.
These gradients could be computed analytically
or by finite-differences, and be used to (hopefully) accelerate
the convergence of the iterations,
e.g.\@, through a Newton--Raphson scheme. 
In practice, given the high-dimensional nature of the present problem,
this is no easy endeavour. We did not explore it further in the present work.

\textit{Joint updates}. In Eq.~\eqref{diag:Iter},
we update simultaneously $G$ and $\Gamma$.
In the view of (possibly) accelerating convergence,
and limiting the costly evaluation of the one-loop diagram from Eq.~\eqref{diag:one_loop_iter},
one could search, in turn, for a fixed-point at constant $G$,
or at constant $\Gamma$.
As such, one could iterate, at a given time,
over only one of the two equations in Eq.~\eqref{diag:Iter}.
In practice, since we successfully managed to get
the simultaneous updates from Eq.~\eqref{diag:Iter} to appropriately converge,
we did not explore this aspect further.

\subsection{Symmetries}
\label{sec:Symmetries}

In order to alleviate the numerical complexity of the evaluation
of Eq.~\eqref{diag:Iter},
it is essential to leverage the symmetries of isotropic \VRR\ as efficiently as possible.
Here, we briefly summarise the key symmetries,
and refer to Appendix~\ref{app:Symmetries} for many more details.

\subsubsection{Isotropy}
\label{sec:Isotropy}

The \MSR\ scheme 
preserves isotropy under renormalisation.
Isotropy allows one to recast the two-point propagator as
\begin{equation}
G_{\1\2} = \delta_{a_{1}}^{a_{2}} \, G^{L}_{\ell_{1}} (\xi_{1} , \xi_{2}) ,
\label{eq:iso_G}
\end{equation}
with the extended coordinates ${ \1 \!=\! (\eps , t , a)}$, where ${a\!=\!(\ell,m)}$
and ${ \xi \!=\! (\eps , t) }$.
Similarly, the renormalised vertex satisfies
\begin{equation}
\Gamma_{\1\2\3} = E^{\Delta M}_{a_1 a_2 a_3} \, \Gamma^{L}_{\ell_1 \ell_2 \ell_3} ( \xi_{1} , \xi_{2} , \xi_{3}) .
\label{eq:iso_Gamma}
\end{equation}
In that expression, we used the excluded Elsasser coefficients,
${ E^{\Delta M} }$ (Eq.~\ref{eq:prop_delta_EL}).
Hence, the closure reduces to solving Eq.~\eqref{diag:Iter} for ${[G^{L}, \Gamma^{L}]}$.
Importantly, none of these functions involve any dependence on the indices $m$.
Finally, following Eq.~\eqref{eq:iso_Gamma}, we only need to keep \textit{triangles}, ${(\ell_{1}, \ell_{2}, \ell_{3})}$,
such that ${ E^{\Delta M}_{a_{1} a_{2} a_{3}} \!\neq\! 0 }$.

\subsubsection{Time stationarity}
\label{sec:TimeStationarity}
Similarly to isotropy, time-stationarity is maintained under renormalisation.
Thus,
for any time $T$, the two-point propagator satisfies
\begin{equation}
G^{L} (t_{1} \!+\! T , t_{2} \!+\! T) = G^{L} (t_{1} , t_{2}) ,
\label{eq:stat_G}
\end{equation}
where, for clarity, we display only the time dependence in $G^L$.
In the same way,
the renormalised vertex satisfies 
\begin{equation}
\Gamma^{L} (t_{1} \!+\! T , t_{2} \!+\! T , t_{3} \!+\! T) = \Gamma^{L} (t_{1} , t_{2} , t_{3}) .
\label{eq:stat_Gamma}
\end{equation}
Overall, time stationarity greatly reduces the memory imprint
of the numerical implementation.

\subsubsection{Fluctuation-dissipation theorem}
\label{sec:FDT}

Since isotropic \VRR\ is a thermodynamical equilibrium~\citep{Roupas+2017,Fouvry+2019,Magnan+2022}, 
the \FDT\ holds~\citep{Kraichnan1959a,Krommes2002},
hence relating correlation (Eq.~\ref{eq:def_C_generic}) and response functions (Eq.~\ref{eq:def_R}).
More precisely, for isotropic \VRR\ with a single population,
we generically have
${ C_{12} \!=\! \delta_{a_{1}}^{a_{2}} \, C_{\ell_{1}} (t_{1} \!-\! t_{2}) }$
and ${ R_{12} \!=\! \delta_{a_{1}}^{a_{2}} \, R_{\ell_{1}} (t_{1} \!-\! t_{2}) }$,
verifying the identity
\begin{equation}
C_{\ell} (t) = C_{0} \, R_{\ell} (|t|) ,
\label{eq:FDT}
\end{equation}
with ${R_{\ell}(t\!<\!0)\!=\!0}$ and ${R_{\ell}(0)\!=\!1}$.
Here, we introduced 
\begin{equation}
C_{0} = N/(4 \pi) ,
\label{eq:def_C0}
\end{equation}
which stems from the initial conditions
(i.e.\ at zero time difference),
and is derived in Appendix~\ref{app:IC_2pt}.

Hence, $G^{L}$ is fully determined from the isotropic response $R_{\ell}$, i.e.\ its ${(+-)}$ component (see Eq.~\ref{eq:components_G}). In~\citetalias{Flores+2025},
we proved that the \DIA\ (bare) prediction from Section~\ref{sec:ZeroLoop}
exactly satisfies the \FDT\@.

\subsubsection{Time and harmonic discretisation}
\label{sec:Discretisation_time_harm}

To evaluate the diagrams in Eq.~\eqref{diag:Iter}, we discretise both time and harmonic indices.
The response function $R_{\ell}$ is stored via the vector
\begin{equation}
\bR_{\ell} = \big[ R_{\ell} (0) , R_{\ell} (\DT) , ... , R_{\ell} (\DT\!\!\times\!\!\NSTEPS) \big] ,
\label{eq:def_vec_R}
\end{equation}
with the timestep $\DT$ and the cutoff time ${\TMAX\!=\!\DT\!\!\times\!\!\NSTEPS}$,
beyond which we set ${R_{\ell}\!=\!0}$.
Harmonic sums are truncated at ${\ell\leq\LMAX}$, assuming ${R_{\ell}\!=\!0}$ for any ${\ell\!>\!\LMAX}$.
We proceed similarly for the renormalised vertex, ${ \Gamma^{L}_{\ell_{1} \ell_{2} \ell_{3}} (t_{1} , t_{2} , t_{3}) }$.
It is set to zero whenever the separation between any two times exceeds $\TMAX$,
and is restricted to ${0\!\leq\! \ell_{1} , \ell_{2} , \ell_{3} \!\leq\! \LMAX  }$.
Since we restrict the \VRR\ model to an  ${\ell\!=\!2}$ interaction (Section~\ref{sec:VRR_dyn}), $\LMAX$ does not need to be very large to describe small-$\ell$ correlations.
The case of a multipole interaction spectrum
is studied in Appendix~\ref{app:InteractionSpectrum}.
We verify the convergence with respect to $\DT$, $\LMAX$, and $\NSTEPS$
in Appendix~\ref{app:CVparam}.

\subsubsection{Antisymmetry}
\label{sec:Antisymmetry}

The bare vertex $\gamma_{\1\2\3}$ is fully symmetric (Appendix~\ref{app:Bare_gamma}), and so is $\Gamma_{\1\2\3}$.
Using isotropy (Eq.~\ref{eq:iso_Gamma}) and the antisymmetry of $E^{\Delta M}$ (Eq.~\ref{eq:def_excl_E}),
${\Gamma^{L}_{\ell_{1}\ell_{2}\ell_{3}}(\xi_{1},\xi_{2},\xi_{3})}$ is antisymmetric under any exchange ${ (\ell_i,\xi_{i}) \!\leftrightarrow\! (\ell_j,\xi_{j}) }$.
This property allows one to sort the harmonic indices ${(\ell_{1},\ell_{2},\ell_{3})}$,
and reduce the total number of triangles that need to be stored.
This is detailed in Appendix~\ref{app:Symmetries}.

\subsection{Iteration}
\label{sec:Iteration}

We are now set to implement numerically
the \MSR\ iteration.
Given some state ${ [G^{(i)} , \Gamma^{(i)}] }$,
our goal is to compute a new state,
${ [G^{(i+1)} , \Gamma^{(i+1)}] }$,
following Eq.~\eqref{diag:Iter}.
To do so,
we need to perform three consecutive operations, namely
(i) computing the new dressed vertex, $\Gamma^{(i+1)}$,
following Eq.~\eqref{diag:one_loop_iter};
(ii) computing the new inverse propagator, ${ [G^{(i+1)}]^{-1} }$,
following Eq.~\eqref{diag:Dyson_iter};
(iii) inverting the propagator to obtain ${ G^{(i+1)} }$.
Focusing on the one-loop closure from Eq.~\eqref{diag:Iter},
we now briefly detail the numerical complexity of each of these steps.
We emphasise, in particular, venues to accelerate
these evaluations by introducing appropriate reduced tensors.
Further details on this section are provided in Appendix~\ref{app:Iteration}.

\subsubsection{Updating the dressed vertex}
\label{sec:UpdateGamma}

In order to compute the right-hand side of Eq.~\eqref{diag:one_loop_iter},
the most challenging part is to compute its one-loop diagram.
Indeed, it must be evaluated for all possible values
of the external coordinates, or \textit{legs},
i.e.\ all values of ${ \1 \!=\! (\eps , t , \ell , m) }$,
while summing over all the internal ones.

Owing to the symmetries from Section~\ref{sec:Symmetries},
computing directly all the needed one-loop diagrams in Eq.~\eqref{diag:one_loop_iter}
would require ${ \mO (\NSTEPS^{8} \, \LMAX^{6}) }$ operations (Appendix~\ref{app:UpdateGamma}).
This is a prohibitive scaling.
To further leverage symmetries, let us introduce the tensor
\begin{equation}
\Lambda_{\1\2\3} = \Gamma_{\1\2\3'} G_{\3'\3} ,
\label{eq:def_Lambda}
\end{equation}
which is diagrammatically given by
\begin{equation}
\begin{aligned}
\Lambda_{\1\2\3} &\; = \;
\begin{tikzpicture}[baseline=(current bounding box.center)]
\begin{feynman}
\vertex [blob, minimum size=12pt, line width=1pt] (v) at (2.3,0) {};
\vertex (a) at (1.75,0);
\vertex (b) at (2.75,0.5);
\vertex (c) at (2.75,-0.5);
\draw[plain, thick] (a) -- (v) node[pos=0.3, above=-0.05]  {$\scalebox{0.8}{$\1$}$};
\draw[plain, thick] (v) -- (b) node[pos=0.45, above] {$\scalebox{0.8}{$\2$}$};
\draw[double, thick] (v) -- (c) node[pos=0.35, below=-0.025] {$\scalebox{0.8}{$\3$}$};
\end{feynman}
\end{tikzpicture} 
\; = \; 
\begin{tikzpicture}[baseline=(current bounding box.center)]
\begin{feynman}
\vertex [blob, minimum size=12pt, line width=1pt] (v) at (2.3,0) {};
\vertex [dot, minimum size=4pt, line width=1pt] (vdot) at (2.5,-0.225) {};
\vertex (a) at (1.75,0);
\vertex (b) at (2.75,0.5);
\vertex (c) at (2.75,-0.5);
\draw[plain, thick] (a) -- (v) node[pos=0.3, above=-0.05]  {$\scalebox{0.8}{$\1$}$};
\draw[plain, thick] (v) -- (b) node[pos=0.45, above] {$\scalebox{0.8}{$\2$}$};
\draw[plain, thick] (v) -- (c) node[pos=0.35, below=-0.025] {$\scalebox{0.8}{$\3$}$};
\end{feynman}
\end{tikzpicture}\, .
\end{aligned}
\label{diag:def_Lambda}
\end{equation}
The second diagram in Eq.~\eqref{diag:def_Lambda} defines
the diagrammatic representation of $\Lambda$.
The black dot on its third leg marks the coordinate contracted with $G$, 
and indicates that the three external legs are not interchangeable.
Just like $\Gamma_{\1\2\3}$, $\Lambda_{\1\2\3}$ satisfies isotropy (Eq.~\ref{eq:iso_Gamma}) and time-stationarity (Eq.~\ref{eq:stat_Gamma}),
but is symmetric only in its first two coordinates.

Armed with this new tensor, we can rewrite the one-loop diagram
from Eq.~\eqref{diag:one_loop} into
\begin{equation}
\begin{aligned}
\begin{tikzpicture}[baseline=(current bounding box.center)]
\begin{feynman}
\vertex [blob, minimum size=12pt, line width=1pt] (v1) at (-0.75,0) {};
\vertex [blob, minimum size=12pt, line width=1pt] (v2) at (0.3,0.6) {};
\vertex [blob, minimum size=12pt, line width=1pt] (v3) at (0.3,-0.6) {};
\vertex (in) at (-1.3,0);
\vertex (out1) at (0.8,0.9);
\vertex (out2) at (0.8,-0.9);
\diagram* {
(in) -- [plain, thick] (v1),
(v1) -- [double, thick] (v2),
(v1) -- [double, thick] (v3),
(v2) -- [double, thick] (v3),
(v2) -- [plain, thick] (out1),
(v3) -- [plain, thick] (out2)
};
\end{feynman}
\end{tikzpicture}
= \,\,
\begin{tikzpicture}[baseline=(current bounding box.center)]
\begin{feynman}
\vertex [blob, minimum size=12pt, line width=1pt] (v1) at (-0.75,0) {};
\vertex [dot, minimum size=4pt, line width=1pt] (vdot1) at (-0.5,0.15) {};
\vertex [blob, minimum size=12pt, line width=1pt] (v2) at (0.3,0.6) {};
\vertex [dot, minimum size=4pt, line width=1pt] (vdot2) at (0.3,0.3) {};
\vertex [blob, minimum size=12pt, line width=1pt] (v3) at (0.3,-0.6) {};
\vertex [dot, minimum size=4pt, line width=1pt] (vdot3) at (0.04,-0.45) {};
\vertex (in) at (-1.3,0);
\vertex (out1) at (0.8,0.9);
\vertex (out2) at (0.8,-0.9);
\diagram* {
(in) -- [plain, thick] (v1),
(v1) -- [plain, thick] (v2),
(v1) -- [plain, thick] (v3),
(v2) -- [plain, thick] (v3),
(v2) -- [plain, thick] (out1),
(v3) -- [plain, thick] (out2)
};
\end{feynman}
\end{tikzpicture}\, .
\end{aligned}
\label{diag:one_loop_with_Lambda}
\end{equation}
To perform the iteration from Eq.~\eqref{diag:one_loop_iter},
we can now proceed in two successive steps:
(i) compute and store $\Lambda$ (Eq.~\ref{diag:def_Lambda});
(ii) compute the updated one-loop diagram involving $\Lambda$ (Eq.~\ref{diag:one_loop_with_Lambda}).
Step (ii) is the most numerically expensive with a total complexity of ${\mO(\NSTEPS^{5} \, \LMAX^{6})}$ operations (Appendix~\ref{app:UpdateGamma}). 
This is a significant improvement over the direct evaluation of Eq.~\eqref{diag:one_loop_iter}. In practice, 
computing Eq.~\eqref{diag:one_loop_with_Lambda} remains the main bottleneck of our \MSR\ iteration scheme.

\subsubsection{Updating the inverse propagator}
\label{sec:UpdateGinv}

We can now turn our interest to updating the inverse two-point propagator,
as given by Eq.~\eqref{diag:Dyson_iter}.
This requires the computation of the loop diagram
associated with the self-energy (Eq.~\ref{diag:Sigma}).
Leveraging once again isotropy (Section~\ref{sec:Isotropy}) and
time-stationarity (Section~\ref{sec:TimeStationarity}),
computing directly the self-energy involves ${ \mO (\NSTEPS^{5} \, \LMAX^{3}) }$ operations.

Fortunately, introducing once again the tensor $\Lambda$ (Eq.~\ref{diag:def_Lambda})
with its bare equivalent, $\lambda$, defined as
\begin{equation}
\begin{aligned}
\lambda_{\1\2\3} & \; = \;
\begin{tikzpicture}[baseline=(current bounding box.center)]
\begin{feynman}
\vertex [dot, minimum size=8pt, line width=1pt] (v) at (2.3,0) {};
\vertex (a) at (1.75,0);
\vertex (b) at (2.75,0.5);
\vertex (c) at (2.75,-0.5);
\draw[plain, thick] (a) -- (v) node[pos=0.3, above=-0.05]  {$\scalebox{0.8}{$\1$}$};
\draw[plain, thick] (v) -- (b) node[pos=0.45, above] {$\scalebox{0.8}{$\2$}$};
\draw[double, thick] (v) -- (c) node[pos=0.35, below=-0.025] {$\scalebox{0.8}{$\3$}$};
\end{feynman}
\end{tikzpicture} 
\; = \;
\begin{tikzpicture}[baseline=(current bounding box.center)]
\begin{feynman}
\vertex [dot, minimum size=8pt, line width=1pt] (v) at (2.3,0) {};
\vertex [dot, minimum size=4pt, line width=1pt] (vdot) at (2.46,-0.18) {};
\vertex (a) at (1.75,0);
\vertex (b) at (2.75,0.5);
\vertex (c) at (2.75,-0.5);
\draw[plain, thick] (a) -- (v) node[pos=0.3, above=-0.05]  {$\scalebox{0.8}{$\1$}$};
\draw[plain, thick] (v) -- (b) node[pos=0.45, above] {$\scalebox{0.8}{$\2$}$};
\draw[plain, thick] (v) -- (c) node[pos=0.35, below=-0.025] {$\scalebox{0.8}{$\3$}$};
\end{feynman}
\end{tikzpicture}\, ,
\end{aligned}
\label{diag:def_lambda}
\end{equation}
one can rewrite the diagram of the self-energy from Eq.~\eqref{diag:Dyson_iter}
into
\begin{equation}
\begin{aligned}
\begin{tikzpicture}[baseline=(current bounding box.center)]
\begin{feynman}
\vertex (a) at (0.25,0);
\vertex (b) at (1,0);
\vertex (d) at (3,0);
\vertex [dot, minimum size=8pt, line width=1pt] (v1) at (0.75,0) {};
\vertex [blob, minimum size=12pt, line width=1pt] (v2) at (2.5,0) {};
\draw[plain, thick] (a) -- (v1) ;
\draw[plain, thick] (d) -- (v2) ;
\diagram* {
(v1) -- [double, thick, quarter left, looseness=1] (v2) ,
(v1) -- [double, thick, quarter right, looseness=1] (v2)
};
\end{feynman}
\end{tikzpicture}
\; = \;
\begin{tikzpicture}[baseline=(current bounding box.center)]
\begin{feynman}
\vertex (a) at (0.25,0);
\vertex (b) at (1,0);
\vertex (d) at (3,0);
\vertex [dot, minimum size=8pt, line width=1pt] (v1) at (0.75,0) {};
\vertex [dot, minimum size=4pt, line width=1pt] (vdot1) at (0.925,0.17) {};
\vertex [blob, minimum size=12pt, line width=1pt] (v2) at (2.5,0) {};
\vertex [dot, minimum size=4pt, line width=1pt] (vdot2) at (2.275,-0.2) {};
\draw[plain, thick] (a) -- (v1) ;
\draw[plain, thick] (d) -- (v2) ;
\diagram* {
(v1) -- [plain, thick, quarter left, looseness=1] (v2) ,
(v1) -- [plain, thick, quarter right, looseness=1] (v2)
};
\end{feynman}
\end{tikzpicture} \, .
\end{aligned}
\label{diag:Sigma_with_Lambda}
\end{equation}
As such, by reducing the number of internal sums,
the computation of the self-energy, hence of $G^{-1}$,
now requires ${ \mO (\NSTEPS^{3} \, \LMAX^{3}) }$.
This is largely sub-dominant compared to the computation
of the one-loop diagram in Section~\ref{sec:UpdateGamma}.

\subsubsection{Inverting the propagator}
\label{sec:InvertGinv}

We recall that the scheme of Eq.~\eqref{diag:Iter},
together with the isotropy property from Section~\ref{sec:Isotropy},
provides us with the isotropic renormalised vertex $\Gamma^L$,
and the \textit{inverse} isotropic two-point propagator, ${[G^L]^{-1}}$.
However, it still needs to be inverted to obtain $G^L$.
As mentioned in Section~\ref{sec:FDT},
the isotropic two-point propagator, $G^{L}$, can be
reconstructed from the isotropic response function, $R_\ell$.
As a result, when iterating upon Eq.~\eqref{diag:Dyson_iter},
we may limit ourselves to solely computing the $(-+)$ component
of $[G^{L}]^{-1}$.
The inversion from $R^{-1}_{\ell}$ to $R_{\ell}$
can then be efficiently obtained by direct substitution,
with a complexity in ${ \mO (\NSTEPS^{2} \, \LMAX) }$.
We detail the associated algorithm in Appendix~\ref{app:InvertGinv}.

All the various elements of Section~\ref{sec:Numerical}
have been implemented in an efficient \texttt{julia} code
which is publicly distributed~\citep{github_MSR}.

\section{Results}
\label{sec:Application}

We are now set to carry out the \MSR\ closure
for \VRR\ in the isotropic limit.
All the figures presented in this section
were obtained by implementing the \MSR\ iterative
one-loop renormalisation scheme from Eq.~\eqref{diag:Iter}.
In practice, the control parameters for the numerical implementation
were set to
${ \LMAX \!=\! 7 }$, i.e.\ the maximum harmonic number;
${ \DT \!= \Tc/80}$, i.e.\ the timestep of discretisation;
${ \TMAX \!=\! \Tc}$, i.e.\ the maximum depth in time;
${ \ITER \!=\! 10}$, i.e.\ the number of iterations of Eq.~\eqref{diag:Iter}.
Since the statistical properties of the isotropic \VRR\ system are invariant with respect to the total number of particles $N$
(see Section~\ref{sec:Invariance})
we do not need to specify its value to obtain the MSR predictions.

In Appendix~\ref{app:CVparam},
we check for the appropriate convergence
of our numerical scheme
with respect to each of these control parameters.
With the present choices, the total number of degrees of freedom
that are adjusted by the iterative renormalisation scheme
is ${ \!\sim\! 5.6\!\times\!10^2}$ for $G$
and ${ \!\sim\! 3.7\!\times\!10^6 }$ for $\Gamma$.
Performing one iteration of Eq.~\eqref{diag:Iter}
typically requires ${ \!\sim\! 62\,\mathrm{h}}$
of computation time on 128 cores.
Ensuring convergence of the \MSR\ renormalisation scheme
at one-loop order is a demanding numerical task.

Armed with these predictions,
we now focus
on the two-point correlation function (Section~\ref{sec:Prediction_2_point});
the renormalised interaction vertex (Section~\ref{sec:Prediction_Gamma});
the three-point correlation function, or skewness, (Section~\ref{sec:Prediction_3_point}).
We compare the \MSR\ predictions at bare and one-loop order
with direct measurements in $N$-body simulations.
We refer to Appendix~\ref{app:NumericalSimulations} for details
on the setup of the $N$-body simulations.

\subsection{Two-point correlation}
\label{sec:Prediction_2_point}

As stated in Section~\ref{sec:Isotropy},
for isotropic \VRR, the two-point correlation function from Eq.~\eqref{eq:def_C_generic}
only depends on $\ell$, the isotropic harmonic number.
We may write
\begin{equation}
C_{12} \!=\! \delta_{a_{1}}^{a_{2}} \, C_{\ell_{1}} (t_{1} \!-\! t_{2}),
\label{eq:def_iso_C}
\end{equation}
where ${a\!=\!(\ell,m)}$. Physically, $\ell$ scales like ${1/ \theta }$,
with $\theta$ the angular separation between two particles.
We thus focus on determining the isotropic correlation ${t \! \mapsto \! C_\ell(t)}$.

As visible in Fig.~\ref{fig:Sphere},
in \VRR\@, coherent structures emerge and disappear continuously
in phase space.
From $C_{\ell}$, one can extract the typical lifetimes
of such coherent \textit{vortices}, and that for every scale $\ell$.
Accurately predicting $C_{\ell}$
is therefore essential to characterise the spectrum
both in space and time of the fluctuations.

\begin{figure}[htbp!]
\begin{center}
\includegraphics[width=1.0\linewidth]{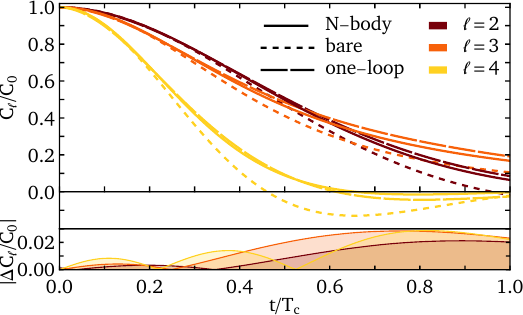}
\caption{Top panel: Normalised isotropic two-point correlation function, ${C_\ell}$,
as predicted by \MSR\@, at bare order (dotted lines -- Eq.~\ref{diag:bare})
and one-loop order (dashed lines -- Eq.~\ref{diag:one_loop}),
and measured in $N$-body simulations (full lines),
for various harmonics $\ell$ (different colors).
Bottom panel: Absolute error between the one-loop \MSR\ prediction
and the $N$-body measurements.
Here, time has been rescaled by the coherence time, $\Tc$ (Appendix~\ref{app:IC_TC}).
The \MSR\ one-loop prediction improves upon the bare prediction
in recovering the late-time exponential behavior of the correlation functions.
}
\label{fig:Money_plot}
\end{center}
\end{figure}

In Fig.~\ref{fig:Money_plot},
we illustrate the time-dependence of $C_\ell$,
for different scales $\ell$,
as a function of the rescaled time, ${ t / \Tc }$,
where $\Tc$ is the system's coherence time defined in Appendix~\ref{app:IC_TC}.
In that figure, the amplitude is also
rescaled with respect to the initial amplitude,
$C_{0}$, as introduced in Eq.~\eqref{eq:def_C0}.
In Fig.~\ref{fig:Money_plot},
we compare the predictions from the bare and one-loop \MSR\ scheme
with $N$-body measurements. Such measurements, and the errors associated, 
are detailed in Appendix~\ref{app:NumericalSimulations}. 
The two-point correlation functions measured in $N$-body simulations
are converged to better than 1\%.

In Fig.~\ref{fig:Money_plot},
we find that the one-loop \MSR\ prediction
is an ${80\%}$ improvement over the bare (\DIA\@) prediction,
first investigated in~\citetalias{Flores+2025}.
For ${ \ell \!=\! 4 }$,
we find that the maximum absolute error
of the bare prediction is ${ \sim\! 0.15 }$,
while it drops to ${ \sim\! 0.03 }$ for the one-loop prediction.
In particular, the one-loop closure captures much better
the late-time exponential behaviour
of the correlation functions,
i.e.\ the decay of the correlations
during the diffusive regime.
In addition, for ${ \ell \!=\! 4 }$,
the one-loop prediction correctly recovers
the oscillation of the correlation function
that predicts a negative correlation for ${ t / \Tc \!\gtrsim\! 0.6 }$.
The improvement from Fig.~\ref{fig:Money_plot}
is one of the key result of the present work.
It confirms that the \MSR\ formalism,
when pushed beyond bare order,
provides a convergent and accurate prediction
of the time evolution of the two-point correlation function, for isotropic \VRR.

\subsection{Renormalised interaction vertex}
\label{sec:Prediction_Gamma}

The \MSR\ formalism does not limit itself to 
predicting the two-point correlation function, $C_{\ell}$.
It also estimates the renormalised vertex, $\Gamma$,
i.e.\ the target of the statistical closure (Eq.~\ref{eq:func_Gamma}).
As such, by investigating in detail the structure of $\Gamma$,
one can assess further, a posteriori, the sanity of the \MSR\ prediction,
in particular by comparing the dressing of $\Gamma$ with respect to the bare vertex $\gamma$.

\begin{figure}[htbp!]
\centering
\includegraphics[width=1.0\linewidth]{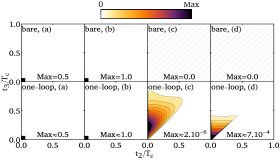}
\caption{ Top (resp.\ Bottom) panels: Dependence of the bare (resp.\ renormalised) vertex, ${ \gamma^{L}_{\1\2\3} (0 , t_{2} , t_{3}) }$ (resp.\ ${ \Gamma^{L}_{\1\2\3} (0 , t_{2} , t_{3}) }$),
for different triangle configurations (different panels)
as a function of the times ${ (t_{2} , t_{3}) }$,
following the specification from Table~\ref{tab:Table_triangles}.
The amplitude of each panel is normalised by the maximum amplitude of panel (b).
For each panel, we represent 11 contours
spaced linearly between the minimum and maximum amplitude
in the considered domain.
For triangles (a) and (b), the black square was added to emphasise
that these vertices are fully dominated by their amplitude at ${ t_{2} \!=\! t_{3} \!=\! 0 }$.
Dashed regions correspond to exactly vanishing bare vertices.
Interestingly, the fixed-point implementation of the \MSR\ renormalisation scheme
leads to very diverse behaviours of the renormalised vertex. 
}
\label{fig:Maps_Gamma}
\end{figure}

As a result of isotropy (Section~\ref{sec:Isotropy}),
the interaction vertex, $\gamma$ (resp.\ $\Gamma$), can be fully characterised
by its isotropic component, $\gamma^{L}$ (resp.\ $\Gamma^{L}$).
More precisely, following Eq.~\eqref{eq:iso_Gamma}, $\gamma^{L}_{\1\2\3}$ and
$\Gamma^{L}_{\1\2\3}$ depend only on
the considered triangle, ${ (\ell_{1} , \ell_{2} , \ell_{3}) }$,
the triplet of spin indices, ${ (\eps_{1} , \eps_{2} , \eps_{3}) }$,
and the triplet of times, ${ (t_{1} , t_{2} , t_{3}) }$.
Owing to time stationarity (Eq.~\ref{eq:stat_Gamma}),
we can set ${ t_{1} \!=\! 0 }$
and limit ourselves to investigating the dependence
on the times ${ (t_{2} , t_{3}) }$.
To easily compare the renormalised vertex, $\Gamma$,
with the bare vertex, $\gamma$,
it is interesting to pick the set of spins ${ (\eps_{1} , \eps_{2} , \eps_{3}) \!=\! (+ \! + \! -) }$,
given the exclusion rules satisfied by $\gamma$
(see Appendix~\ref{app:Bare_gamma}).
In practice, in the following figure,
we consider four different triangle configurations,
as detailed in Table~\ref{tab:Table_triangles}.
\begin{table}[h]
\centering
\begin{tabular}{|l|c|c|c|c|}
\hline
\textbf{Panel} & (a) & (b) & (c) & (d) 
\\
\hline
\textbf{Triangle} ${ (\ell_1 \ell_2 \ell_3) }$ & (122) & (232) & (133) & (333) 
\\
\hline
\textbf{Time} ${ (t_{1} , t_{2} , t_{3}) }$ & \multicolumn{4}{|c|}{${ (0 , t_{2} , t_{3}) }$}
\\
\hline
\textbf{Spin} ${ (\eps_1 \eps_2 \eps_3) }$ -- $\gamma^{L}_{\1\2\3} \, \scalebox{0.8}{\&} \, \Gamma^{L}_{\1\2\3}$ & \multicolumn{4}{|c|}{${ (+ \! + \! -) }$}
\\
\hline
\textbf{Spin} ${ (\eps_1 \eps_2 \eps_3) }$ -- $G^{L}_{\1\2\3}$ & \multicolumn{4}{|c|}{${ (+ \! + \! +) }$} 
\\
\hline
\end{tabular}
\caption{List of the triangles, ${ (\ell_{1} , \ell_{2} , \ell_{3}) }$,
along with the associated choice for the times and spin indices
for which we represent the bare (resp.\ renormalised) vertex $\gamma^{L}$ (resp.\ $\Gamma^{L}$)
in Fig.~\ref{fig:Maps_Gamma}.
The same triangles are also used in Section~\ref{sec:Prediction_3_point},
when investigating the skewness.
}
\label{tab:Table_triangles}
\end{table}
In Fig.~\ref{fig:Maps_Gamma},
we present the bare and one-loop predictions for $\Gamma^{L}$
for different triangle configurations, as a function of two times.
This is a rich figure
that deserves to be commented upon in detail.

\textit{Dressing}. First, we recall that in the fixed-point approach from Eq.~\eqref{diag:Iter},
the renormalised vertex, $\Gamma^{L}$,
is initialised to the bare vertex, $\gamma^{L}$,
i.e.\ a value that only contains information
about the system's equations of motion.
At the end of the fixed-point scheme,
$\Gamma^L$ deviates from $\gamma^L$, i.e.\ it gets \textit{dressed}.
The goal of this section is to characterise such dressing.

\textit{Scale}. In Fig.~\ref{fig:Maps_Gamma}, we normalise the amplitude
of each panel by the maximum amplitude of the bare vertex (across all triangle configurations).
In practice, this amplitude is given by that of $\gamma^{L}_{\mathrm{(b)}}$,
i.e.\ the black square in panel (b) of Fig.~\ref{fig:Maps_Gamma}.
This sets a consistent scale across all panels, ensuring that ${\gamma^L\!\sim\!1}$ throughout.

\textit{Non-vanishing bare vertex}. Panels ``bare,~(a)'' and ``bare,~(b)''
exhibit a narrow distribution for $\gamma^{L}$.
This is because these particular combinations of triangle and spin triplets (Table~\ref{tab:Table_triangles})
are associated with a non-zero bare vertex, $\gamma^{L}$ (see Appendix~\ref{app:Bare_gamma}).
In practice, ${ \gamma^{L} (t_{1} , t_{2} , t_{3}) }$,
is local in time, i.e.\ it is non-zero only for ${ t_{1} \!=\! t_{2} \!=\! t_{3} }$
(Eq.~\ref{app:Bare_gamma}).
Since we imposed ${ t_{1} \!=\! 0 }$ in Fig.~\ref{fig:Maps_Gamma},
the bare vertex contributes only at ${ t_{2} \!=\! t_{3} \!=\! 0 }$.
In Fig.~\ref{fig:Maps_Gamma}, this is highlighted with a black square.
For the same triangles,
we find that panels ``one-loop,~(a)'' and ``one-loop,~(b)''
also exhibit a very narrow distribution for $\Gamma^{L}$.
As such, for these two triangle configurations, $\Gamma^{L}$ remains fully dominated by its initial condition, $\gamma^{L}$, and the iterative corrections to $\Gamma^{L}$ remain small, of relative order ${ \leq\!10^{-6} }$.

\textit{Vanishing bare vertex}. Let us now turn our interest to panels (c) and (d) in Fig.~\ref{fig:Maps_Gamma}.
For these triangle configurations, the bare vertex, $\gamma^{L}$, exactly vanishes at all times, as shown in panels ``bare,~(c)'' and ``bare,~(d)''.
Consequently, any non-zero signal in $\Gamma^{L}$ at the end
of the one-loop iteration is associated
with a self-consistent dressed contribution
that emerges through the fixed-point search.
Such corrections only occur at one-loop order in the \MSR\@ formalism,
since at bare order,
no renormalisation of the interaction vertex is performed (Eq.~\ref{diag:bare}).

\textit{Amplitude}. It is interesting to compare the amplitude of $\Gamma^L$
with respect to that of $\gamma^L$, in order to characterise the amount by which $\Gamma^L$ gets dressed.
With the convention ${\gamma^{L}_{\mathrm{(b)}}\!=\!1}$ at ${t_2\!=\!t_3\!=\!0}$, we find that the maximum amplitude of $\Gamma^{L}$
in panel ``one-loop,~(d)'' is of the order ${7\!\!\times\!\!10^{-4}}$. This is a crucial point: the one-loop renormalisation procedure of Eq.~\eqref{diag:Iter} ensures that the corrections to $\gamma^L$ remain small, bounded, and well-behaved. Consequently, the renormalised vertex $\Gamma^L$ exhibits only \textit{small} deviations from the bare vertex.
Heuristically, one may write ${|\Gamma^L \!-\! \gamma^L| / \gamma^L \!\ll\!1}$.
These small corrections reflect the self-consistent aspect of the dressing,
improving the accuracy of the predictions for the correlations.

\textit{Causality}. We note that only the upper diagonal
of panel ``one-loop,~(d)'' is populated, i.e.\ ${ \Gamma^{L} (0 , t_{2} , t_{3}) }$
is non-zero only for ${ t_{3} \!\geq\! t_{2} }$.
This is because computing $\Gamma^L$
involves the two-point propagator, $G$,
and, hence involves the response function, $R$ (Eq.~\ref{eq:components_G}).
Since the response function is causal (see Eq.~\ref{eq:def_R}),
it can lead to sharp, but physical, discontinuities in time.

\textit{Locality}. We note that the dressing of $\Gamma^{L}$
in panel ``one-loop,~(d)''
occurs in a time-localised fashion. Phrased differently,
$\Gamma^{L}$ \textit{turns on} only within a limited region
in the ${ (t_{2} , t_{3})}$-plane. Recalling that in this figure
we imposed ${ t_{1} \!=\! 0 }$, this means
that the time-stationary vertex $\Gamma^{L}$
has only a finite depth in time.
In practice, we find that the typical decay time for $\Gamma^{L}$
is of order $\Tc$, the system's coherence time,
just like for the two-point correlation function from Fig.~\ref{fig:Money_plot}.
This was to be expected, since \VRR\ being a fully degenerate
and non-linear system (see Eq.~\ref{eq:Master_Equation_short}),
there is only one single typical timescale,
namely the coherence time, $\Tc$ (Appendix~\ref{app:IC_TC}).

\textit{Numerical measurements}. As argued,
$\Gamma$ is the target of the \MSR\ statistical closure.
Yet, $\Gamma$ is challenging to measure directly in $N$-body simulations.
Indeed, following its definition from Eq.~\eqref{eq:def_Gamma_3_point},
to measure $\Gamma$ numerically, one would need to measure
both the inverse of the two-point propagator, $G_{\1\2}^{-1}$,
and the three-point propagator, $G_{\1\2\3}$.
To do so, two difficulties need to be overcome:
(i) One can somewhat straightforwardly
measure the two-point correlation function, $C_{12}$
(as in Fig.~\ref{fig:Money_plot})
and, leveraging the \FDT\@, infer the full $G_{\1\2}$.
Yet, this tensor still needs to be inverted numerically.
This could prove unstable.
(ii) Similarly, assuming that one uses a large enough sample
of realisations, measuring the three-point correlation $C_{123}$
is within reach (Fig.~\ref{sec:Prediction_3_point}).
Yet, to access the full propagator $G_{\1\2\3}$,
one needs to appropriately determine the contributions
from all the other triplets of spin indices.
This would require exploring further the signature of the \FDT\
for three-point function, a delicate endeavour.
Of course, should one be in a position to circumvent these difficulties,
measuring directly $\Gamma$ in $N$-body simulations
is surely a promising prospect to assess the effective level
of dressing undergone by the system.
This will be the topic of a future work.

\subsection{Three-point correlation}
\label{sec:Prediction_3_point}

We now turn our interest to the characterisation
of  three-point correlations in the system. These are particularly important since they offer
direct insight into the system's level of non-Gaussianity.
Here, the starting point is Eq.~\eqref{diag:G3},
which states that the knowledge of both the renormalised vertex, $\Gamma_{\1\2\3}$,
and the two-point propagator, $G_{\1\2}$,
is sufficient to determine the three-point propagator, $G_{\1\2\3}$.
In addition, as detailed in Appendix~\ref{app:Prediction_3_point},
for ${ (\eps_{1} , \eps_{2} , \eps_{3}) \!=\! (+ \! + \! +) }$,
$G^{\tplus \tplus \tplus}_{123}$ is exactly the three-point correlation function of the system, ${\ml \vphi_{1} \, \vphi_{2} \, \vphi_{3} \mr}$.
Following Eq.~\eqref{diag:G3},
one can readily check that $G_{\1\2\3}$ satisfies
the exact same symmetries as $\Gamma_{\1\2\3}$ (Eq.~\ref{eq:iso_Gamma}).
As such, in the following, we focus our interest
on $G_{\1\2\3}^{L}$, the isotropic component of the three-point propagator.
More precisely, we can write the relation
\begin{equation}
\ml \vphi_{1} \, \vphi_{2} \, \vphi_{3} \mr = E^{\Delta M}_{a_{1} a_{2} a_{3}} \, C^{L}_{\ell_{1} \ell_{2} \ell_{3}} (t_{1} , t_{2} , t_{3}) ,
\label{eq:intro_CL}
\end{equation}
where $C^{L}_{\ell_{1} \ell_{2} \ell_{3}}$ follows directly from $G^{L}_{\1\2\3}$ (Eq.~\ref{eq:link_CL}).

In order to assess quantitatively the amplitude
of non-Gaussianities,
we follow Eq.~\eqref{eq:intro_skew},
and introduce the isotropic skewness
\begin{equation}
S^{L}_{\ell_{1} \ell_{2} \ell_{3}} (t_{1} , t_{2} , t_{3}) =  C^{L}_{\ell_{1} \ell_{2} \ell_{3}} (t_{1} , t_{2} , t_{3}) / C^{\pth}_0,
\label{eq:iso_skew}
\end{equation}
with $C_{0}$ the initial amplitude of the two-point correlation
from Eq.~\eqref{eq:def_C0}. 
We are now set to compare the \MSR\ bare and one-loop predictions
for the isotropic skewness, $S^{L}_{\ell_{1} \ell_{2} \ell_{3}}$,
with measurements in direct $N$-body simulations.
In Appendix~\ref{app:NumericalSimulations},
we detail our procedure to measure accurately the isotropic three-point correlation from Eq.~\eqref{eq:iso_skew}
in such simulations.
We detail in Table~\ref{tab:Table_triangles} the exact set
of triangle configurations considered.

\begin{figure}[htbp!]
\centering
\includegraphics[width=1.0\linewidth]{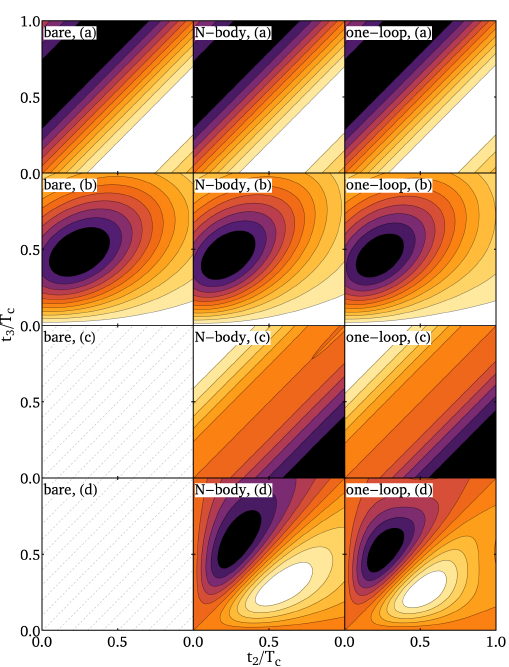}
\caption{Dependence of the three-point isotropic skewness, ${ S^{L}_{\ell_{1} \ell_{2} \ell_{3}} (0 , t_{2} , t_{3}) }$ (Eq.~\ref{eq:iso_skew}),
for different triangles (different lines),
as a function of the two times ${ (t_{2} , t_{3}) }$,
following the specification from Table~\ref{tab:Table_triangles},
and using the same convention as in Fig.~\ref{fig:Maps_Gamma}.
The left column corresponds to the bare \MSR\ (i.e.\@, \DIA\@) prediction,
the centre column to the $N$-body measurement,
and the right column to the one-loop \MSR\ prediction.
Dashed regions correspond to exactly vanishing skewnesses.
For all triangles,
the one-loop prediction offers a satisfactory match
with the $N$-body measurements.
}
\label{fig:Maps_full}
\end{figure}

\begin{figure}[htbp]
\centering
\includegraphics[width=1.0\linewidth]{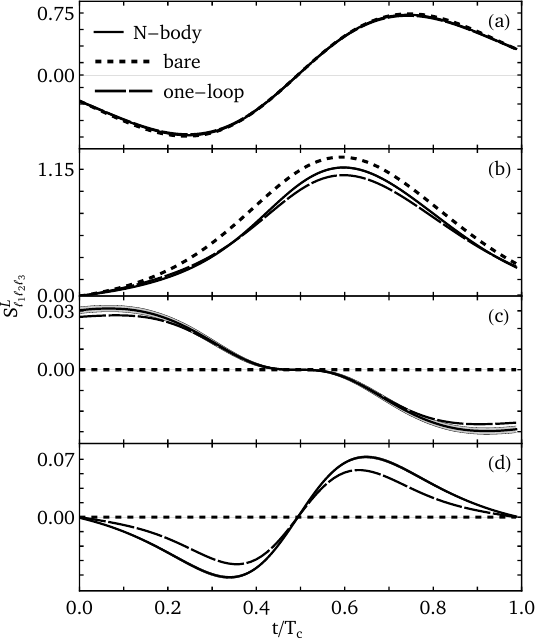} 
\caption{Slice of the isotropic skewness
from Fig.~\ref{fig:Maps_full},
representing ${ t \!\mapsto\! S^{L}_{\ell_{1} \ell_{2} \ell_{3}} (0, t , \Tc\!-\!t) }$.
The dotted (resp.\ dashed) line corresponds
to the bare (resp.\ one-loop) \MSR\ prediction,
while the full line corresponds to the $N$-body measurements.
The shaded regions correspond to the uncertainties associated with such measurements,
as detailed in Appendix~\ref{app:NumericalSimulations}.
The one-loop prediction presents a satisfactory quantitative match to
the numerical measurements, notably for triangles (c) and (d),
for which \MSR\ predicts at bare order an exactly vanishing skewness.
}
\label{fig:G_3_profile_full}
\end{figure}

In Fig.~\ref{fig:Maps_full}, we show the amplitude of $S^{L}$
as a function of time,
for different triangle configurations (see Table~\ref{tab:Table_triangles}),
using the same convention as in Fig.~\ref{fig:Maps_Gamma}.
In this figure,
we present the bare and one-loop \MSR\ predictions
along with the associated $N$-body measurements.
Figure~\ref{fig:Maps_full} is the \textit{main result} of this paper.
As a complement, in Fig.~\ref{fig:G_3_profile_full} we present the amplitude of each panel from Fig.~\ref{fig:Maps_full} measured along the anti-diagonal direction, namely the function ${ t \!\mapsto\! S^{L}(0, t , \Tc\!-\!t) }$.
Figure~\ref{fig:G_3_profile_full} provides a more quantitative
comparison between the \MSR\ theoretical predictions
and the $N$-body measurements.
Let us now comment in detail these two figures.

\textit{Non-vanishing bare prediction}. For triangles (a) and (b),
the bare (\DIA\@) prediction of $S^{L}$ is non-zero.
In this case, we find that the one-loop prediction for the skewness remains close to the bare one.
This is consistent with the trend that was already noted in Fig.~\ref{fig:Maps_Gamma},
where we found that for these triangles,
$\Gamma^{L}$ was fully dominated by the bare vertex, $\gamma^{L}$.
In Fig.~\ref{fig:G_3_profile_full}, we note that both predictions reproduce accurately the $N$-body measurements for triangle (a),
with the one-loop prediction achieving a precision of ${4\%}$.
Interestingly, for triangle (b), the one-loop prediction also improves
(marginally) upon the bare prediction.

\textit{Vanishing bare prediction}. In contrast, for triangles (c) and (d),
the bare prediction for $S^{L}$ exactly vanishes.
This is a consequence of the bare vertex
being exactly zero for these triangles (see Fig.~\ref{fig:Maps_Gamma}),
so that Eq.~\eqref{diag:G3} predicts a vanishing skewness.
For these triangles,
the renormalised vertex $\Gamma^{L}$
is entirely built from higher-order corrections.
This results in a non-vanishing one-loop prediction for the skewness.
In Fig.~\ref{fig:G_3_profile_full}, we find that the one-loop \MSR\ prediction
reproduces accurately the $N$-body measurements.
For triangle (d), the one-loop prediction captures the overall shape
of $S^{L}$, but underestimates its amplitude by ${ \sim\! 20\% }$. 
Panels (c) and (d) clearly demonstrate the improvement of the one-loop \MSR\ prediction over the bare one, 
since the bare order incorrectly predicts
a vanishing skewness for these triangles.

\textit{Symmetry}. In Fig.~\ref{fig:Maps_full},
we note that for the triangles (a), (c) and (d),
${ S_{\ell_{1} \ell_{2} \ell_{3}} (0 , t_{2} , t_{3}) }$
is antisymmetric
with respect to the diagonal ${ t_{2} \!=\! t_{3} }$.
This is because these triangles are such that ${ \ell_{2} \!=\! \ell_{3} }$
(Table~\ref{tab:Table_triangles}). 
As a result, following the symmetry from Eq.~\eqref{eq:intro_CL},
the isotropic three-point correlation must be antisymmetric
with respect to the exchange ${ t_{2} \!\leftrightarrow\! t_{3} }$.
Triangle (b) does not exhibit such an antisymmetry
because it satisfies ${ \ell_{2} \!\neq\! \ell_{3} }$.

\textit{Amplitude}. In Fig.~\ref{fig:G_3_profile_full}, we note that the amplitude of the isotropic  skewness $S^{L}$ strongly depends on the triangle configuration. Since the initial conditions of isotropic \VRR\ are quasi-Gaussian (Section~\ref{sec:Invariance}), we have ${ S^{L} \!=\! 0 }$
for ${t_1\!=\!t_2\!=\!t_3\!=\!0}$.
The triangles (a) and (b) are associated with a non-vanishing bare prediction.
As a result, for these triangles,
non-Gaussianities grow rapidly,
reaching order unity within one coherence time, $\Tc$.
This reflects the regime of strong turbulence,
in which dynamically-generated non-Gaussianities become large.
For such triangles, the bare \MSR\ prediction already captures
most of the signal.
By contrast, the triangles (c) and (d)
are associated with a vanishing bare prediction.
In that case, the amplitude of the skewness
is found to be 10 to 40 times smaller
than that of triangles (a) and (b).
These weaker signals originate
from the dressing of non-Gaussianities,
as already visible in Fig.~\ref{fig:Maps_Gamma}.
As such, the one-loop renormalisation scheme
generates additional, but small, contributions
that gradually fill in the three-point correlations.
In summary, the bare \MSR\ prediction reliably captures strong non-Gaussianities,
while the one-loop renormalisation scheme refines this description and accounts for additional weak non-Gaussianities
at other scales. 
Phrased differently, higher-order closures account for further
(small) non-Gaussianities beyond the bare prediction.

\section{Discussion}
\label{sec:Discussion}

In this section, we continue our discussion on the \MSR\ scheme
by considering, in turn,
the invariance of isotropic \VRR\ with respect to $N$ (Section~\ref{sec:Invariance}),
the failure of a \textit{naive} one-loop prediction (Section~\ref{sec:Naive}),
and the prohibitive numerical cost associated
with the two-loop \MSR\ prediction (Section~\ref{sec:TwoLoopComplexity}).

\subsection{Invariance with respect to $N$}
\label{sec:Invariance}

Throughout this paper, we have noted
that the initial conditions in isotropic \VRR\ are quasi-Gaussian,
with non-Gaussianities becoming large over one coherence time.
This section aims to quantify these statements more precisely.

In our $N$-body simulations, the initial conditions
for isotropic \VRR\ are generated as follows:
(i) for each realisation, the $N$ particles are drawn independently
and uniformly over the unit sphere; (ii) we generate a large ensemble of such independent realisations.
This procedure determines the initial spectrum of the isotropic two-point correlation function.
Specifically, following Eq.~\eqref{eq:def_C0},
one has ${ C_{\ell} (t \!=\! 0) \!=\! N / (4 \pi) }$, independently of $\ell$.
This is consistent with white noise, as discussed in Appendix~\ref{app:IC_2pt}. 

As emphasised in Section~\ref{sec:Prediction_3_point},
to characterise the degree of non-Gaussianity in the system,
it is natural to investigate the three-point skewness.
Following Eq.~\eqref{eq:iso_skew},
we define it generically via
\begin{equation}
S_{123} = C_{123} / C_{0}^{\pth} ,
\label{eq:def_S_generic}
\end{equation}
where we recall that ${1\!=\!(a,t)\!=\!(\ell,m,t)}$,
i.e.\ Eq.~\eqref{eq:def_S_generic} does not rely on any assumption of isotropy.

As obtained in Appendix~\ref{app:IC_3pt},
for isotropic \VRR\@, the initial condition
for the three-point skewness is
\begin{equation}
S_{a_1 a_2 a_3}(t_1\!=\!t_2\!=\!t_3\!=\!0) = A_{a_1 a_2 a_3} / C_{0}^{\phh},
\label{eq:cond_init_3_point}
\end{equation}
where the coefficients $A_{a_1 a_2 a_3}$ are independent of $N$.
Importantly, since ${ C_{0} \!\propto\! N }$ (Eq.~\ref{eq:def_C0}),
the initial skewness scales like ${ N^{-1/2} }$.
Thus, for $N$ large enough,
we may consider that the initial conditions are quasi-Gaussian.

As detailed in Eq.~\eqref{eq:def_alphaL},
the coefficients $A_{a_1 a_2 a_3}$ obey 
\begin{equation}
A_{a_1 a_2 a_3} = 0 \quad \text{if} \quad \ell_1 \!+\! \ell_2 \!+\! \ell_3 \quad \text{is odd} .
\label{eq:def_alpha_short}
\end{equation}
This exclusion rule contrasts with that of the excluded Elsasser coefficients,
$E^{\Delta M}_{a_{1} a_{2} a_{3}}$ (Eq.~\ref{eq:def_excl_E}),
which satisfy
\begin{equation}
E^{\Delta M}_{a_1 a_2 a_3} = 0 \quad \text{if} \quad \ell_1 \!+\! \ell_2 \!+\! \ell_3 \quad \text{is even}.
  \label{eq:def_EM_short}
\end{equation}
These Elsasser coefficients enter the \MSR\ prediction
for the three-point skewness (Eq.~\ref{eq:iso_skew}) via
\begin{equation}
S_{123} = E^{\Delta M}_{a_1 a_2 a_3} S^L_{\ell_1 \ell_2 \ell_3}(t_1,t_2,t_3) .
\label{eq:def_iso_S}
\end{equation}
From that equation, we note that,
for isotropic \VRR\@,
\MSR\ predicts an exactly vanishing skewness for even triangles, ${ (\ell_{1} , \ell_{2} , \ell_{3}) }$,
and a non-trivial skewness for odd triangles.
This is the reason why, in Section~\ref{sec:Prediction_3_point},
we limited ourselves to only presenting \MSR\ predictions
for the three-point skewness of odd triangles (see Table~\ref{tab:Table_triangles}).
At this stage, one could fear that Eqs.~{\eqref{eq:cond_init_3_point}-\eqref{eq:def_alpha_short}} are in conflict with Eqs.~{\eqref{eq:def_EM_short}-\eqref{eq:def_iso_S}}, with regard to the three-point skewness of even triangles.
Indeed, for even triangles,
these first two equations seem to indicate a non-vanishing three-point skewness
in the $N$-body measurements,
while the last two equations predict, within the \MSR\ framework, a vanishing three-point skewness.
Fortunately, both results are reconciled once one accounts
for both the time evolution and $N$-dependence of the present measurements.

Let us now examine in detail these two different behaviors
in our $N$-body simulations. Given the different exclusion
properties of even and odd triangles (see Eqs.~\ref{eq:def_alpha_short} and~\ref{eq:def_EM_short}),
we ought to be careful in the measurement of the three-point skewness
in numerical simulations (see Appendix~\ref{app:NumericalSimulations}).
In practice, we measure
\begin{subequations}
\begin{align}
S^{L, \even}_{\ell_1 \ell_2 \ell_3}(t_1,t_2,t_3) {} & = \big\langle S_{123}/A_{a_1 a_2 a_3} \big\rangle_{m} ,
\label{eq:def_iso_S_even}
\\
S^{L, \odd}_{\ell_1 \ell_2 \ell_3}(t_1,t_2,t_3) {} & = \big\langle S_{123}/E^{\Delta M}_{a_1 a_2 a_3} \big\rangle_{m}.
\label{eq:def_iso_S_odd}
\end{align}
\label{eq:def_iso_S_evenodd}\end{subequations}
In that expression, for some given even (resp.\ odd) triangle ${ (\ell_{1} , \ell_{2} , \ell_{3}) }$,
${ \langle \,\cdot\, \rangle_{m} }$ amounts to averaging over all ${ (m_{1} , m_{2} , m_{3}) }$
such that ${ A_{a_{1} a_{2} a_{3}} }$ (resp.\ ${ E^{\Delta M}_{a_{1} a_{2} a_{3}} }$)
is non-zero. The definition of $S^{L, \odd}_{\ell_{1} \ell_{2} \ell_{3}}$ in Eq.~\eqref{eq:def_iso_S_odd}
is identical to the definition of the isotropic three-point skewness, $S^{L}_{\ell_{1} \ell_{2} \ell_{3}}$, in Eq.~\eqref{eq:iso_skew}.
We also point that because $S^{L , \even}_{\ell_{1} \ell_{2} \ell_{3}}$ and $S^{L , \odd}_{\ell_{1} \ell_{2} \ell_{3}}$
differ in their normalisation,
one should not compare, at face value, their respective amplitude.
Let us now investigate, in turn,
the dependence of $S^{L , \even}$
and ${S^{L , \odd}}$, as one varies $N$.

\begin{figure}[htbp]
\centering
\includegraphics[width=1.0\linewidth]{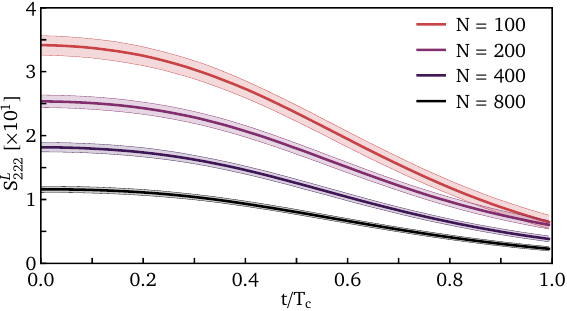} 
\caption{Profile of the three-point isotropic skewness ${ t \!\mapsto\! S^{L, \even}_{\ell_{1} \ell_{2} \ell_{3}} (0,t,t) }$ (Eq.~\ref{eq:def_iso_S_even})
as measured in $N$-body simulations,
for the even triangle ${(\ell_{1} , \ell_{2} , \ell_{3}) \!=\! (2,2,2)}$
and different values of $N$ (different colors).
The shaded region illustrates the associated uncertainties,
see Appendix~\ref{app:NumericalSimulations} for details.
As $N$ increases, the maximum amplitude of the skewness
of an even triangle decreases like $N^{-1/2}$.
}
\label{fig:G_3_even_profile}
\end{figure}

\textit{Even triangles}. As emphasised in Eq.~\eqref{eq:cond_init_3_point},
for zero time separation, i.e.\ for ${ t_{1} \!=\! t_{2} \!=\! t_{3} \!=\! 0 }$,
$S^{L , \even}$ has a non-zero value,
which scales like ${ N^{-1/2} }$.
Then, as the time separation grows,
$S^{L , \even}$ decays.
This is illustrated in Fig.~\ref{fig:G_3_even_profile},
where we represent the $N$-body measurement of ${ t \!\mapsto\! S^{L, \even}_{\ell_{1} \ell_{2} \ell_{3}} (0,t,t) }$ for the even triangle ${ (\ell_1, \ell_2, \ell_3)\!=\!(2,2,2) }$,
as one varies $N$.
In this figure, we find that the amplitude of $S^{L , \even}$
is suppressed like ${ N^{-1/2} }$ as one increases $N$.
Hence, for any given even triangle, we may assume that $S^{L , \even}$ vanishes once $N$ large is enough.
This allows us to reconcile the nonzero initial amplitude of the skewness of even triangles seen in $N$-body simulations (Eq.~\ref{eq:cond_init_3_point}),
with the \MSR\ formalism predicting that their skewness must vanish identically at all times (Eq.~\ref{eq:def_iso_S}).
For $N$ large enough, both statements are in agreement.

\begin{figure}[htbp]
\centering
\includegraphics[width=1.0\linewidth]{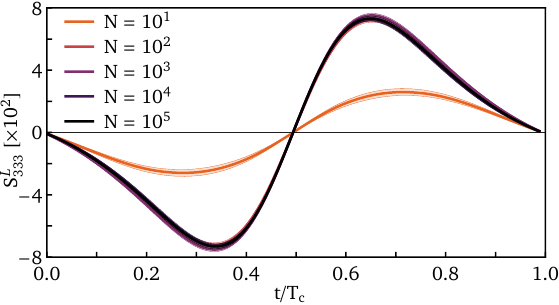} 
\caption{Profile of the three-point isotropic skewness ${ t \!\mapsto\! S^{L, \odd}_{\ell_{1} \ell_{2} \ell_{3}} (0,t,\Tc\!-\!t) }$ (Eq.~\ref{eq:def_iso_S_odd})
as measured in $N$-body simulations,
for the odd triangle ${(\ell_{1} , \ell_{2} , \ell_{3}) \!=\! (3,3,3)}$
and different values of $N$,
using the same convention as in Fig.~\ref{fig:G_3_even_profile}.
For $N$ large enough, the three-point skewness
of an odd triangle converges, and becomes invariant with respect to $N$.
}
\label{fig:G_3_profile}
\end{figure}

\textit{Odd triangles}. Following Eq.~\eqref{eq:cond_init_3_point},
for zero time separation,
i.e.\ for ${ t_{1} \!=\! t_{2} \!=\! t_{3} \!=\! 0 }$,
one has ${ S^{L , \odd}\!=\! 0 }$.
However, as time separation grows,
$S^{L , \odd}$ develops a non-zero amplitude.
This is illustrated in Fig.~\ref{fig:G_3_profile},
where we present the $N$-body measurement of the three-point skewness,
for the odd triangle ${ (\ell_1,\ell_2,\ell_3) \!=\! (3,3,3) }$,
via the function ${ t \!\mapsto\! S^{L, \odd}_{\ell_{1} \ell_{2} \ell_{3}} (0,t,\Tc \!-\! t) }$,
as one varies $N$.
This figure shows that the profile
of $S^{L , \odd}$ for the triangle ${ (3,3,3) }$
becomes independent of $N$ once ${N \!\gtrsim\! 100}$.
Beyond this threshold, the three-point skewness
becomes invariant with respect to $N$.
Phrased differently, on that scale,
we may assume that the initial statistics of the fluctuations
are sufficiently Gaussian.
We anticipate that smaller scales,
i.e.\ larger $\ell$,
require larger $N$, i.e.\ a denser population,
to approach initial Gaussian statistics.

\begin{figure}[htbp]
\centering
\includegraphics[width=1.0\linewidth]{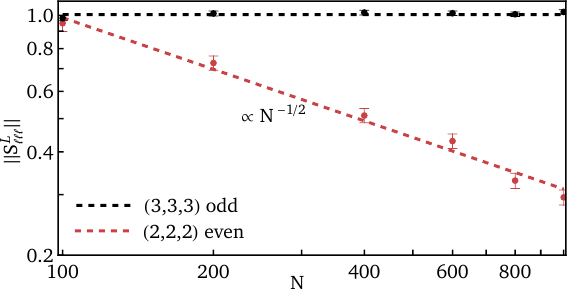} 
\caption{$N$-body measurement of the norm of the skewness, as a function of $N$, as defined in Eq.~\eqref{eq:def_norm}.
The red (resp.\ black) dots correspond to the even (resp.\ odd) triangle ${ (\ell_{1} , \ell_{2} , \ell_{3}) \!=\! (2,2,2) }$
(resp.\ ${ (3,3,3) }$), with their associated uncertainties (see Appendix~\ref{app:NumericalSimulations}).
Both set of points have been rescaled
by their value at ${ N \!=\! 100 }$.
The dashed lines are there to guide the eye
and correspond to ${N \!\mapsto\! (N / 100)^{- 1/2}}$ (red)
and ${ N \!\mapsto\! 1 }$ (black).
As $N$ increases, the amplitude of the skewness for the even triangle
decreases like ${ N^{-1/2} }$,
while it stays constant for the odd triangle.
}
\label{fig:slope}
\end{figure}

\textit{Scaling of the norm}. In order to assess these points further,
in Fig.~\ref{fig:slope}, we present the dependence
in $N$-body simulations of the \textit{norm} of the skewness
for both even and odd triangles,
as one varies $N$.
Here, the norm is defined as
\begin{equation}
\left\lVert S^{L , \even / \odd}_{\ell_{1} \ell_{2} \ell_{3}} \right\rVert \!=\! \bigg[ \!\! \int_{0}^{\Tc} \!\!\!\!\! \rd t_{2} \rd t_{3} \, \big| S^{L , \even / \odd}_{\ell_{1}  \ell_{2} \ell_{3}} (0 , t_{2} , t_{3}) \big|^{2} \bigg]^{1/2}.
\label{eq:def_norm}
\end{equation}
In Fig.~\ref{fig:slope},
we find that, as $N$ increases,
the norm of the three-point skewness
(i) remains constant for the odd triangle;
(ii) decreases for the even triangle as ${ N^{-1/2} }$.
We thus recover that the \MSR\ prediction is valid in the ${N \!\gg\! 1}$ regime,
i.e.\ in the limit of sufficiently Gaussian initial conditions,
for which the three-point skewness of even triangles vanishes asymptotically.

To conclude this section, for the range of scales considered here,
namely ${ \ell \!\leq\! 4}$,
we empirically find that choosing ${ N\!=\! 10^4 }$ in the $N$-body simulations
ensures that (i) configurations associated with even triangles
are sufficiently damped (Fig.~\ref{fig:G_3_even_profile}),
while (ii) configurations associated with odd triangles have converged to
their $N$-invariant regime (Fig.~\ref{fig:G_3_profile}).

It would be of prime interest to explore the generalisation
of the \MSR\ closure scheme to non-Gaussian
initial conditions, and assess the validity of this approach
in such a regime~\citep[see, e.g.\@,][]{Rose1985}.
This will be the topic of future work.

\subsection{Naive one-loop prediction}
\label{sec:Naive}

As illustrated in Eq.~\eqref{diag:one_loop},
the one-loop \MSR\ prediction proceeds
by expanding the renormalised vertex, $\Gamma$,
in \textit{powers} of itself.
In~\citetalias{Flores+2025} (section~{IV.D} therein),
we investigated an alternative, \textit{naive}, closure
that involved rather expanding $\Gamma$ in \textit{powers} of the bare vertex, $\gamma$.
At one-loop order, this expansion reads
\begin{equation}
\begin{tikzpicture}[baseline=(current bounding box.center)]
\begin{feynman}
\vertex [blob, minimum size=12pt, line width=1pt] (v) at (0,0) {};
\vertex (a) at (-0.55,0);
\vertex (b) at (0.45,0.5);
\vertex (c) at (0.45,-0.5);
\diagram* {
(a) -- [plain, thick] (v) -- [plain, thick] (b),
(v) -- [plain, thick] (c)
};
\end{feynman}
\node at (0.8,0) {$=$};
\begin{feynman}
\vertex [dot, minimum size=8pt, line width=1pt] (v) at (1.8,0) {};
\vertex (a) at (1.25,0);
\vertex (b) at (2.25,0.5);
\vertex (c) at (2.25,-0.5);
\diagram* {
(a) -- [plain, thick] (v) -- [plain, thick] (b),
(v) -- [plain, thick] (c)
};
\end{feynman}
\node at (2.75,0) {$+$};
\begin{scope}[xshift=0.525\linewidth]
\begin{feynman}
\vertex [dot, minimum size=8pt, line width=1pt] (v1) at (-0.75,0) {};
\vertex [dot, minimum size=8pt, line width=1pt] (v2) at (0.3,0.6) {};
\vertex [dot, minimum size=8pt, line width=1pt] (v3) at (0.3,-0.6) {};
\vertex (in) at (-1.3,0);
\vertex (out1) at (0.8,0.9);
\vertex (out2) at (0.8,-0.9);
\diagram* {
(in) -- [plain, thick] (v1),
(v1) -- [double, thick] (v2),
(v1) -- [double, thick] (v3),
(v2) -- [double, thick] (v3),
(v2) -- [plain, thick] (out1),
(v3) -- [plain, thick] (out2)
};
\end{feynman}
\end{scope}
\end{tikzpicture} \, .
\label{diag:one_loop_naive}
\end{equation}

Because it shares the exact same functional form as the genuine
one-loop \MSR\ prediction (Eq.~\ref{diag:one_loop}),
Eq.~\eqref{diag:one_loop_naive} straightforwardly lends itself
to the same numerical scheme as the one presented in Section~\ref{sec:Numerical}.
In particular, because the bare vertex is local in time (Eq.~\ref{eq:def_gamma_next}),
the evaluation of the one-loop diagram in Eq.~\eqref{diag:one_loop_naive}
is much faster than that of Eq.~\eqref{diag:one_loop},
since most of the contributions from the time integrals vanish.
This simple time-dependence is also the reason
why in eq.~{(G2)} of~\citetalias{Flores+2025},
it was possible to obtain an explicit integro-differential equation
satisfied by the two-point correlation function for the naive one-loop prediction.
We refer to Fig.~\ref{fig:CV_Naive} in Appendix~\ref{app:NumericalSimulations} for confirmation that both implementations---the numerical scheme from Section~\ref{sec:Numerical} applied to Eq.~\eqref{diag:one_loop_naive}
and the solution of eq.~{(G2)} from~\citetalias{Flores+2025}---are in tight agreement.

\begin{figure}[htbp]
\centering
\includegraphics[width=1.0\linewidth]{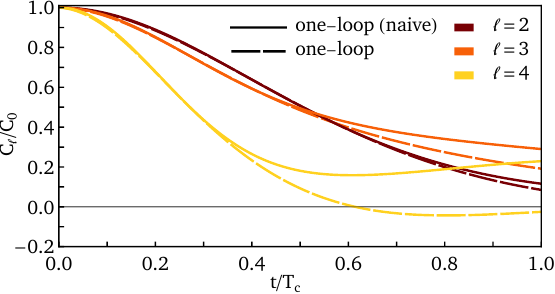} 
\caption{Same as Fig.~\ref{fig:Money_plot},
but this time comparing the one-loop \MSR\ prediction
from Eq.~\eqref{diag:one_loop},
with its naive version from Eq.~\eqref{diag:one_loop_naive}.
The naive expansion diverges on long timescales.
}
\label{fig:NaivevsMSR}
\end{figure}

In Fig.~\ref{fig:NaivevsMSR}, we compare the one-loop \MSR\ prediction (Eq.~\ref{diag:one_loop})
for the two-point correlation function 
with its naive analog (Eq.~\ref{diag:one_loop_naive}).
In this figure, we find that the naive one-loop prediction
from Eq.~\eqref{diag:one_loop_naive} is not convergent.
Indeed, the two-point correlation function, $C_{\ell}$,
diverges at late times.
In addition, this divergence gets stronger as one increases $\ell$,
as already illustrated in fig.~{10} of~\citetalias{Flores+2025}.
On the contrary, the \MSR\ one-loop prediction from Eq.~\eqref{diag:one_loop}
does not exhibit such divergences.

The failure of Eq.~\eqref{diag:one_loop_naive} is rooted in the lack of self-consistency in $\Gamma$,
since the vertices inside the loop are frozen at the bare $\gamma$.
This approach implicitly treats $\gamma$ as a small expansion parameter, which causes the series to diverge in the strongly turbulent regime of \VRR, where ${\gamma\!\sim\!\mathcal{O}(1)}$. 
Consequently, adding higher-order naive loops to Eq.~\eqref{diag:one_loop_naive} would only worsen this divergence.
The correct procedure is to determine $\Gamma$ self-consistently. This yields a non-linear fixed-point equation---such as Eq.~\eqref{diag:one_loop}---which does not require $\gamma$ to be small for the solution to exist.
In this self-consistent framework, the \textit{loop order}---e.g.\@, one-loop or two-loop---labels the complexity of the fixed-point equation by the number of irreducible diagrams included. 
This is fundamentally different from the perturbative expansion of Eq.~\eqref{diag:one_loop_naive}.

\subsection{Two-loop prediction?}
\label{sec:TwoLoopComplexity}

In Section~\ref{sec:Application},
we focused our interest on the one-loop \MSR\ prediction,
as given by Eq.~\eqref{diag:one_loop}.
Naturally, one may ask whether the \MSR\ numerical scheme can be extended
to two-loop order, as expressed in Eq.~\eqref{diag:two_loop}.
Here, we briefly discuss what such an implementation would involve.

At two-loop order, the structure of the numerical renormalisation scheme 
remains formally identical to that at one-loop (Eq.~\ref{diag:Iter}). 
The Dyson equation (Eq.~\ref{diag:Dyson}) remains unchanged, 
but the self-consistent expansion of the renormalised vertex $\Gamma$ must be updated. 
Thus, the fixed-point scheme of Eq.~\eqref{diag:Iter} applies at two-loop order
by replacing Eq.~\eqref{diag:one_loop_iter}
with Eq.~\eqref{diag:two_loop}. All symmetries of Section~\ref{sec:Symmetries} (isotropy, time stationarity, triangular exclusions, etc.) remain true, 
with Eq.~\eqref{eq:contract_W6} giving the two-loop contraction formula for the Elsasser coefficients. 
In short, increasing the order of the loop-expansion \textit{simply} amounts to considering an ever more intricate
self-consistent relation satisfied by $\Gamma$.

The total complexity of evaluating all two-loop diagrams in Eq.~\eqref{diag:two_loop} scales as ${\mO(\NSTEPS^{14}\LMAX^{9})}$.
This is absolutely computationally prohibitive.
In practice, following the same idea as in Eq.~\eqref{eq:def_Lambda},
this difficulty can be (partially) reduced by the leverage
of additional symmetries.
Let us introduce the tensor
\begin{equation}
\Psi_{\1\2\3} = \Gamma_{\1'\2'\3} \, G_{\1'\1} \, G_{\2'\2} .
\label{eq:def_Psi}
\end{equation}
Diagrammatically, it reads
\begin{equation}
\begin{aligned}
\Psi_{\1\2\3} & \; = \;
\begin{tikzpicture}[baseline=(current bounding box.center)]
\begin{feynman}
\vertex [blob, minimum size=12pt, line width=1pt] (v) at (2.3,0) {};
\vertex (a) at (1.75,0);
\vertex (b) at (2.75,0.5);
\vertex (c) at (2.75,-0.5);
\draw[double, thick] (a) -- (v) node[pos=0.3, above=-0.05]  {$\scalebox{0.8}{$\1$}$};
\draw[double, thick] (v) -- (b) node[pos=0.45, above] {$\scalebox{0.8}{$\2$}$};
\draw[plain, thick] (v) -- (c) node[pos=0.35, below=-0.025] {$\scalebox{0.8}{$\3$}$};
\end{feynman}
\end{tikzpicture} 
\; = \;
\begin{tikzpicture}[baseline=(current bounding box.center)]
\begin{feynman}
\vertex [blob, minimum size=12pt, line width=1pt] (v) at (2.3,0) {};
\vertex [dot, minimum size=4pt, line width=1pt] (vdot1) at (2.0,0) {};
\vertex [dot, minimum size=4pt, line width=1pt] (vdot2) at (2.5,0.22) {};
\vertex (a) at (1.7,0);
\vertex (b) at (2.75,0.5);
\vertex (c) at (2.75,-0.5);
\draw[plain, thick] (a) -- (v) node[pos=0.3, above=-0.05]  {$\scalebox{0.8}{$\1$}$};
\draw[plain, thick] (v) -- (b) node[pos=0.45, above] {$\scalebox{0.8}{$\2$}$};
\draw[plain, thick] (v) -- (c) node[pos=0.35, below=-0.025] {$\scalebox{0.8}{$\3$}$};
\end{feynman}
\end{tikzpicture}\, .
\end{aligned}
\label{diag:def_Psi}
\end{equation}
We point out that $\Psi_{\1\2\3}$
satisfies the exact same symmetries as $\Lambda_{\1\2\3}$,
introduced in Eq.~\eqref{diag:def_Lambda}.

Using this new tensor, one can rewrite the two-loop diagram
from Eq.~\eqref{diag:two_loop} into
\begin{equation}
\begin{aligned}
\begin{tikzpicture}[baseline=(current bounding box.center)]
\begin{feynman}
\vertex [blob, minimum size=12pt, line width=1pt] (v1) at (3,0) {};
\vertex [blob, minimum size=12pt, line width=1pt] (v2) at (3.75,0.6) {};
\vertex [blob, minimum size=12pt, line width=1pt] (v3) at (3.75,-0.6) {};
\vertex [blob, minimum size=12pt, line width=1pt] (v4) at (4.75,0.6) {};
\vertex [blob, minimum size=12pt, line width=1pt] (v5) at (4.75,-0.6) {};
\vertex (in) at (2.45,0);
\vertex (out1) at (5.25,0.6);
\vertex (out2) at (5.25,-0.6);
\diagram* {
(in) -- [plain, thick] (v1),
(v1) -- [double, thick] (v2),
(v1) -- [double, thick] (v3),
(v3) -- [double, thick] (v4),
(v2) -- [double, thick] (v5),
(v2) -- [double, thick] (v4),
(v3) -- [double, thick] (v5),
(v4) -- [plain, thick] (out1),
(v5) -- [plain, thick] (out2)
};
\end{feynman}
\end{tikzpicture} 
\; = \;
\begin{tikzpicture}[baseline=(current bounding box.center)]
\begin{feynman}
\vertex [blob, minimum size=12pt, line width=1pt] (v1) at (3,0) {};
\vertex [dot, minimum size=4pt, line width=1pt] (vdot1) at (3.225,0.2) {};
\vertex [dot, minimum size=4pt, line width=1pt] (vdot2) at (3.225,-0.2) {};
\vertex [blob, minimum size=12pt, line width=1pt] (v2) at (3.75,0.6) {};
\vertex [dot, minimum size=4pt, line width=1pt] (vdot3) at (4.45,0.6) {};
\vertex [dot, minimum size=4pt, line width=1pt] (vdot4) at (4.55,0.365) {};
\vertex [blob, minimum size=12pt, line width=1pt] (v3) at (3.75,-0.6) {};
\vertex [dot, minimum size=4pt, line width=1pt] (vdot5) at (4.45,-0.6) {};
\vertex [dot, minimum size=4pt, line width=1pt] (vdot6) at (4.55,-0.365) {};
\vertex [blob, minimum size=12pt, line width=1pt] (v4) at (4.75,0.6) {};
\vertex [blob, minimum size=12pt, line width=1pt] (v5) at (4.75,-0.6) {};
\vertex (in) at (2.45,0);
\vertex (out1) at (5.25,0.6);
\vertex (out2) at (5.25,-0.6);
\diagram* {
(in) -- [plain, thick] (v1),
(v1) -- [plain, thick] (v2),
(v1) -- [plain, thick] (v3),
(v3) -- [plain, thick] (v4),
(v2) -- [plain, thick] (v5),
(v2) -- [plain, thick] (v4),
(v3) -- [plain, thick] (v5),
(v4) -- [plain, thick] (out1),
(v5) -- [plain, thick] (out2)
};
\end{feynman}
\end{tikzpicture}\, .
\end{aligned}
\label{diag:two_loop_with_Psi}
\end{equation}
To evaluate this diagram, one could then proceed in two successive steps:
(i) compute and store $\Psi$ (Eq.~\ref{diag:def_Psi});
(ii) compute and store the two-loop diagram from Eq.~\eqref{diag:two_loop_with_Psi}.
Step (ii) yields a total of ${\mO(\NSTEPS^{8}\LMAX^{9})}$ operations.
This is a significant improvement over the direct evaluation of Eq.~\eqref{diag:two_loop}.

Yet, even with this improvement, evaluating Eq.~\eqref{diag:two_loop_with_Psi} is vastly more computationally intensive than the one-loop diagram of Eq.~\eqref{diag:one_loop_with_Lambda}, which scales as ${\mO(\NSTEPS^{5}\LMAX^{6})}$.
Using the parameters of Section~\ref{sec:Application},
one iteration of Eq.~\eqref{diag:Iter} at one-loop required ${ \!\sim\!62\,\mathrm{h} }$ on 128 cores.
With the same parameters, one iteration at two-loop order
would require roughly ${10^{10}\,\mathrm{h} \!\sim\! 1 \, \mathrm{Myr}}$
on 128 cores.
Needless to say, this is surely not within any reasonable reach.

As described in Section~\ref{sec:Application},
the one-loop \MSR\ prediction already captures
the main statistical properties of the system with accuracy.
It yields significant improvements over the bare prediction,
in agreement with simulations (see, e.g.\@, Fig.~\ref{fig:Maps_full}). 
Given this success,
while the two-loop \MSR\ prediction is expected to remain finite and well-behaved
(see, e.g.\@, corrections in Fig.~\ref{fig:G_3_profile_full}),
any additional corrections stemming from it are likely to be minor.

Even for the \textit{simple} \VRR\ toy model,
evaluating the \MSR\ high-order prediction becomes rapidly infeasible.
This difficulty worsens significantly when considering, for instance,
non-isotropic systems or multi-population models.
This highlights a severe limitation of the \MSR\ closure scheme at high order.
Overcoming this challenge requires either new numerical methods for finding the fixed-point solution ${[G, \Gamma]}$,
or turning to alternative statistical closures with manageable complexity.
This will be the topic of future work.

\section{Conclusion}
\label{sec:Conclusion}

In this work, we used isotropic \VRR\ as a testbed for the \MSR\ renormalisation 
scheme~\citep{MSR1973}, in the fully non-linear regime of Eq.~\eqref{eq:nonlinear}. 
We developed a numerical method for solving the \MSR\ closure equations (Section~\ref{sec:MSR_general}), 
allowing us to compute both bare and one-loop order predictions for 
the two- and three-point correlation functions in space and time. 
The core of our approach is a fixed-point iterative scheme (Section~\ref{sec:Numerical}), 
whose efficiency is greatly improved by leveraging the symmetries intrinsic to \VRR. 
In Section~\ref{sec:Application}, we presented the results for the two- and three-point correlation functions,
as well as for the renormalised interaction vertex. 
We found that the one-loop corrections substantially improve upon the bare predictions, 
yielding between ${4\%}$ and ${20\%}$ agreement with $N$-body measurements of the correlations. 
Our main result is summarised in Fig.~\ref{fig:Maps_full}, which demonstrates that in the strongly coupled regime of \VRR, 
non-Gaussianities become important within one coherence time. 
While the bare closure already captures strong signals, 
the one-loop closure provides systematic improvements by introducing
additional non-Gaussian contributions.
We also discussed some further aspects of the \MSR\ scheme (Section~\ref{sec:Discussion}),
in particular, the invariance with respect to $N$,
and the intrinsic numerical difficulty associated with the two-loop order prediction.
Let us now conclude by listing a few possible venues
for future research that follow naturally from this work.

\textit{Time integrals}. Our numerical implementation
of the \MSR\ scheme requires performing a large number
of time integrals (see, e.g.\@, Eq.~\ref{diag:one_loop_iter}).
These integrals dominate the numerical cost.
In practice, one could try to better leverage time stationarity
by performing the integrals in the associated Fourier space,
rather than in time.
Indeed, in Fourier space, the property from Eq.~\eqref{eq:stat_G} (resp.\ Eq.~\ref{eq:stat_Gamma})
leads to one (resp.\ two) Dirac in Fourier space,
hence allowing, in principle, for much hastened numerical evaluations.
In practice, the effective implementation in Fourier space
would still remain delicate,
as it would require dealing appropriately with causal functions~\citep[see, e.g.\@, section~{4} in][]{Canet+2011}.
This will be the topic of future explorations.

\textit{Functional renormalisation group}. In the present work,
we focused our effort on implementing the renormalisation techniques
from~\cite{MSR1973}. On that front, it would be particularly interesting
to investigate the use of methods stemming from
\textit{functional renormalisation group}~\citep[see, e.g.\@,][]{Delamotte2012,Dupuis+2021}.
In particular, the so-called \textit{vertex expansion} from that framework
offers deep similarities with the equations faced here.
Such approaches were recently developed in~\cite{Tarpin+2019,Canet2022,Fontaine+2023}
to predict the two-point correlation function
of decaying turbulence in various geometries.
This would be no light numerical endeavour
since the effective numerical implementation of the functional renormalisation
group requires the effective integration of an intricate non-linear
differential equation with respect to the scale of the regulator introduced.
Focusing on \VRR\@, comparing quantitatively the lowest-order prediction
of the vertex expansion with the present \MSR\ results
will be the topic of future work.

\textit{Turbulence cascade}. The master equation of \VRR\@, Eq.~\eqref{eq:Master_Equation_short},
drives in phase space the equivalent of a turbulent cascade
from small $\ell$ (i.e.\ large scales) towards large $\ell$ (i.e.\ small scales).
Here, the \VRR\ system is not submitted to any external forcing,
i.e.\ one is in the regime of \textit{decaying turbulence}~\citep[see, e.g.\@,][]{Krommes2002}.
In that context, we expect that methods from critical balance theory~\citep[see, e.g.\@,][]{Goldreich+1995},
in particular when applied to phase space cascade~\cite[see][for a thorough application to the Vlasov--Poisson system]{Ginat+2025,Nastac+2025},
should prove powerful to characterise the decay of fluctuations toward small scales,
where fluctuations are \textit{dissipated} by the system's finite number of particles.
This will be the topic of future research.

\textit{Astrophysical applications}. Coming back to the astrophysical implications
of \VRR\@, one should leverage the present renormalisation methods
in more realistic regimes.
Notably, one should:
(i) Go beyond the isotropic limit,
i.e.\@ consider anisotropic distributions of orientations
as could be expected around SgrA*~\citep[see, e.g.\@,][]{Roupas+2017}.
(ii) Go beyond the single population assumption,
by considering annuli with different mass, semi-major axis and eccentricity.
This would be particularly interesting since
the presence of different stellar populations
is paramount to drive anisotropic segregations~\citep[see, e.g.\@,][]{Szolgyen+2018,Magnan+2022,Mathe+2023,Wang+2023}.
(iii) Include the contribution from the Lense--Thirring precession
driven by the central \BH\@'s spin~\citep[see, e.g.\@,][]{Fragione+2022},
hence adding a linear term in Eq.~\eqref{eq:Master_Equation_short}.
Yet, for the observed S stars~\citep{Gillessen+2017},
the Lense--Thirring precession is rather weak,
so that the \VRR\ non-linearities are expected to remain dominant.
Hence, in this strongly turbulent regime, we anticipate that a self-consistent closure will remain necessary.
(iv) Go beyond the double orbit-average assumption underlying \VRR\@.
For example, orbit-averaging only over the Keplerian motion,
one would need to consider a population of Keplerian ellipses~\citep[see, e.g.\@,][]{Tremaine2020a,Tremaine2020b,Gruzinov+2020}.
The dynamics would still be quadratically non-linear,
i.e.\ formally similar to Eq.~\eqref{eq:generic_equation}.
Yet, this would be no light endeavour since there are more intricate symmetries
and a much larger phase space to consider.
Ultimately, one goal of characterising correlation functions
is to constrain the efficiency with which observed stellar discs
around SgrA*~\citep{Paumard+2006,Bartko+2009,Lu+2009,Yelda+2014,vonFellenberg+2022} can spontaneously dilute on cosmic times~\citep[see, e.g.\@,][]{Panamarev+2022,Fouvry+2023}.

\textit{Harmonic cores}. \VRR\ is not the only self-gravitating system
whose dynamics is inherently non-linear. Of prime importance
is the case of constant density cores, e.g.\@, at the center of dwarf galaxies.
In that case, the self-consistent potential is quadratic,
so that particles follow harmonic orbits that all share the exact same orbital frequency,
making the system \textit{over-resonant}.
This leads to peculiar dynamical behaviours
such as an inefficient dynamical friction~\citep{Read+2006,Zelnikov+2016},
dynamical buoyancy~\citep{Banik+2022},
or lack of orbital resonances~\cite{Kaur+2022}.
Leveraging its similarities with \VRR\@,
it would be rewarding to apply renormalisation techniques
to the harmonic case, so as to better characterise
the associated correlations.

\textit{Liquid crystals}. The Hamiltonian of \VRR\
with a quadrupolar interaction is formally identical
to that of the Maier--Saupe model of liquid crystals~\citep[see, e.g.\@,][]{Maier+1958,Plischke+2006,Roupas+2017}.
In particular, in the disordered limit,
we expect for the isotropic assumption (Section~\ref{sec:Isotropy}) to hold,
a welcome simplification.
On the contrary, in the ordered limit
or in the presence of an external forcing (see Eq.~\ref{eq:generic_equation}),
the emergence of preferred directions would break isotropy.
This would make the renormalisation much more challenging numerically.
Finally, given that liquid crystals comprise a much larger number
of constituents than stars around a supermassive black hole,
we expect for the scale invariance with respect to $N$ (Section~\ref{sec:Invariance})
to hold particularly well.
Exploring the applicability of our \MSR\ renormalisation scheme
to the dynamics of liquid crystals is a promising venue
for future work.

\textit{Four-point function.}
In this work, we focused our interest in computing the two- and three-point
correlation functions of the isotropic \VRR\ system, which already capture
much of the relevant statistical information, such as the lifetime of coherent structures and the degree of non-Gaussianity.
A natural extension would be to compute the four-point function. 
This could, in principle, be achieved by reasoning in a similar way, by using the three-point propagator $G_{\1\2\3}$ as a starting point,
and introducing a four-point interaction vertex yet to be renormalised. This is explored in eq.~{(A6)} of~\cite{MSR1973} 
and fig.~{20} of~\cite{Krommes2002}.
We expect this to be no easy numerical endeavor.

\textit{Kardar--Parisi--Zhang (KPZ) equation.}
The KPZ equation describes the stochastic growth of random surfaces~\citep[see, e.g.\@,][]{Barabasi+1995}.
Its mathematical structure closely resembles that of Eq.~\eqref{eq:generic_equation}.
In addition, the KPZ system shares several of the symmetries exploited in this work~\citep[see, e.g.\@,][]{Canet2025}.
As such, the present \MSR\ numerical scheme
could be adapted to study stochastic processes described by the KPZ equation.
This could prove particularly fruitful in 1D,
where the interaction kernels from Eq.~\eqref{eq:generic_equation}
satisfy stringent exclusion constraints.
This suggests that the numerical cost of the \MSR\ fixed-point scheme would be significantly lower than for \VRR\@.
Given the existence of exact scaling results in 1D~\citep{Prahofer+2004},
our framework could be directly benchmarked against these solutions.
Finally, as the KPZ equation has been extensively studied through the lens of the \textit{Functional Renormalisation Group}~\citep[see, e.g.\@,][]{Canet2022, Canet2025},
applying our \MSR\ approach would offer a valuable opportunity
to compare and cross-validate these related theoretical frameworks.

\begin{acknowledgments}
This work is partially supported by the grant Segal ANR-19-CE31-0017 
of the French Agence Nationale de la Recherche
and by the Idex Sorbonne Universit\'e.
This research was supported in part by grant NSF PHY-2309135
to the Kavli Institute for Theoretical Physics (KITP).
We are grateful to S.\ Rouberol for the smooth running
of the Infinity cluster where the simulations were performed.
We warmly thank C.\ Pichon and A.\ El Rhirhayi for helpful feedback.
We are also grateful to the referees for their insightful suggestions.
\end{acknowledgments}

\appendix

\section{Bare dynamics}
\label{app:Bare}

\subsection{Bare vertex}
\label{app:Bare_gamma}

\subsubsection{Definition}
\label{app:gamma_def}

In Eq.~\eqref{eq:Master_Equation_short}, we introduced the coupling coefficient $\gamma_{123}$, 
which encapsulates all the information about the \VRR\ interaction.
Following eq.~{(A14)} in~\citetalias{Flores+2025}, it reads
\begin{equation}
\gamma_{1 2 3} = \gamma_{a b c} \, \deltaD ( t_{a} , t_{b} ) \, \deltaD ( t_{a} , t_{c} ) ,
\label{eq:def_gamma_next}
\end{equation}
with
\begin{align}
\gamma_{a b c} = E_{a b c} \, \big\{ {} & \mJ_{\ell_c} \!-\! \mJ_{\ell_b} \big\} ,
\label{eq:def_gamma_next_next}
\end{align}
where ${1\!=\!(t, a)}$ and ${a\!=\!(\ell,m)}$. 
In Eq.~\eqref{eq:def_gamma_next_next}, 
the Elsasser coefficients $E_{abc}$ are defined in Eq.~\eqref{eq:def_Elsasser}.
The coupling coefficients $\mJ_\ell$, derived from the Hamiltonian of \VRR, 
are given in eq.~{(A2)} of~\citetalias{Flores+2025}. Remarkably, the $\mJ_\ell$ coefficients vanish for $\ell$ odd. 
In this work, we restrict ourselves to the single-population case, 
i.e.\ the coefficient $\mJ_\ell$ is identical for all stars, 
and therefore independent of their masses, eccentricities, and semi-major axes.
Moreover, we limit the model to the ${\ell\!=\!2}$ interaction, i.e.\ $\mJ_\ell$ vanishes for all ${\ell\!\neq\!2}$. The case of a multipole interaction spectrum is studied in Appendix~\ref{app:InteractionSpectrum}.
We also note that $\gamma_{123}$ is symmetric with respect to its last two indices.

In Eq.~\eqref{eq:Evol_G_3}, we further introduced the generalised coordinate ${\1\!=\!(\eps,t,a)}$, and defined
\begin{align}
\gamma_{\1 \2 \3} = {} & \half \big\{ \gamma_{1 2 3} \!+\! \gamma_{1 3 2} \big\} \, \delta_{\eps_{1}}^{-} \, \delta_{\eps_{2}}^{+} \delta_{\eps_{3}}^{+}
\nonumber
\\
+ {} & \half \big\{ \gamma_{2 1 3} \!+\! \gamma_{2 3 1} \big\} \, \delta_{\eps_{2}}^{-} \, \delta_{\eps_{1}}^{+} \, \delta_{\eps_{3}}^{+}
\nonumber
\\
+ {} & \half \big\{ \gamma_{3 2 1} \!+\! \gamma_{3 1 2} \big\} \, \delta_{\eps_{3}}^{-} \, \delta_{\eps_{2}}^{+} \, \delta_{\eps_{1}}^{+} .
\label{eq:def_gamma}
\end{align}
The coupling coefficient $\gamma_{\1\2\3}$ is referred to as the \textit{bare interaction vertex}.
The introduction of $\eps$ ensures that $\gamma_{\1\2\3}$ becomes fully symmetric under any permutations of ${(\1,\!\2,\!\3)}$. 
This symmetry is crucial for enabling the diagrammatic representation of the vertices, as illustrated in Eq.~\eqref{diag:full}.

In Eq.~\eqref{eq:Evol_G_3}, we also introduced the Pauli-like delta matrix 
${\varsigma_{\1\2}\!=\!\varsigma_{\eps_1 \eps_2} \delta_{12} \!=\! \varsigma_{\eps_1 \eps_2} \delta^{a_2}_{a_1} \deltaD(t_1,t_2)}$, where
\begin{align}
\varsigma_{\tplus\tplus} {} & = \;\;\; 0 ; \quad \;\;\; \varsigma_{\tplus\tmin} = 1 ;
\nonumber
\\
\varsigma_{\tmin\tplus} {} & = -1 ; \quad \;\;\; \varsigma_{\tmin\tmin} = 0 .
\label{eq:components_Pauli}
\end{align}

\subsubsection{Isotropy}
\label{app:gamma_iso}

Using the definition from Eq.~\eqref{eq:def_gamma_next_next}, together with the symmetries of the Elsasser coefficients (see Appendix~\ref{app:Elsasser}), 
we can write, for a single-population system,
\begin{equation}
\gamma_{\1\2\3} = E^{\Delta M}_{a_1 a_2 a_3} \, \gamma^{L}_{\ell_1 \ell_2 \ell_3} ( \xi_{1} , \xi_{2} , \xi_{3}) ,
\label{eq:iso_gamma}
\end{equation}
where ${\xi\!=\!(\eps,t)}$ and the excluded anisotropic coefficient ${E^{\Delta M}_{a_1 a_2 a_3}}$ is given in Eq.~\eqref{eq:def_excl_E}.
The explicit expression for the isotropic bare vertex $\gamma^{L}$ is then
\begin{align}
\gamma^{L}_{\ell_1 \ell_2 \ell_3} [\xi_{1} , \xi_{2} , \xi_{3}] = {} & E^{L}_{a_1 a_2 a_3} \, \deltaD (t_{1} , t_{2}) \, \deltaD (t_{1} , t_{3})
\nonumber
\\
\times  \big\{ {} & \big( \mJ_{\ell_3} \!\!-\!\! \mJ_{\ell_2} \big) \, \delta_{\eps_{1}}^{-} \, \delta_{\eps_{2}}^{+} \delta_{\eps_{3}}^{+}
\nonumber
\\
+ {} & \big( \mJ_{\ell_1} \!\!-\!\! \mJ_{\ell_3} \big) \, \delta_{\eps_{1}}^{+} \, \delta_{\eps_{2}}^{-} \delta_{\eps_{3}}^{+} 
\nonumber
\\
+ {} & \big( \mJ_{\ell_2} \!\!-\!\! \mJ_{\ell_1} \big) \, \delta_{\eps_{1}}^{+} \, \delta_{\eps_{2}}^{+} \delta_{\eps_{3}}^{-} \big\} .
\label{eq:exp_gammaL}
\end{align}

\subsubsection{Stationarity}
\label{app:gamma_stat}

From Eqs.~\eqref{eq:def_gamma_next} and~\eqref{eq:iso_gamma}, we note that the isotropic bare vertex $\gamma^L$ is local in time,
 i.e.\ it is zero unless ${t_1\!=\!t_2\!=\!t_3}$. Moreover, it is time-stationary, meaning that for any time $T$
\begin{equation}
\gamma^{L} (t_{1} \!+\! T , t_{2} \!+\! T , t_{3} \!+\! T) = \gamma^{L} (t_{1} , t_{2} , t_{3}) .
\label{eq:stat_gamma}
\end{equation}

In practice, when implementing Eq.~\eqref{diag:one_loop_iter} numerically, 
it remains to spell out how the Dirac delta in time,
as in Eq.~\eqref{eq:def_gamma_next}, should be discretised.
In order to comply with the Riemann sum approximation
from Eq.~\eqref{eq:shape_int}, we write
\begin{align}
\!\! \int_{- \infty}^{+ \infty} \!\!\!\! \rd t \, \deltaD (t) \, F (t) {} & = F (0)
\nonumber
\\
{} & \simeq \DT \sum_{n} \deltaD (n) \, F(n) ,
\label{eq:approx_int_delta}
\end{align}
with the notation ${ F (n) \!=\! F (n \, \DT) }$.
Hence, as expected, an appropriate discretisation
of the Dirac delta is simply ${ \deltaD (n) \!=\! \delta_{n}^{0} / \DT }$.
We use this formula to represent the Dirac deltas
appearing in Eq.~\eqref{eq:def_gamma_next}.

\subsection{Bare propagator}
\label{app:Bare_g}

\subsubsection{Isotropy}
\label{app:iso_g}
From Eq.~\eqref{eq:def_g} and the identity ${g^{-1}_{\1\2} g_{\2\3}\!=\!\delta_{\1\3}}$, we find that the bare propagator can be recast as
\begin{equation}
g_{\1\2} = \delta_{a_{1}}^{a_{2}} \, g^{L}_{\ell_{1}} (\xi_{1} , \xi_{2}),
\label{eq:iso_g}
\end{equation}
where ${\xi\!=\!(\eps,t)}$. 
In particular, its non-zero components are
\begin{subequations}
\begin{align}
g^{L}_{\ell_{1}} [(+, t_1), (-,t_2)] = \Theta(t_1 \!-\! t_2),\\
g^{L}_{\ell_{1}} [(-, t_1), (+,t_2)] = \Theta(t_2 \!-\! t_1),
\end{align}
\label{eq:iso_g_components}\end{subequations}
where $\Theta$ is the usual Heaviside function,
with the convention ${\Theta(t\!=\!0)\!=\!1}$.

\subsubsection{Stationarity}
\label{app:stat_g}
From Eq.~\eqref{eq:iso_g_components}, we readily find that the bare propagator is time-stationary, 
i.e.\ ${\forall \, T}$
\begin{equation}
g^{L} (t_{1} \!+\! T , t_{2} \!+\! T) = g^{L} (t_{1} , t_{2}) .
\label{eq:stat_g}
\end{equation}
Taking the bare response function $r_\ell$ to be the ${(+ -)}$ component of the bare propagator,
we can finally write, ${\forall \, \ell}$
\begin{equation}
r_\ell(t) = \Theta(t).
\label{eq:def_bare_resp}
\end{equation}

\section{Elsasser coefficients}
\label{app:Elsasser}

\subsection{Definition}
\label{app:Elsasser_Definition}

The Elsasser coefficients appearing in Eq.~\eqref{eq:def_gamma_next_next}
are independent of the considered orbits.
They stem directly from the geometry of the problem,
namely from the unit sphere.
Following the same convention as in~\cite{Fouvry+2019},
they are defined as
\begin{equation}
E_{abc} = E^{L}_{abc} \, E^{M}_{abc} ,
\label{eq:def_Elsasser}
\end{equation}
where ${ E^{L}_{abc} }$ is independent of the indices ${ (m_{a} , m_{b} , m_{c}) }$.
In order to emphasise the symmetries, let us introduce $\tau$
an arbitrary transposition of two indices.
Then, we have the symmetries
\begin{subequations}
\begin{align}
E^{M}_{\tau(abc)} {} & = - E^{M}_{abc} ,
\\
E^{L}_{\tau(abc)} {} & = + E^{L}_{abc} ,
\\
E_{\tau(abc)} {} & = - E_{abc} .
\end{align}
\label{eq:sym_E}\end{subequations}
In practice, the Elsasser coefficients satisfy some stringent exclusion rules.
We therefore define the \textit{triangular} symbol
\begin{equation}
\delta^{\Delta}_{abc} = 
\begin{cases}
\displaystyle 1 \; \text{if} \;
|\ell_{a} \!-\! \ell_{b}| < \ell_{c} < \ell_{a} \!+\! \ell_{b}
\; \text{and} \;
\ell_{a} \!+\! \ell_{b} \!+\! \ell_{c}
\; \text{is odd} ,
\\
\displaystyle 0 \quad \text{otherwise} .
\end{cases}
\label{eq:def_delta_Elsasser}
\end{equation}
This symbol is fully symmetric, i.e.\ we have ${ \delta_{\tau(abc)}^{\Delta} \!=\! \delta_{abc}^{\Delta} }$.
Importantly, we can rewrite the Elsasser coefficients as
\begin{equation}
E_{abc} = \delta_{abc}^{\Delta} \, E^{L}_{abc} \, E^{M}_{abc} .
\label{eq:prop_delta_EL}
\end{equation}
As a consequence, it is natural for us to introduce the excluded anisotropic Elsasser coefficients
\begin{equation}
E^{\Delta M}_{abc} = \delta_{abc}^{\Delta} \, E_{abc}^{M} , 
\label{eq:def_excl_E}
\end{equation}
which are fully antisymmetric in their indices.
It is for the excluded anisotropic Elsasser coefficients that we will write explicit contraction rules.
With this, we can rewrite the total Elsasser coefficients as
\begin{equation}
E_{abc} = E^{L}_{abc} \, E^{\Delta M}_{abc} .
\label{eq:rewrite_E_to}
\end{equation}

\subsection{Contraction rules}
\label{app:Contraction_Elsasser}

The excluded Elsasser coefficients, $E^{\Delta M}_{abc}$ (Eq.~\ref{eq:def_excl_E}),
satisfy a few important contraction rules.
These play a crucial role in ensuring the isotropic symmetries
of ${ (G , \Gamma) }$, as introduced in Section~\ref{sec:Isotropy}.

Following~\cite{Varshalovich1988}, for the diagram of the self-energy in Eq.~\eqref{diag:Dyson_iter},
we use the relation
\begin{equation}
\sum_{\mathclap{m_{2} , m_{3}}} E^{\Delta M}_{a_{1} a_{2} a_{3}} E^{\Delta M}_{a_{1'} a_{2} a_{3}} = \delta_{a_{1}}^{a_{1'}} \, \delta^{\Delta}_{\ell_{1}\ell_{2}\ell_{3}} \, \frac{1}{2 \ell_{1} \!+\! 1} .
\label{eq:contract_W2}
\end{equation}
Importantly, we note that the right-hand side of this expression depends on ${ (m_{1} , m_{1'}) }$
only through ${ \delta_{a_{1}}^{a_{1'}} \!=\! \delta_{\ell_{1}}^{\ell_{1'}} \delta_{m_{1}}^{m_{1'}} }$.

For the one-loop diagram in Eq.~\eqref{diag:one_loop_iter},
we follow~\citep{Varshalovich1988}
\begin{align}
\sum_{\mathclap{m_{4},m_{5},m_{6}}} E^{\Delta M}_{a_{1}a_{4}a_{5}} E^{\Delta M}_{a_{2}a_{4}a_{6}} E^{\Delta M}_{a_{3}a_{5}a_{6}} 
= E^{\Delta M}_{a_{1}a_{2}a_{3}} W^{(3)}[\ell_1,...,\ell_6],
\label{eq:contract_W3}
\end{align}
with 
\begin{align}
W^{(3)}[\ell_1,...,\ell_6] {} & = (-1)^{1 + \ell_{4} + \ell_{5} + \ell_{6}}
\begin{Bmatrix}
\ell_{1} & \ell_{2} & \ell_{3}
\\
\ell_{6} & \ell_{5} & \ell_{4}
\end{Bmatrix} 
\nonumber
\\
{} & \times \delta^{\Delta}_{\ell_{1}\ell_{4}\ell_{5}} \delta^{\Delta}_{\ell_{2}\ell_{4}\ell_{6}} \delta^{\Delta}_{\ell_{3}\ell_{5}\ell_{6}},
\end{align}
where we introduced the Wigner ${6j}$-symbols in the right-hand side.
Importantly, we note that the right-hand side of this expression depends on ${ (m_{1},m_{2},m_{3}) }$
only through the factor ${ E^{\Delta M}_{a_{1}a_{2}a_{3}} }$.

Armed with the relations from Eqs.~\eqref{eq:contract_W2} and~\eqref{eq:contract_W3},
it appears now clearly
how the isotropy of $g$ (Eq.~\ref{eq:iso_g}) and $\gamma$ (Eq.~\ref{eq:iso_gamma})
is transmitted to $G$, $\Sigma$ and $\Gamma$,
when performing the sum over the indices $m$
in the iteration from Eq.~\eqref{diag:Iter}.
We point out that this symmetry argument also holds at bare order,
i.e.\ when imposing the relation from Eq.~\eqref{diag:bare}.
This is detailed in Appendix~\ref{app:Symmetries}.

Similarly, at two-loop order (Eq.~\ref{diag:two_loop}), the same symmetry holds.
Indeed, following~\cite{Varshalovich1988}, one has
\begin{equation}
\sum_{m_4,...,m_9} \! \! \! \! \!E^{\Delta M}_{145} E^{\Delta M}_{468} E^{\Delta M}_{579} E^{\Delta M}_{278} E^{\Delta M}_{369}
= E^{\Delta M}_{123} \, W^{(5)} [\ell_{1} , ... , \ell_{9}] ,
\label{eq:contract_W6}
\end{equation}
where we used the shortened notation ${ E^{\Delta M}_{123} \!=\! E^{\Delta M}_{a_{1} a_{2} a_{3}} }$.
In Eq.~\eqref{eq:contract_W6}, the right-hand side is given by
\begin{align}
W^{(5)} [\ell_{1} , ... , \ell_{9}] {} &= (-1)^{\ell_{3} + \ell_{6} + \ell_{9}} \,
\begin{Bmatrix}
\ell_{1} & \ell_{3} & \ell_{2}
\\
\ell_{4} & \ell_{6} & \ell_{8}
\\
\ell_{5} & \ell_{9} & \ell_{7}
\end{Bmatrix}
\nonumber
\\
{} & \times \delta^{\Delta}_{145} \, \delta^{\Delta}_{468} \, \delta^{\Delta}_{579} \, \delta^{\Delta}_{278} \, \delta^{\Delta}_{369} ,
\label{eq:value_W5}
\end{align}
where we introduced the Wigner $9j$-symbol.

\section{Closures}
\label{app:Expansions}

In this section, we present a heuristic derivation of the one-loop and two-loop closures from Eqs.~\eqref{diag:one_loop} and~\eqref{diag:two_loop}. 

To derive a self-consistent expansion, we first recast Eq.~\eqref{eq:func_Gamma} by expressing the bare vertex $\gamma$ in terms of the renormalised vertex $\Gamma$ and propagator $G$, i.e.\ ${\gamma\!=\!\gamma[\Gamma, \,G]}$. We explicitly write
\begin{equation}
\gamma_{\1\2\3} = \Gamma_{\1\2\3} - \frac{\delta \Sigma_{\1\2}}{\delta G_{\5\6}} \, G_{\5\5'} G_{\6\6'} \Gamma_{\3\5'\6'} ,
\label{eq:func_Gamma_bis}
\end{equation}
where we recall (Eq.~\ref{diag:Sigma})
\begin{equation}
\label{eq:Sigma_bis}
\Sigma_{\1\2} = \half \, \gamma_{\1\3\4} G_{\3\3'} G_{\4\4'} \Gamma_{\2\3'\4'} ,
\end{equation}
and the summation over repeated coordinates ${\1\!=\!(\eps, 1) \!=\!(\eps, t, \ell, m)}$ is implied. 
Consequently, these coordinates may be relabelled freely throughout the derivation.
Importantly, we also recall that $\gamma_{\1\2\3}$, $\Gamma_{\1\2\3}$, and $G_{\1\2}$ are fully symmetric. 
To recover the loop expansions from Eqs.~\eqref{diag:one_loop} and~\eqref{diag:two_loop},
we iteratively generate the loop contributions
by accounting for the derivatives appearing in ${\delta \Sigma / \delta G}$ at different loop orders.

\subsection{One-loop order}
\label{app:One-loop}

For the one-loop order expansion, one inserts the bare condition for the vertex ${ \gamma[\Gamma,\,G] \!=\! \Gamma }$ (Eq.~\ref{diag:bare})
into the self-energy $\Sigma$ from Eq.~\eqref{eq:Sigma_bis}.
At this order, we get
\begin{equation}
\label{eq:Sigma_one_loop}
\Sigma_{\1\2} = \half \, \Gamma_{\1\3\4} G_{\3\3'} G_{\4\4'} \Gamma_{\2\3'\4'} .
\end{equation}
Since the condition ${\delta\gamma / \delta G \!=\!0}$ must be verified,
at this order, we set ${\delta\Gamma / \delta G \!=\!0}$ (Eq.~\ref{diag:bare}). By differentiating Eq.~\eqref{eq:Sigma_one_loop}, this yields
\begin{equation}
\label{eq:deriv_Sigma}
\frac{\delta \Sigma_{\1\2}}{\delta G_{\5\6}} = \Gamma_{\1\4\5} \, G_{\4\4'} \, \Gamma_{\2\4'\6}.
\end{equation}
We then insert Eq.~\eqref{eq:deriv_Sigma} into Eq.~\eqref{eq:func_Gamma_bis} to get 
\begin{equation}
\label{eq:one_loop_expl}
\gamma_{\1\2\3} = \Gamma_{\1\2\3} - \Gamma_{\1\4\5} \, G_{\4\4'} \, \Gamma_{\2\4'\6} \, G_{\5\5'} \, G_{\6\6'} \, \Gamma_{\3\5'\6'} .
\end{equation}
This is exactly Eq.~\eqref{diag:one_loop}.

\subsection{Two-loop order}
\label{app:Two-loop}

To recover the two-loop order expansion, we proceed similarly.
We insert the one-loop result from Eq.~\eqref{eq:one_loop_expl} into Eq.~\eqref{eq:Sigma_bis}.
The self-energy $\Sigma$ becomes 
\begin{align}
\Sigma_{\1\2} = {} & \half \, \Gamma_{\1\3\4} G_{\3\3'} G_{\4\4'} \Gamma_{\2\3'\4'}
\label{eq:Sigma_two_loop}
\\ 
- \, {} & \half \, \Gamma_{\1\5\6} G_{\5\5'} \Gamma_{\3\5'\7} G_{\6\6'} G_{\7\7'} \Gamma_{\4\6'\7'} G_{\3\3'} G_{\4\4'} \Gamma_{\2\3'\4'} . 
\nonumber
\end{align}
We now differentiate Eq.~\eqref{eq:one_loop_expl}
and impose ${\delta\gamma / \delta G \!=\!0}$.
Computing the differential
and neglecting contributions from ${ \delta \Gamma / \delta G }$
in the last term of Eq.~\eqref{eq:one_loop_expl},
we find
\begin{align}
\label{eq:deriv_Gamma_two_loop}
\frac{\delta \Gamma_{\1\2\3}}{\delta G_{\5\6}} & {} = \Gamma_{\1\5\7} G_{\7\7'} \Gamma_{\2\7'\8} G_{\8\8'} \Gamma_{\3\8'\6}
\nonumber
\\
& {} + \Gamma_{\1\5\7} G_{\7\7'} \Gamma_{\3\7'\8} G_{\8\8'} \Gamma_{\2\8'\6}
\nonumber
\\
& {}  + \Gamma_{\2\5\7} G_{\7\7'} \Gamma_{\1\7'\8} G_{\8\8'} \Gamma_{\3\8'\6} .
\end{align}
We now insert Eq.~\eqref{eq:deriv_Gamma_two_loop} into ${\delta \Sigma / \delta G}$
(Eq.~\ref{eq:Sigma_two_loop}). After extensive calculations, we obtain
\begin{align}
\frac{\delta \Sigma_{\1\2} }{\delta G_{\5\6}} & {} =  \Gamma_{\1\4\5} G_{\4\4'} \Gamma_{\2\4'\6} 
\label{eq:deriv_Sigma_two_loop}
\\           
& {}  + \half \Gamma_{\1\4\9} G_{\4\4'} \Gamma_{\5\4'\7} G_{\7\7'} \Gamma_{\2\7'\8} G_{\8\8'} \Gamma_{\6\8'\9'} G_{\9\9'} . 
\nonumber
\end{align}
Importantly, in Eq.~\eqref{eq:deriv_Sigma_two_loop}, we only keep ${\mathcal{O}(\Gamma^4 G^4)}$ terms,
i.e.\ the ones generating two-loop diagrams. 
Among the eleven such terms, ten cancel exactly, leaving only the contribution that yields the \textit{irreducible} two-loop diagram.
Finally, by inserting Eq.~\eqref{eq:deriv_Sigma_two_loop} into Eq.~\eqref{eq:func_Gamma_bis}, we obtain
\begin{align}
\label{eq:two_loop_expl}
\gamma_{\1\2\3} & {} = \Gamma_{\1\2\3} - \Gamma_{\1\4\5} \, G_{\4\4'} \, \Gamma_{\2\4'\6} \, G_{\5\5'} \, G_{\6\6'} \, \Gamma_{\3\5'\6'} 
\nonumber
\\
& {} - \half \Gamma_{\1\4\9} G_{\4\4'} \Gamma_{\5\4'\7} G_{\7\7'} \Gamma_{\2\7'\8} G_{\8\8'} \Gamma_{\6\8'\9'} G_{\9\9'} 
\nonumber
\\
 & {} \times G_{\5\5'} \, G_{\6\6'} \, \Gamma_{\3\5'\6'} .
\end{align}
This is exactly Eq.~\eqref{diag:two_loop}.

Remarkably, this self-consistent expansion procedure
recovers only the contributions from irreducible diagrams.
This is because any diagram that could be simplified by cutting a single internal line
is already accounted for by the dressed propagators and vertices~\cite{MSR1973}.

\section{Numerical implementation of \MSR\@}
\label{app:Numerical}

This appendix provides additional details
regarding the numerical implementation of the \MSR\ renormalisation scheme.
Each of the coming sections complements
the associated one in the main text (Section~\ref{sec:Numerical}).

\subsection{Symmetries}
\label{app:Symmetries}

\subsubsection{Isotropy}
\label{app:Isotropy}

We are interested in \VRR\ in the isotropic limit. 
Within the iterative scheme of Eq.~\eqref{diag:Iter}, the isotropy conditions from
Eqs.~\eqref{eq:iso_gamma} and~\eqref{eq:iso_g} are consistently transmitted to 
the dressed vertex $\Gamma$ and the two-point propagator $G$, as shown in 
Eqs.~\eqref{eq:iso_Gamma} and~\eqref{eq:iso_G}. 
Moreover, from Eq.~\eqref{eq:Sigma}, this property is also satisfied by the self-energy $\Sigma$.
This follows directly from the contraction rules of the Elsasser coefficients, as we now detail.

\textit{Vertex isotropy}. The one-loop diagram in Eq.~\eqref{diag:one_loop_iter} involves a summation 
over three Elsasser coefficients and three isotropic propagators. According to Eq.~\eqref{eq:contract_W3},
this summation ensures that the structure of Eq.~\eqref{eq:iso_gamma} is preserved when renormalising $\Gamma$.  

\textit{Propagator isotropy}. The loop diagram corresponding to the self-energy in Eq.~\eqref{diag:Dyson_iter} 
involves a summation over two Elsasser coefficients and two isotropic 
propagators. According to Eq.~\eqref{eq:contract_W2}, this reproduces the structure of Eq.~\eqref{eq:iso_g} for the propagator $G$.  

The \MSR\ scheme guarantees that isotropy, once imposed at the bare order, remains true under renormalisation.
Exploiting isotropy allows us to focus exclusively on the isotropic quantities
${ [ G^{L} , \Gamma^{L} , g^{L} , \gamma^L , \Sigma^L ] }$.
Performing the \MSR\ closure then reduces to finding self-consistent solutions for ${ [G^{L} , \Gamma^{L}] }$.
The restriction to isotropic functions represents a substantial simplification,
since the angular dependence is captured only through $\ell$.

\subsubsection{Time stationarity}
\label{app:TimeStationarity}

In the same manner, the iterative scheme of Eq.~\eqref{diag:Iter} ensures that the time-stationarity from
Eqs.~\eqref{eq:stat_gamma} and~\eqref{eq:stat_g} is transmitted to $\Gamma$ and $G$, as illustrated in 
Eqs.~\eqref{eq:stat_Gamma} and~\eqref{eq:stat_G}. 

The stationarity of $G$ is fully consistent with the fact that an isotropic distribution
of orientations on the unit sphere is a thermodynamical equilibrium~\citep[see, e.g.\@,][]{Roupas+2017,Magnan+2022}.
As a result, the system’s statistics is expected
to be time-stationary~\citep[see, e.g.\@,][]{Fouvry+2019}.

\textit{Vertex stationarity}. Crucially, under the iterative scheme of Eq.~\eqref{diag:one_loop_iter}, 
the isotropic renormalised vertex $\Gamma^L$ becomes dressed in time, 
i.e.\ it loses the time locality of Eq.~\eqref{eq:def_gamma_next} 
through the time integrals appearing in the one-loop diagram (see Fig.~\ref{fig:Maps_Gamma}). 
Nevertheless, $\Gamma^L$ preserves the stationarity condition of Eq.~\eqref{eq:stat_gamma}, 
since the time integrals involve only time-stationary vertices and propagators, 
ensuring that the resulting vertex remains time-stationary.

\textit{Propagator stationarity}. In a similar way, the loop diagram in Eq.~\eqref{diag:Dyson_iter} 
involves time integrals over time-stationary vertices and propagators, ensuring that the condition
from Eq.~\eqref{eq:stat_g} is transmitted to the two-point propagator $G$.

Given the time symmetry from Eq.~\eqref{eq:stat_G},
instead of storing the full dependence
${(t_{1}, t_{2}) \mapsto G^{L}(t_{1}, t_{2})}$,
it suffices to store only ${t \mapsto G^{L}(0, t)}$.
The same symmetry applies to both $g^{L}$ and $\Sigma^{L}$.
Similarly, from Eq.~\eqref{eq:stat_Gamma}, rather than storing the full time-dependence
${ (t_{1} , t_{2} , t_{3}) \!\mapsto\! \Gamma^{L} (t_{1} , t_{2} , t_{3})}$,
we may limit ourselves to only storing ${ (t_{2} , t_{3}) \!\mapsto\! \Gamma^{L} (0 , t_{2} , t_{3}) }$.
Time stationarity therefore significantly reduces the memory imprint of storing time dependencies.

\subsubsection{Time discretisation}
\label{app:TimeDiscretisation}
In order to compute the diagrams appearing in Eq.~\eqref{diag:Iter}, 
the internal time integrals associated with the propagator legs must be discretised.
To do so, we introduce a timestep $\DT$.
It allows us to approximate time integrals by simple Riemann sums via
\begin{equation}
\!\! \int_{- \infty}^{+ \infty} \!\! \rd t \, F (t) = \DT \sum_{n = - \infty}^{+ \infty} F (n) ,
\label{eq:shape_int}
\end{equation}
with the shortened notation ${ F (n) \!=\! F (n \, \DT) }$,
that will be used throughout this section.
In practice, we only store a finite number of timesteps,
i.e.\ a finite range of time separation.
Therefore, the limits of every time integral to be performed
are determined by the constraints of the time range
associated with the functions involved in the integrand.
Phrased differently, all the sums over the time indices
involve a finite number of terms.

More precisely, the response function,
$R_{\ell}$, is effectively stored as in Eq.~\eqref{eq:def_vec_R}.
We recall that, owing to causality ,
for any ${ n \!<\! 0 }$, we systematically have
${ R_{\ell} (n) \!=\! R_{\ell} (n \, \DT) \!=\! 0 }$.
The representation from Eq.~\eqref{eq:def_vec_R}
assumes that for any ${ n \!>\! \NSTEPS }$,
one has ${ R_{\ell} (n) \!=\! 0 }$.
Phrased differently, we assume that the response function
vanishes for ${ t \!\geq\! \TMAX \!=\! \DT \, \NSTEPS }$.
We proceed similarly to discretise the self-energy, $\Sigma^{L}$,
relying on the time-stationarity from Eq.~\eqref{eq:stat_G}.
In Appendix~\ref{app:CVparam}, we check the convergence of the 
prediction with respect to $\DT$, and also verify that it is free from boundary effects related to $\TMAX$.

For the dressed vertex, $\Gamma^{L}$,
the time discretisation is slightly more cumbersome.
In practice, given Eq.~\eqref{eq:stat_Gamma},
we limit ourselves to only storing, for ${0 \!\le\! n, m \!\le\! \NSTEPS}$,
\begin{align}
\Gamma^{L}(0, n, m), \, \Gamma^{L}(n, 0, m), \,\, \mathrm{and} \,\, \Gamma^{L}(n, m, 0).
\label{eq:Gamma_matrix}
\end{align}
As such, with the representation from Eq.~\eqref{eq:Gamma_matrix},
one has ${ \Gamma^L (n,m,p ) \!=\! \Gamma^{L} (n \, \DT , m \, \DT , p \, \DT) \!=\! 0 }$
for any triplets ${ (n,m,p) }$ such that
${ \Max[n,m,p] \!-\! \Min [n,m,p] \!>\! \NSTEPS }$.

For the bare vertex, $\gamma^{L}$, the exclusions are even more stringent
since it is local in time (Eq.~\ref{eq:def_gamma_next}).
As a consequence, ${ \gamma^{L} (n,m,p) }$
may be non-zero only for ${ n \!=\! m \!=\! p }$,
as detailed in Appendix~\ref{app:Bare_gamma}.

\subsubsection{Triangular exclusion}
\label{app:Triangular}

In practice, the infinite sums over harmonic numbers $\ell$ appearing in the diagrams
from Eq.~\eqref{diag:Iter} must be truncated. 
To do so, we limit ourselves to considering the response functions,
$R_{\ell}$, for ${0 \!\leq\! \ell \!\leq\! \LMAX}$, thus assuming that ${ R_{\ell} }$ vanishes exactly for all ${ \ell \!>\! \LMAX }$.
Similarly, for the renormalised vertex,
we restrict ourselves to only considering vertices,
${ \Gamma^{L}_{\ell_{1} \ell_{2} \ell_{3}}}$,
such that ${ 0 \!\leq\! \ell_{1} , \ell_{2} , \ell_{3} \!\leq\! \LMAX }$.
This restriction also applies to the bare vertex, $\gamma^{L}_{\ell_{1} \ell_{2} \ell_{3}}$.
Moreover, since our model includes only ${\ell\!=\!2}$ interactions, 
for any \textit{triangle} ${(\ell_1, \ell_2, \ell_3)}$, $\ell_1$ interacts mainly with ${\ell_2\!=\! 2}$ and ${\ell_3\!=\! \ell_1 \! \pm 1}$
and vice versa. Hence, $\LMAX$ can be kept small.
In Appendix~\ref{app:CVparam}, we check the convergence of the prediction with respect to $\LMAX$.
The case of a multipole interaction spectrum is studied in Appendix~\ref{app:InteractionSpectrum}.

In addition, the excluded Elsasser coefficients,
$E^{\Delta M}_{a_{1}a_{2}a_{3}}$ (Eq.~\ref{eq:iso_Gamma}),
allow for additional stringent restrictions on the set
of triangles ${ (\ell_{1} , \ell_{2} , \ell_{3}) }$,
that may be associated with non-zero interaction vertices,
$\gamma^{L}_{\ell_{1} \ell_{2} \ell_{3}}$ and $\Gamma^{L}_{\ell_{1} \ell_{2} \ell_{3}}$.
More precisely, following the definition from Eq.~\eqref{eq:prop_delta_EL},
we can limit ourselves to only considering triangles ${ (\ell_{1} , \ell_{2} , \ell_{3}) }$
such that ${ \delta^{\Delta}_{\ell_{1} \ell_{2} \ell_{3}} \!\neq\! 0 }$,
with this triangular exclusion defined in Eq.~\eqref{eq:sym_E}.
Phrased differently, within the set of triangles
${ (\ell_{1} , \ell_{2} , \ell_{3}) }$ with ${ 0 \!\leq\! \ell_{1} , \ell_{2} , \ell_{3} \!\leq\! \LMAX }$,
only a small subset of them are associated with a non-vanishing exclusion.
For example, for ${ \LMAX \!=\! 7 }$,
among the ${ (\LMAX \!+\! 1)^{3} / 6 \!\simeq\! 85 }$ possible distinct triangles,
only $30$ of them are associated with ${ \delta_{\ell_{1} \ell_{2} \ell_{3}}^{\Delta} \!\neq\! 0 }$.

\subsubsection{Antisymmetry}
\label{app:Antisymmetry}

As a result of antisymmetry, when evaluating ${\Gamma^{L}_{\ell_{1} \ell_{2} \ell_{3}} (\xi_{1} , \xi_{2} , \xi_{3})}$ with ${ \xi \!=\! (\eps , t) }$,
up to a minus sign,
one can always sort its harmonic arguments,
i.e.\ one can assume that ${ \ell_{1} \!\leq\! \ell_{2} \!\leq\! \ell_{3} }$.
Once this sorting performed, only four type of triangles may arise,
namely
(i) ${ (\ell_{1} , \ell_{1} , \ell_{1}) }$;
(ii) ${ (\ell_{1} , \ell_{1} , \ell_{2}) }$ with ${ \ell_{1} \!<\! \ell_{2} }$;
(iii) ${ (\ell_{1} , \ell_{2} , \ell_{2}) }$ with ${ \ell_{1} \!<\! \ell_{2} }$;
(iv) ${ (\ell_{1} , \ell_{2} , \ell_{3}) }$ with ${ \ell_{1} \!<\! \ell_{2} \!<\! \ell_{3} }$.
For each of these triangles, one can further leverage antisymmetry,
and, possibly, reduce the number of triplets of ${ (\eps_{1} , \eps_{2} , \eps_{3}) }$
that ought to be effectively considered.
Although cumbersome to implement,
combining the triangular exclusion (Section~\ref{app:Triangular}) 
with the present antisymmetry is key to 
(i) reducing the memory required to store all the vertices and 
(ii) lowering the computational cost of each iteration in the fixed-point search of Eq.~\eqref{diag:Iter}. 
We refer to~\cite{github_MSR} for more detail on our effective implementation.

\subsection{Iteration}
\label{app:Iteration}

\subsubsection{Updating the dressed vertex}
\label{app:UpdateGamma}

The numerical cost of evaluating all the possible one-loop diagrams from  Eq.~\eqref{diag:one_loop_iter}
is significantly reduced by symmetries.
Owing to isotropy (Section~\ref{sec:Isotropy} and~\ref{app:Isotropy}),
the sums over $m$ are immediately carried out.
Owing to time stationarity (Eq.~\ref{eq:stat_Gamma}),
the external legs depend only on two times.
As a result, there are ${ \mO (\NSTEPS^{2} \, \LMAX^{3}) }$ possible choices
for the external legs. 
Here, we do not mention the complexity related to $\eps$, 
since there is only a finite number of possible combinations for it.
For the internal legs, following Eq.~\eqref{eq:iso_G},
each summation over the two-point propagator requires summing
over a single harmonic index, but over two time coordinates.
As a result, there are ${ \mO (\NSTEPS^{6} \, \LMAX^{3}) }$ different values to be summed upon.
Joined together, such a naive evaluation
would then require ${ \mO (\NSTEPS^{8} \, \LMAX^{6}) }$ operations.

We note that the tensor $\Lambda_{\1\2\3}$ from Eq.~\eqref{diag:def_Lambda}, satisfies isotropy
(i.e.\ $\Lambda$ becomes $\Lambda^L$, Eq.~\ref{eq:iso_Gamma}),
time stationarity (Eq.~\ref{eq:stat_Gamma}),
as well as the triangular exclusion (Section~\ref{app:Triangular}).
However, contrary to $\Gamma^{L}$,
$\Lambda^{L}$
is antisymmetric only in its two first coordinates.
Provided one makes minute changes compared to Section~\ref{sec:Antisymmetry} and~\ref{app:Antisymmetry},
this reduced antisymmetry relation can also
be leveraged to reduce the memory imprint of the tensors $\Lambda^{L}$.
We refer to~\cite{github_MSR} for details
on our effective implementation.

For step (i) in Section~\ref{sec:UpdateGamma}, there are ${\mO(\NSTEPS^{2} \, \LMAX^{3})}$ possible external legs, 
with the inner summation requiring ${\mO(\NSTEPS)}$ computations, 
yielding a total complexity of ${\mO(\NSTEPS^{3} \, \LMAX^{3})}$ for all $\Lambda$.
For step (ii), Eq.~\eqref{diag:one_loop_with_Lambda} shows the same number of external legs as previously, 
with an inner summation of ${\mO(\NSTEPS^{3} \, \LMAX^{3})}$, 
giving a total complexity of ${\mO(\NSTEPS^{5} \, \LMAX^{6})}$ for the updated one-loop diagram.

\subsubsection{Inverting the propagator}
\label{app:InvertGinv}

Given the representation of the response function from Eq.~\eqref{eq:def_vec_R},
and owing to causality (Eq.~\ref{eq:def_R}), it is useful to introduce
the Toeplitz matrix
\begin{equation}
\msfR_{\ell} = 
\begin{bmatrix}
R_{0} & 0 & \dots & 0
\\
R_{1} & R_{0} & \dots & 0
\\
\vdots & \vdots & \ddots & \vdots
\\
R_{\scalebox{0.55}{\NSTEPS}} & R_{\scalebox{0.55}{\NSTEPS}-1} & \dots & R_{0} 
\end{bmatrix},
\label{eq:shape_mat_R}
\end{equation}
where, in this section, we use ${ R_{n} \!=\! R_{\ell} (t \!=\! n \, \DT ) }$,
for ${0\!\leq \!n\!\leq\NSTEPS }$,
with $\DT$ the considered timestep,
and drop the index $\ell$ for concision.
Phrased differently, the matrix $\msfR_{\ell}$
can directly be constructed from the vector $\bR_{\ell}$
introduced in Eq.~\eqref{eq:def_vec_R}.

Following Eq.~\eqref{eq:components_G},
the isotropic two-point propagator,
once discretised in time, can be represented via the matrix
\begin{equation}
\msfG_{\ell}^{L} 
=
\begin{bmatrix}
\msfC_{\ell} & \msfR_{\ell}
\\
\msfR_{\ell}^{\rT} & 0
\end{bmatrix} ,
\label{eq:mat_shape_G}
\end{equation}
where $\msfC_{\ell}$ is the matrix representation of the two-point correlation function,
satisfying the \FDT\ from Eq.~\eqref{eq:FDT}.

Inverting the matrix from Eq.~\eqref{eq:mat_shape_G}, we find
\begin{equation}
\big[ \msfG_{\ell}^{L} \big]^{-1} =
\begin{bmatrix}
0 & (\msfR_{\ell}^{-1})^{ \rT}
\\
\msfR_{\ell}^{-1} & - \msfR_{\ell}^{-1} \msfC_{\ell} (\msfR_{\ell}^{-1})^{ \rT} 
\end{bmatrix} ,
\label{eq:inv_G} 
\end{equation}
where $\msfR^{-1}_{\ell}$ is the inverse of the matrix in Eq.~\eqref{eq:shape_mat_R}.
Thus, we may limit ourselves to computing the ${(-+)}$ component of Eq.~\eqref{diag:Dyson_iter}, namely ${ [\msfG^{L}_{\ell}]^{-1}_{(\tmin \tplus)} \!=\! \msfR_{\ell}^{-1} }$.

We now detail the algorithm associated with the inversion from $\msfR^{-1}_{\ell}$ to $\msfR_{\ell}$.
Following the notation from Eq.~\eqref{eq:shape_mat_R},
for all ${ 0 \!\leq\! \ell \!\leq\! \LMAX }$,
evaluating the Dyson equation (Eq.~\ref{diag:Dyson_iter})
provides us with the vector
${ \bR_{\ell}^{-1} \!=\! [R_{0}^{-1} , R_{1}^{-1}, ... , R_{\NSTEPS}^{-1}] }$,
with the shortened notation ${ R_{n}^{-1} \!=\! R_{\ell}^{-1} (t \!=\! n \, \DT) }$.
Our goal is now to compute the associated inverse vector
${ \bR_{\ell} \!=\! [R_{0} , R_{1}, ... , R_{\NSTEPS}] }$.

By definition, we have
\begin{equation}
\int \!\! \rd t_{2} \, R_{\ell} (t_{1} \!-\! t_{2}) \, R_{\ell}^{-1} (t_{2} \!-\! t_{3}) = \deltaD [t_{1} , t_{3}] .
\label{eq:def_prod_inv_R}
\end{equation}
Discretising this relation with respect to time, we obtain
\begin{equation}
\sum_{n_{2}} \DT \, \big[ \msfR_{\ell} \big]_{n_{1} n_{2}} \, \big[ \msfR_{\ell}^{-1} \big]_{n_{2} n_{3}} = \frac{1}{\DT} \, \msfI_{n_{1} n_{3}} ,
\label{eq:disc_prod_inv_R}
\end{equation}
with $\msfI$ the identity matrix,
and the matrix $\msfR$ introduced in Eq.~\eqref{eq:shape_mat_R}.
From Eq.~\eqref{eq:disc_prod_inv_R}, we obtain the matrix relation
\begin{equation}
\msfR_{\ell} \, \msfR_{\ell}^{-1} = \frac{1}{\DT^{2}} \, \msfI .
\label{eq:link_inv_R}
\end{equation}
As such, up to a factor ${ 1 / \DT^{2} }$,
the matrices $\msfR_{\ell}$ and $\msfR_{\ell}^{-1}$
are true inverses of each other.

Because they are both lower triangular matrices (see Eq.~\ref{eq:shape_mat_R}),
following~\cite{Commenges+1984},
the inversion ${ \msfR_{\ell}^{-1} \!\mapsto\! \msfR_{\ell} }$
can be performed efficiently using a direct substitution method.
More precisely, the algorithm reads
\begin{subequations}
\begin{align}
R_{0} {} & = \frac{1}{\DT^{2}} \, \frac{1}{R_{0}^{-1}} ,
\label{eq:init_Toep}
\\
R_{n}^{} {} & = - \frac{1}{R_{0}^{-1}} \sum_{k = 1}^{n} R_{n-k} R_{k}^{-1}
\;\; \text{for} \;\; n = 1, ..., \NSTEPS .
\label{eq:recc_Toep}
\end{align}
\label{eq:inv_Toep}\end{subequations}
For a given $\ell$, the complexity of the inversion
${ \bR^{-1}_{\ell} \!\mapsto\! \bR_{\ell} }$ scales like ${ \mO (\NSTEPS^{2}) }$.
Since it must be made for all harmonics $\ell$,
independently from one another,
overall, the inversion of the two-point propagator
requires ${ \mO (\NSTEPS^{2} \, \LMAX) }$ operations.

With the present discretisation,
the bare response function, ${ r_{\ell} (t) }$ (Eq.~\ref{eq:def_bare_resp}),
and its inverse, ${ r_{\ell}^{-1} (t) }$,
are also both lower triangular Toeplitz matrices.
They are fully characterised by the vectors
\begin{subequations}
\begin{align}
\br_{\ell} {} & = [1, 1, ... , 1],
\\
\br_{\ell}^{-1} {} & = \frac{1}{\DT^{2}} [1 , - 1 , 0 , ... , 0] .
\end{align}
\label{eq:disc_r_rinv}\end{subequations}

\section{Initial conditions}
\label{app:IC}

\subsection{Two-point function}
\label{app:IC_2pt}

In this Appendix, we compute the amplitude of the initial two-point correlation function, ${C_{ab} (t_{a} \!=\! 0 , t_{b} \!=\! 0)\!=\! C^{(0)}_{ab}}$, 
with ${a\!=\!(\ell,m)}$ and assuming ${ \ell_{a} , \ell_{b} \!>\! 0 }$.
Following Eqs.~\eqref{eq:def_Fd} and~\eqref{eq:expansion_vphid},
the stochastic field, ${ \vphi_{a} (t) }$, is given instantaneously by
\begin{equation}
\vphi_{a} (t) = \sum_{i = 1}^{N} Y_{a} [\hbL_{i} (t)].
\label{eq:calc_phia}
\end{equation}
For ${ t_{a} \!=\! t_{b} \!=\! 0 }$, following Eq.~\eqref{eq:def_C_generic},
we find that the two-point correlation function reads
\begin{align}
C_{ab}^{(0)} {} & =  \sum_{i = 1}^{N} \sum_{j = 1}^{N} \big\langle Y_{a} [\hbL_{i} (0)] \, Y_{b} [\hbL_{j} (0)] \big\rangle 
\nonumber
\\
{} & - \sum_{i = 1}^{N} \sum_{j = 1}^{N} \big\langle Y_{a} [\hbL_{i} (0)] \big\rangle \, \big\langle Y_{b} [\hbL_{j} (0)] \big\rangle ,
\label{eq:calc_correl_init}
\end{align}
where the average is over the particles' initial conditions.
At the initial time, the particles are drawn independently from one another,
and their initial position on the unit sphere, ${ \hbL_{i} (0) }$,
is drawn according to the uniform probability distribution function, ${ P (\hbL) \!=\! 1 / (4 \pi) }$.
As a consequence, Eq.~\eqref{eq:calc_correl_init} simply becomes
\begin{align}
C_{ab}^{(0)} = {} & N \!\! \int \!\! \frac{\rd \hbL}{4 \pi} \, Y_{a} (\hbL) \, Y_{b} (\hbL) 
\nonumber
\\
- \, {} & N \bigg[ \!\! \int \!\! \frac{\rd \hbL}{4 \pi} \, Y_{a} (\hbL) \bigg] \bigg[ \!\! \int \!\! \frac{\rd \hbL}{4 \pi} \, Y_{b} (\hbL) \bigg] .
\label{eq:calcer_moment_1}
\end{align}
Having assumed ${ \ell_{a} , \ell_{b} \!>\! 0 }$, the last term in Eq.~\eqref{eq:calcer_moment_1} vanishes.
Our normalisation convention for the real spherical harmonics~\citepalias{Flores+2025}
imposes ${ \!\int\! \rd \hbL \, Y_{a} Y_{b} \!=\! \delta_{a}^{b} }$,
so that we find
\begin{equation}
C_{ab}^{(0)} = \delta_{a}^{b} \, \frac{N}{4 \pi} = \delta_{a}^{b} \, C_0.
\label{eq:calcest_moment_1}
\end{equation}
This initial condition is the one imposed in the \FDT\ relation of Eq.~\eqref{eq:FDT}.

\subsection{Coherence time}
\label{app:IC_TC}

According to eq.~{(19)} of~\cite{Fouvry+2019}, in the single-population ${\ell\!=\!2}$ interaction model of \VRR, the coherence time, i.e.\ the characteristic timescale for the decay of correlations,
is given by
\begin{equation}
\Tc = \frac{4 \sqrt{5}}{\sqrt{3}} \sqrt{\frac{a^3_\star}{G \MBH}} \frac{\MBH}{M_\star} \sqrt{N},
\label{eq:def_Tc}
\end{equation}
where ${a_\star}$ is the semi-major axis of a star, ${M_\star\!=\!Nm_\star}$ the total stellar mass, and $\MBH$ the mass of the \BH\@.
We point out that $\Tc$ scales like $N^{1/2}$.

From Eq.~\eqref{eq:def_Tc}, we can define the isotropic Gaussian correlation~\citep[see eq.~{(18)} in][]{Fouvry+2019}
\begin{equation}
C^\mathrm{G}_\ell (t) = C_0 \exp{\big[- \tfrac{A_\ell}{2} (t / \Tc)^2\big]},
\end{equation}
where ${A_\ell\!=\!\ell(\ell\!+\!1)}$.
Taking into account (i) isotropy from Eq.~\eqref{eq:iso_G}, (ii) the \FDT\ from Eq.~\eqref{eq:FDT}, 
and (iii) the propagator components from Eq.~\eqref{eq:components_G},
we can construct the Gaussian two-point propagator $G_{\mathrm{G}}$ from Section~\ref{sec:FixedPoint}.

\subsection{Three-point function}
\label{app:IC_3pt}

Similarly to Eq.~\eqref{eq:calc_correl_init}, assuming ${ \ell_{a} , \ell_{b}, \ell_{c}\!>\! 0 }$, we can write the initial three-point correlation function as
\begin{equation}
C_{abc}^{(0)} = \sum_{i,j,k = 1}^{N} \big\langle Y_{a} [\hbL_{i} (0)] \, Y_{b} [\hbL_{j} (0)] \, Y_{c} [\hbL_{k} (0)] \big\rangle.
\label{eq:calc_correl_init_3}
\end{equation}
The average appearing in Eq.~\eqref{eq:calc_correl_init_3} is zero unless ${i \!=\! j \!=\! k }$. Thus, we get
\begin{align}
C_{abc}^{(0)} = {} & N \!\! \int \!\! \frac{\rd \hbL}{4 \pi} \, Y_{a} (\hbL) \, Y_{b} (\hbL) \, Y_{c} (\hbL) = \frac{N}{4 \pi} \, A_{a b c} .
\label{eq:calc_correl_init_3_short}
\end{align}
Following~\cite{Varshalovich1988}, the integral ${ A_{a b c} \!=\! A^L_{a b c} A^M_{a b c} }$ is given by
\begin{equation}
A^L_{a b c} = \sqrt{\frac{(2\ell_a + 1)(2\ell_b + 1)(2\ell_c + 1)}{4\pi}} 
\begin{Bmatrix}
\ell_{a} & \ell_{b} & \ell_{c}
\\
0 & 0 & 0
\end{Bmatrix} ,
\label{eq:def_alphaL}
\end{equation}
and
\begin{equation}
A^M_{a b c} = \;\; \sum_{\mathclap{\eps_a,\eps_b,\eps_c = \pm1}} \;\; \RePart [K_{m_a m_b m_c}^{\eps_a \eps_b \eps_c}]
\begin{Bmatrix}
\ell_{a} & \ell_{b} & \ell_{c}
\\
\eps_a m_a & \eps_b m_b & \eps_c m_c
\end{Bmatrix} .
\end{equation}
Here, using the same convention as in~\cite{Fouvry+2019},
we introduced
\begin{equation}
K_{m_a m_b m_c}^{\eps_a \eps_b \eps_c} = \kappa_{m_a}^{\eps_a} \kappa_{m_b}^{\eps_b} \kappa_{m_c}^{\eps_c},
\end{equation}
as well as 
\begin{equation}
\kappa_m^{\eps} =
\begin{cases}
\frac{i\,(-1)^m}{\sqrt{2}}, & \eps = +1,\; m < 0, \\[1.2ex]
\frac{1}{2}, & \eps = +1,\; m = 0, \\[1.2ex]
\frac{1}{\sqrt{2}}, & \eps = +1,\; m > 0, \\[1.2ex]
-\frac{i}{\sqrt{2}}, & \eps = -1,\; m < 0, \\[1.2ex]
\frac{1}{2}, & \eps = -1,\; m = 0, \\[1.2ex]
\frac{(-1)^m}{\sqrt{2}}, & \eps = -1,\; m > 0~.
\end{cases}
\end{equation}
Importantly, given the exclusion rules for the Wigner $3j$-symbols, the coefficient $A^L_{a b c}$ from Eq.~\eqref{eq:def_alphaL} is zero if ${\ell_a \!+\! \ell_b \!+\! \ell_c}$ is odd, i.e.\ for odd triangles.
This property plays a crucial role in the discussion from Section~\ref{sec:Invariance}.

\section{Three-point correlation function}
\label{app:Prediction_3_point}

In section~{III.B} of~\citetalias{Flores+2025}, we provide the general definition and derivation
of the $n$-point propagator (or cumulant) $G_{\1...\textbf{n}}$, obtained from a cumulant generating function, 
with extended coordinates defined as ${\1\!=\!(\eps,1)\!=\!(\eps,t,\ell,m)}$.
In the present work, we focus only on the two- and three-point propagators, $G_{\1\2}$ and $G_{\1\2\3}$. 
Notably, the ${(+\!+)}$ component of $G_{\1\2}$ corresponds to the two-point correlation function ${C_{12}}$ (see Eq.~\ref{eq:components_G}),
while the ${(+\!+\!+)}$ component of $G_{\1\2\3}$ coincides with the three-point correlation function ${C_{123}}$.
Indeed, one has
\begin{align}
G^{\tplus \tplus \tplus}_{1 \, 2 \, 3} = C_{123} = {} & \ml \vphi_1 \vphi_2 \vphi_3 \mr ,
\label{eq:3_point_correl}
\end{align}
where we recall that ${\ml \vphi \mr \!=\! 0}$ in isotropic \VRR.

In practice, the \MSR\ formalism predicts the three-point correlation to be given by (see Eq.~\ref{eq:def_Gamma_3_point})
\begin{equation}
G_{\1\2\3} = G_{\1 \1'} G_{\2 \2'} G_{\3\3'} \Gamma_{\1'\2'\3'} .
\label{eq:def_G_3_point_app}
\end{equation}
In the isotropic limit, we may then use Eqs.~\eqref{eq:iso_G} and~\eqref{eq:iso_Gamma},
to find 
\begin{equation}
G_{\1\2\3} = E^{\Delta M}_{a_1 a_2 a_3} \, G^{L}_{\ell_1 \ell_2 \ell_3} [\xi_{1} , \xi_{2} , \xi_{3}],
\label{eq:iso_3_point}
\end{equation}
with ${a\!=\!(\ell,m)}$ and ${\xi \!=\!\ (\eps, t ) }$.
Here,
\begin{align}
G^{L}_{\ell_1 \ell_2 \ell_3} [\xi_{1} , \xi_{2} , \xi_{3}] {} & = G^L_{\ell_{1}} [\xi_{1} , \xi_{1'}] \, G^L_{\ell_{2}} [\xi_{2} , \xi_{2'}] G^L_{\ell_3}[\xi_{3} , \xi_{3'}]
\nonumber
\\
{} & \times \, \Gamma^{L}_{\ell_1 \ell_2 \ell_3} [\xi_{1'} , \xi_{2'} , \xi_{3'}] .
\label{eq:def_G3L}
\end{align}
Importantly, $G^{L}$ has the same symmetries as $\Gamma^{L}$,
in particular it is time-stationary (Section~\ref{sec:TimeStationarity}),
and satisfies some triangular exclusions (Section~\ref{sec:Isotropy}).

Joining Eq.~\eqref{eq:3_point_correl} and Eq.~\eqref{eq:iso_3_point},
we thus find that \MSR\ predicts for the three-point correlation 
to behave like
\begin{equation}
\ml \vphi_{1} \, \vphi_{2} \, \vphi_{3} \mr = E^{\Delta M}_{a_{1} a_{2} a_{3}} \, C^{L}_{\ell_{1} \ell_{2} \ell_{3}} (t_{1} , t_{2} , t_{3}) ,
\label{eq:wrap_3pt}
\end{equation}
where the isotropic three-point function is simply
\begin{equation}
C^{L}_{\ell_{1} \ell_{2} \ell_{3}} (t_{1} , t_{2} , t_{3}) \!=\! G^{L}_{\ell_1 \ell_2 \ell_3} [\{ + \!, t_{1} \} , \{ + \!, t_{2} \} , \{ + \!, t_{3} \} ] .
\label{eq:link_CL}
\end{equation}
The \MSR\ bare and one-loop predictions for the three-point correlations presented in Section~\ref{sec:Prediction_3_point}
were obtained through the numerical evaluation of the ${(+\!+\!+)}$ component of Eq.~\eqref{eq:def_G3L}.

\section{Convergence tests for the \MSR\ scheme}
\label{app:CVparam}

Our numerical implementation of the \MSR\ renormalisation scheme involves
a few crucial control parameters, in particular
(i) $\DT$, the timestep of discretisation (Section~\ref{sec:Discretisation_time_harm});
(ii) $\TMAX$, the total depth in time stored for the response function (Section~\ref{sec:Discretisation_time_harm});
(iii) $\LMAX$, the cutoff in the harmonic number $\ell$ (Section~\ref{sec:Discretisation_time_harm});
(iv) $\ITER$, the total number of iterations performed within the fixed-point search (Section~\ref{sec:Iteration}).
Given the numerical cost of the \MSR\ iteration (see Section~\ref{sec:Iteration}),
it is essential to carefully choose these control parameters,
while ensuring an appropriate numerical convergence.
This is the topic of the present section.

\begin{figure}[htbp]
\centering
\includegraphics[width=\linewidth]{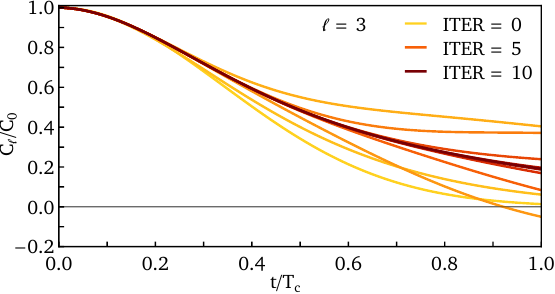}
\caption{Convergence with respect to the number of iterations, \ITER\@,
shown for ${\ell \!=\! 3}$. Colors range from yellow (iteration 0) to red (iteration 10). 
Parameters are set to ${\DT \!=\! \Tc / 80}$, ${\TMAX \!=\! \Tc}$, and ${\LMAX \!=\! 7}$. 
The convergence with respect to \ITER\ is within 1\%.}
\label{fig:CV_iter}
\end{figure}

\begin{figure}[htbp]
\centering
\includegraphics[width=\linewidth]{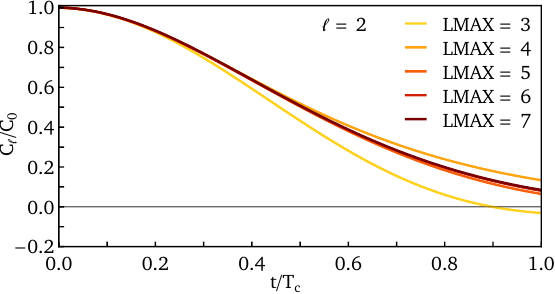}
\caption{Convergence with respect to the maximum harmonic number, \LMAX\@,
for scale ${\ell\!=\!2}$ and varying values ${ \LMAX \!=\!3,4,5,6,7}$.
Parameters are set to ${\DT \!=\! \Tc /80}$, ${\TMAX \!=\! \Tc}$, and ${\ITER \!=\! 10}$. 
The convergence with respect to \LMAX\ for ${\ell\!=\!2}$ is within 2\%.}
\label{fig:CV_LMAX}
\end{figure}

\begin{figure}[htbp]
\centering
\includegraphics[width=\linewidth]{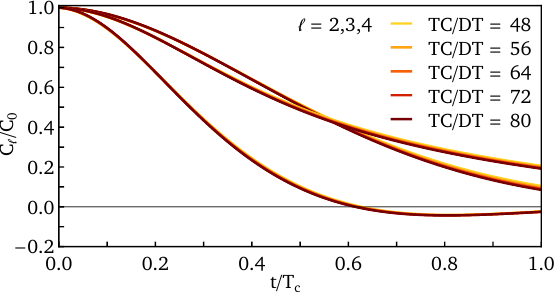}
\caption{Convergence with respect to the timestep \TCDT\@, for scales ${\ell\!=\!2,3,4}$ 
and varying values ${ \TCDT \!=\!48,56,64,72,80 }$.
Parameters are set to ${\LMAX \!=\! 7}$, ${\TMAX \!=\! \Tc}$, and ${\ITER \!=\! 10}$. 
The convergence with respect to \TCDT\ is better than 1\%.}
\label{fig:CV_Nsteps}
\end{figure}

\begin{figure}[htbp]
\centering
\includegraphics[width=\linewidth]{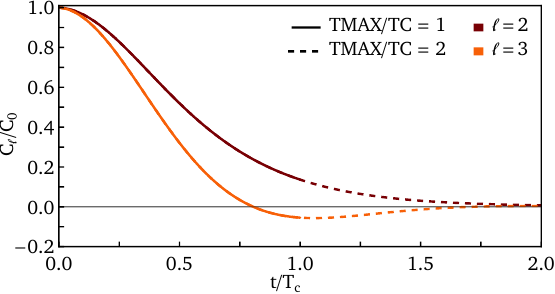}
\caption{Convergence with respect to the time depth \TMAXTC\@, for scales ${\ell\!=\!2,3}$
and varying values ${ \TMAXTC \!=\!1,2 }$.
Parameters are set to ${\LMAX \!=\! 4}$, ${\DT \!=\! \Tc /48}$, and ${\ITER \!=\! 10}$. 
Here, parameters are downgraded to account for the increased temporal depth \TMAX, 
which is more computationally expensive. 
No boundary effects arise from the cutoff in \TMAX.}
\label{fig:CV_TMAX}
\end{figure}

For all \MSR\ predictions (both bare and one-loop) presented in the main text (Section~\ref{sec:Application}), 
we use the parameters ${\DT \!=\!\Tc / 80}$, ${\TMAX \!=\! \Tc}$, ${\LMAX \!=\! 7}$, and ${\ITER \!=\! 10}$.
We now examine the dependence of our results,
namely the isotropic two-point correlation function,
as a function of these parameters.
More precisely, we check for the convergence with respect to 
the number of iterations, $\ITER$ (Fig.~\ref{fig:CV_iter});
the maximum harmonic number, $\LMAX$ (Fig.~\ref{fig:CV_LMAX});
the timestep, $\DT$ (Fig.~\ref{fig:CV_Nsteps});
the time depth, $\TMAX$ (Fig.~\ref{fig:CV_TMAX}).
All these figures confirm that the chosen parameters
provide converged predictions within a few percents.
We find that \LMAX\ is the most limiting parameter.

\begin{figure}[htbp]
\centering
\includegraphics[width=1.0\linewidth]{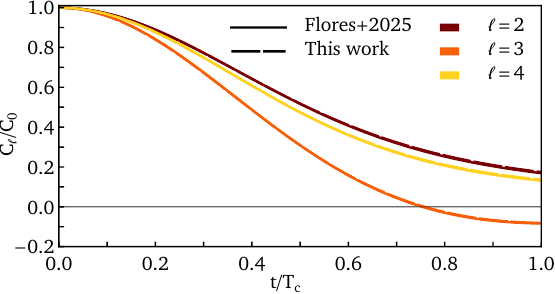} 
\caption{Same as Fig.~\ref{fig:NaivevsMSR},
but here focusing on the naive one-loop prediction from Eq.~\eqref{diag:one_loop_naive}.
The full line is the numerical result obtained
by solving the integro-differential equation satisfied
by the isotropic two-point correlation function~\citepalias[eq.~{(G2)} in][]{Flores+2025}.
The dashed line has been obtained using the generic numerical scheme
from Section~\ref{sec:Numerical}.
Both predictions are in agreement within 2\%.
This is a reassuring sanity check of the present numerical implementation.
}
\label{fig:CV_Naive}
\end{figure}

As a final sanity check of the \MSR\ numerical scheme, 
we verify that it can reproduce both the bare prediction~\citepalias[eq.~{(29)} in][]{Flores+2025} 
and the naive one-loop prediction~\citepalias[eq.~{(G2)} in][]{Flores+2025}
already obtained in~\citetalias{Flores+2025}.
To do so, in the present \MSR\ approach,
it suffices to modify Eq.~\eqref{diag:one_loop_iter} to Eq.~\eqref{diag:bare} for the bare prediction, 
and to Eq.~\eqref{diag:one_loop_naive} for the naive one-loop prediction.
The latter is illustrated in Fig.~\ref{fig:CV_Naive}.

\section{$N$-body measurements}
\label{app:NumericalSimulations}

In this Appendix, we detail our implementation
of the direct $N$-body simulations,
and our approach to measure correlation functions therein.

\subsection{Numerical simulations}
\label{app:Simulations}

\begin{figure}[htbp!]
\centering
\includegraphics[width=1.0\linewidth]{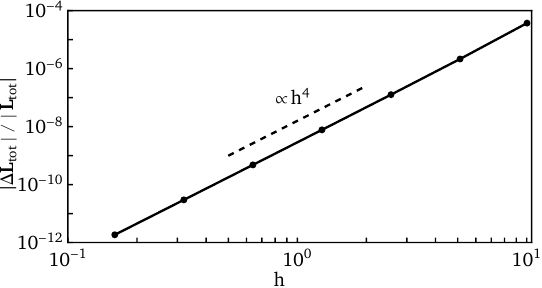}
\caption{Relative error in the total angular momentum, ${ \bLtot \!\propto\! \sum_{i} \hbL_{i} }$,
evaluated at the time ${ t \!=\! 100.24 }$
in one $N$-body simulation with ${ N \!=\! 1\,000 }$,
using the same setup as in Fig.~\ref{fig:Money_plot},
as a function of the timestep $h$.
Relative errors are computed with respect to ${ \bLtot (t \!=\! 0) }$.
The integration scheme used in our $N$-body simulations
is recovered to be fourth-order accurate.
}
\label{fig:Convergence_Nbody}
\end{figure}

Throughout this paper, we investigate the exact same
$N$-body system as in~\citetalias{Flores+2025}.
We refer to appendix~{D} therein for the specific numerical setup.
Compared to~\citetalias{Flores+2025},
we improved the integration scheme by using a fourth-order
Munthe-Kaas scheme~\cite{MuntheKaas1999}.
We refer to eq.~{(45)} in~\cite{Fouvry+2022} for more detail.
In Fig.~\ref{fig:Convergence_Nbody}, we check that this scheme is indeed
fourth-order accurate.

For each simulation, we used a timestep ${ h \!=\! \Tc / 326 }$, with a dump every ${2 h}$,
and integrated the dynamics up to ${ t_{\mathrm{max}} \!\simeq\! 12\, \Tc}$.
Here, $\Tc$ is given by Eq.~\eqref{eq:def_Tc} and depends on the number of particles $N$.
One such simulation required ${\leq\!10\mathrm{s}}$ on a single core for ${N\!=\!10\,000}$, 
while the subsequent calculation of the empirical correlation functions took ${ \leq\!4\mathrm{s} }$.
With these settings, the final relative error in the total energy and
total angular momentum were typically on the order of $10^{-12}$ and $10^{-10}$ respectively.
In practice, we performed $10\,000$ realisations of these simulations.
Including the computation of empirical correlations,
this required about ${20\, \mathrm{min}}$ on 128 cores.

\subsection{Ensemble average}
\label{app:Average}

Once a set of simulations (realisations) is available, we must average over them appropriately to obtain the $N$-body measurements of the correlations. 
We refer to appendix~D of~\citetalias{Flores+2025} for more detail on the measurement of the isotropic two-point correlation from Eq.~\eqref{eq:def_iso_C}.
We now describe our procedure to measure the isotropic three-point correlation from Eq.~\eqref{eq:intro_CL}.

\textit{Averaging over $m$}.
To measure the isotropic component of Eq.~\eqref{eq:intro_CL}, namely
\begin{equation}
 {\ml \vphi_{1} \, \vphi_{2} \, \vphi_{3} \mr = E^{\Delta M}_{a_{1} a_{2} a_{3}} \, C^{L}_{\ell_{1} \ell_{2} \ell_{3}} (t_{1} , t_{2} , t_{3})},
\end{equation}
we average over all triplets ${(m_1, m_2, m_3)}$ for which 
${E^{\Delta M}_{a_{1} a_{2} a_{3}}}$ is non-zero, with ${a \!=\! (\ell, m)}$. 

\textit{Averaging over time}.
Owing to stationarity, we have
\begin{equation}
C^{L}_{\ell_{1} \ell_{2} \ell_{3}} (0, t_{2}, t_{3}) = 
C^{L}_{\ell_{1} \ell_{2} \ell_{3}} (T, t_{2}\!+\!T, t_{3}\!+\!T),
\end{equation}
for any $T$. This allows us to perform a time average over $T$.
This increases the signal-to-noise ratio.

\begin{figure}[htbp]
\centering
\includegraphics[width=1.0\linewidth]{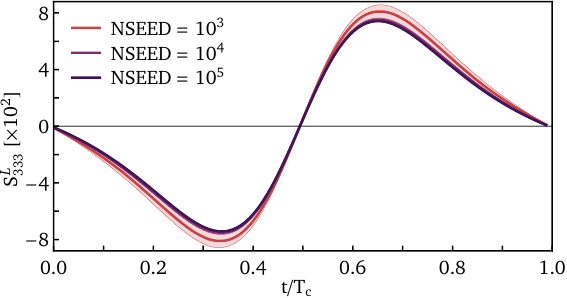} 
\caption{Convergence of the skewness profile with respect to the parameter \NSEED\@, 
for ${ \NSEED \!=\!10^3,10^4,10^5}$, for the triangle ${(\ell_1,\ell_2,\ell_3)\!=\!(3,3,3)}$ and ${N\!=\!1\,000}$. For $\NSEED \!\gtrsim\! 10^{4} $, the measurement is converged
to better than $1\%$.
}
\label{fig:CV_NSEED}
\end{figure}

\textit{Averaging over realisations}.
Finally, we average over the set independent realisations, 
$\NSEED$ in total. We applied a bootstrap resampling procedure 
to estimate the mean and its associated uncertainties, taken as the 16th and 84th percentiles 
(hence corresponding to ${\sim\!1\sigma}$ over the available realisations). 
In Figs.~\ref{fig:Money_plot} to~\ref{fig:slope}, the errors presented correspond to these percentiles.
In Fig.~\ref{fig:CV_NSEED}, we check for the convergence of the skewness profile
with respect to \NSEED\@, for triangle ${(\ell_1,\ell_2,\ell_3)\!=\!(3,3,3)}$ and ${N\!=\!1\,000}$.
For that particular triangle, we find that using ${ \NSEED \!=\! 10^{4} }$
ensures an appropriate convergence of the $N$-body measurement.

\subsection{Figure parameters}
\label{app:Param}

In this section, we specify the values of the parameters used
when performing the $N$-body measurements
of the two- and three-point correlation functions. 

\textit{Figure~\ref{fig:Money_plot}}.
We used ${N\!=\!1\,000}$ and ${\NSEED \!=\! 10^4}$.

\textit{Figure~\ref{fig:Maps_full} and~\ref{fig:G_3_profile_full}}. We used
${N\!=\!10^4}$ and ${\NSEED \!=\! 10^4}$ for triangles (a), (b) and (d);
${N\!=\!10^4}$ and ${\NSEED \!=\! 5\!\times\!10^4}$ for triangle (c).

\textit{Figure~\ref{fig:G_3_even_profile} and~\ref{fig:slope}}.
As $N$ increases, the signal approaches zero,
so that one must increase $\NSEED$ with the value of $N$.
In practice, we used ${ \NSEED \!=\! (N / 100) \!\times\! 10^{4} }$.

\textit{Figure~{\ref{fig:G_3_profile} and~\ref{fig:slope}}}:
As $N$ increases, the signal becomes invariant with respect to $N$,
so that one may choose $\NSEED$ independently of $N$.
In practice, we used ${ \NSEED \!=\! 10^{4} }$.

\section{Interaction Spectrum}
\label{app:InteractionSpectrum}

In our modelling of \VRR\@, we restricted ourselves to the quadrupolar ${\ell \!=\! 2}$ interaction (see Appendix~\ref{app:Bare_gamma}).
As such, the $\mJ_\ell$ coefficients appearing in Eq.~\eqref{eq:def_gamma_next_next}
vanish for ${\ell\! \neq \!2}$. 
Naturally, one could wonder how the results from Section~\ref{sec:Application}
would change if a broader interaction spectrum were considered for $\mJ_{\ell}$,
hence for $\gamma$.
Such a multipole interaction spectrum would better reproduce the astrophysical regime,
where higher multipoles can contribute to the pairwise interaction~\citep[see appendix~{B} in][]{Kocsis+2015}. This is what we explore in this Appendix.

We consider the following scaling for the $\mJ_\ell$ spectrum~\citep{Kocsis+2015} 
\begin{equation}
\label{eq:spectrum}
\mJ_\ell = 
\begin{cases} 
\mJ_2 / (\ell/2)^2 & \text{for }  2 \leq \ell \leq \LINT \text{ and } \ell \text{ even,} \\
0 & \text{otherwise,}
\end{cases}
\end{equation}
with $\LINT$ a new cutoff on the maximum harmonic number
contributing to the pairwise interaction.
In Eq.~\eqref{eq:spectrum}, the dependence in $\ell^{-2}$
corresponds to the shallowest possible interaction spectrum for \VRR\@ (namely circular orbits).
For any other orbits the interaction spectrum is steeper,
with multipoles decaying as $\ell^{-\alpha}$ with ${\alpha\!>\!2}$~\citep[see appendix~{B} in][]{Kocsis+2015}.

In Fig.~\ref{fig:Nbody_LMAX_INT}, we show the $N$-body measurements of the two-point correlation function for different values of $\LINT$.
\begin{figure}[htbp]
\centering
\includegraphics[width=1.0\linewidth]{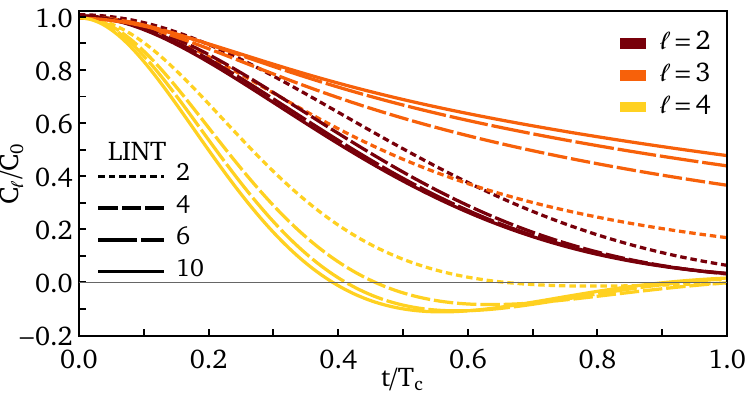} 
\caption{$N$-body measurement of the normalised isotropic two-point correlation function ${C_\ell}$,
following the same convention as in Fig.~\ref{fig:Money_plot}.
Results are shown for various harmonics $\ell$ (different colors) and different cutoffs $\LINT$ (solid--dashed lines, Eq.~\ref{eq:spectrum}).
For each scale $\ell$, the measurement converges towards an asymptotic value
corresponding to ${\LINT \!\to\! \infty}$.
}
\label{fig:Nbody_LMAX_INT}
\end{figure}
Accounting for an interaction spectrum in the $N$-body simulations
requires a generalisation of the code from Appendix~\ref{app:NumericalSimulations}
using efficient recurrence relations to compute the (real) spherical harmonics
and their gradients~\cite[see appendix~{C} in][]{Fouvry+2019}.
The measurements presented in Fig.~\ref{fig:Nbody_LMAX_INT}
were obtained for ${N\!=\! 10^4}$ by averaging over ${10^4}$ realisations
and are converged to better than ${1\%}$.

For scales ${\ell \! =\! 2,3,4}$, the two-point correlation function converges towards an asymptotic value corresponding to ${\LINT \!\to\!\infty}$.
In particular, as $\LINT$ increases,
for even $\ell$,
we find that the correlations decay faster.
As such, as the pairwise interaction gains sensitivity from larger $\ell$ (i.e.\ smaller angular scales),
the vortex-like structures on the sphere (Fig.~\ref{fig:Sphere}) decorrelate more rapidly
and have a shorter lifetime.
For ${\ell\!=\!2}$, the asymptotic value is already mostly reached at ${\LINT \!=\!4}$, beyond which the measurement is insensitive to the cutoff.
This is consistent with the results from~\citep{TakacsKocsis2018},
which showed that only the first few harmonics in the pairwise interaction potential
play a crucial role in setting up the statistical properties of \VRR\@.

\begin{figure}[htbp]
\centering
\includegraphics[width=1.0\linewidth]{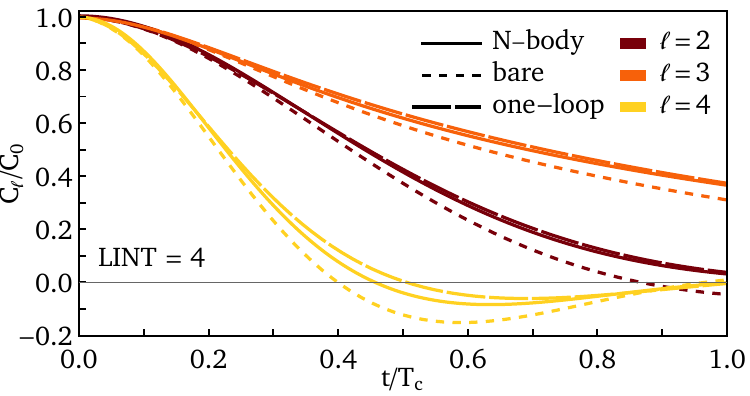} 
\caption{Same as Fig.~\ref{fig:Money_plot}, but for the interaction spectrum from Eq.~\eqref{eq:spectrum} with a ${\LINT\!=\!4}$ cutoff. The one-loop expansion converges and improves upon the bare prediction.
}
\label{fig:Money_plot_Asymp}
\end{figure}

We now study the effect of an interaction spectrum on the \MSR\ bare and one-loop predictions.
The derivation of the \MSR\ equations, Eqs.~\eqref{eq:Eq_perturb}--\eqref{eq:func_Gamma}, does not require specifying the microscopic details of the bare coupling $\gamma$.
Likewise, our numerical scheme from Sec.~\ref{sec:Numerical} does not rely on the choice of interaction spectrum. To compute the associated prediction,
one only needs to modify the definition of $\gamma$ in Eq.~\eqref{diag:one_loop_iter}, and the fixed-point search is expected to remain convergent for any multipole interaction.
This is confirmed in Fig.~\ref{fig:Money_plot_Asymp}, which shows the bare and one-loop predictions for the two-point correlation functions obtained through our numerical fixed-point search, for a ${\LINT \!=\!4}$ cutoff. 
It confirms that the one-loop expansion converges and, importantly,
offers a significant improvement over the bare prediction.

The main limitation here is the numerical cost of the fixed-point search. Including higher multipoles in the interaction necessarily requires a larger cutoff $\LMAX$ in the harmonic discretisation (Sec.~\ref{sec:Discretisation_time_harm}) to satisfy the convergence criteria of Appendix~\ref{app:CVparam}. It also broadens the set of triangles contributing to the bare coupling $\gamma$,
hence further increasing the computational cost.
We verified that for the control parameters ${\LINT\!=\!4}$, ${\LMAX\!=\!8}$, ${\DT\!=\! \Tc/64}$, ${\TMAX\!=\!\Tc}$, and ${\ITER\!=\!10}$ (see Appendix~\ref{app:CVparam}),
all convergence criteria are satisfied for the scales ${\ell\!=\!2}$ and ${\ell\!=\!3}$.
For ${\ell\!=\!4}$, the convergence with respect to $\LMAX$ is slightly less under control.
We believe that this accounts for the minute discrepancy between the prediction and the $N$-body measurement observed in Fig.~\ref{fig:Money_plot_Asymp}.
Based on the trend visible in Fig.~\ref{fig:Nbody_LMAX_INT},
we expect for the one-loop scheme to remain convergent for ${\LINT \! \geq \!4}$.
Unfortunately, this would come at a much higher computational cost,
since $\LMAX$ would need to be increased accordingly.

\begin{figure}[htbp!]
\centering
\includegraphics[width=1.0\linewidth]{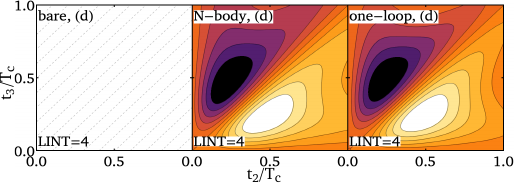}
\caption{Same as triangle (d) in Fig.~\ref{fig:Maps_full}, but using the interaction spectrum from Eq.~\eqref{eq:spectrum} with a ${\LINT\!=\!4}$ cutoff.
The \MSR\ one-loop expansion converges and accurately reproduces the $N$-body measurement.
}
\label{fig:Maps_LMAXINT}
\end{figure}

We now turn to three-point correlation functions.
In practice, relaxing the quadrupolar restriction on $\gamma$
enlarges the set of triangles with a non-vanishing prediction for the bare-order skewness.
Nevertheless, even in the ${\LINT \!\to\! \infty}$ limit, triangles in which all scales $\ell$ are odd continue to vanish, since the coupling coefficients $\mJ_\ell$ are non-zero only for even $\ell$ (Eq.~\ref{eq:spectrum}).
We now focus on one such all-odd triangle.
In Fig.~\ref{fig:Maps_LMAXINT}, we show the three-point skewness for triangle (d) of Fig.~\ref{fig:Maps_full}, computed with an interaction spectrum truncated at ${\LINT\!=\!4}$.
As expected, our scheme yields a one-loop prediction that accurately reproduces the $N$-body measurement. Remarkably, for this triangle, truncating the interaction spectrum at ${\LINT\!=\!4}$ reduces the maximum amplitude of the skewness by $55\%$ compared to the purely quadrupolar case shown in Fig.~\ref{fig:Maps_full}.
This shows that the strength of such non-Gaussianities depends sensitively on the interaction spectrum.
Nonetheless, the overall shape of the skewness remains similar to that of Fig.~\ref{fig:Maps_full}.
This is also the case for triangles (a)--(c) in Fig.~\ref{fig:Maps_full}.


\begin{thebibliography}{85}
\expandafter\ifx\csname natexlab\endcsname\relax\def\natexlab#1{#1}\fi
\expandafter\ifx\csname bibnamefont\endcsname\relax
  \def\bibnamefont#1{#1}\fi
\expandafter\ifx\csname bibfnamefont\endcsname\relax
  \def\bibfnamefont#1{#1}\fi
\expandafter\ifx\csname citenamefont\endcsname\relax
  \def\citenamefont#1{#1}\fi
\expandafter\ifx\csname url\endcsname\relax
  \def\url#1{\texttt{#1}}\fi
\expandafter\ifx\csname urlprefix\endcsname\relax\def\urlprefix{URL }\fi
\providecommand{\bibinfo}[2]{#2}
\providecommand{\eprint}[2][]{\url{#2}}

\bibitem[{\citenamefont{Frisch}(1995)}]{Frisch1995}
\bibinfo{author}{\bibfnamefont{U.}~\bibnamefont{Frisch}},
  \emph{\bibinfo{title}{Turbulence: The Legacy of A. N. Kolmogorov}}
  (\bibinfo{publisher}{Cambridge Univ. Press}, \bibinfo{year}{1995}).

\bibitem[{\citenamefont{{Berera} et~al.}(2013)\citenamefont{{Berera},
  {Salewski}, and {McComb}}}]{Berera+2013}
\bibinfo{author}{\bibfnamefont{A.}~\bibnamefont{{Berera}}},
  \bibinfo{author}{\bibfnamefont{M.}~\bibnamefont{{Salewski}}},
  \bibnamefont{and} \bibinfo{author}{\bibfnamefont{W.~D.}
  \bibnamefont{{McComb}}}, \bibinfo{journal}{Phys. Rev. E}
  \textbf{\bibinfo{volume}{87}}, \bibinfo{eid}{013007} (\bibinfo{year}{2013}).

\bibitem[{\citenamefont{{Nicholson}}(1992)}]{Nicholson1992}
\bibinfo{author}{\bibfnamefont{D.~R.} \bibnamefont{{Nicholson}}},
  \emph{\bibinfo{title}{{Introduction to Plasma Theory}}}
  (\bibinfo{publisher}{Krieger}, \bibinfo{year}{1992}).

\bibitem[{\citenamefont{{Krommes}}(2015)}]{Krommes2015}
\bibinfo{author}{\bibfnamefont{J.~A.} \bibnamefont{{Krommes}}},
  \bibinfo{journal}{JPP} \textbf{\bibinfo{volume}{81}},
  \bibinfo{eid}{205810601} (\bibinfo{year}{2015}).

\bibitem[{\citenamefont{{Lynden-Bell}}(1967)}]{Lynden1967}
\bibinfo{author}{\bibfnamefont{D.}~\bibnamefont{{Lynden-Bell}}},
  \bibinfo{journal}{MNRAS} \textbf{\bibinfo{volume}{136}}, \bibinfo{pages}{101}
  (\bibinfo{year}{1967}).

\bibitem[{\citenamefont{{Binney} and {Tremaine}}(2008)}]{Binney+2008}
\bibinfo{author}{\bibfnamefont{J.}~\bibnamefont{{Binney}}} \bibnamefont{and}
  \bibinfo{author}{\bibfnamefont{S.}~\bibnamefont{{Tremaine}}},
  \emph{\bibinfo{title}{{Galactic Dynamics: Second Edition}}}
  (\bibinfo{publisher}{Princeton Univ. Press}, \bibinfo{year}{2008}).

\bibitem[{\citenamefont{{Bernardeau} et~al.}(2002)\citenamefont{{Bernardeau},
  {Colombi}, {Gazta{\~n}aga}, and {Scoccimarro}}}]{Bernardeau+2002}
\bibinfo{author}{\bibfnamefont{F.}~\bibnamefont{{Bernardeau}}},
  \bibinfo{author}{\bibfnamefont{S.}~\bibnamefont{{Colombi}}},
  \bibinfo{author}{\bibfnamefont{E.}~\bibnamefont{{Gazta{\~n}aga}}},
  \bibnamefont{and}
  \bibinfo{author}{\bibfnamefont{R.}~\bibnamefont{{Scoccimarro}}},
  \bibinfo{journal}{Phys. Rep.} \textbf{\bibinfo{volume}{367}},
  \bibinfo{pages}{1} (\bibinfo{year}{2002}).

\bibitem[{\citenamefont{Bernardeau et~al.}(2012)\citenamefont{Bernardeau,
  Van~de Rijt, and Vernizzi}}]{Bernardeau+2012}
\bibinfo{author}{\bibfnamefont{F.}~\bibnamefont{Bernardeau}},
  \bibinfo{author}{\bibfnamefont{N.}~\bibnamefont{Van~de Rijt}},
  \bibnamefont{and} \bibinfo{author}{\bibfnamefont{F.}~\bibnamefont{Vernizzi}},
  \bibinfo{journal}{Phys. Rev. D} \textbf{\bibinfo{volume}{85}},
  \bibinfo{pages}{063509} (\bibinfo{year}{2012}).

\bibitem[{\citenamefont{{Kocsis} and {Tremaine}}(2015)}]{Kocsis+2015}
\bibinfo{author}{\bibfnamefont{B.}~\bibnamefont{{Kocsis}}} \bibnamefont{and}
  \bibinfo{author}{\bibfnamefont{S.}~\bibnamefont{{Tremaine}}},
  \bibinfo{journal}{MNRAS} \textbf{\bibinfo{volume}{448}},
  \bibinfo{pages}{3265} (\bibinfo{year}{2015}).

\bibitem[{\citenamefont{{Krommes}}(2002)}]{Krommes2002}
\bibinfo{author}{\bibfnamefont{J.~A.} \bibnamefont{{Krommes}}},
  \bibinfo{journal}{Phys. Rep.} \textbf{\bibinfo{volume}{360}},
  \bibinfo{pages}{1} (\bibinfo{year}{2002}).

\bibitem[{\citenamefont{Diamond et~al.}(2010)\citenamefont{Diamond, Itoh, and
  Itoh}}]{Diamond+2010}
\bibinfo{author}{\bibfnamefont{P.~H.} \bibnamefont{Diamond}},
  \bibinfo{author}{\bibfnamefont{S.-I.} \bibnamefont{Itoh}}, \bibnamefont{and}
  \bibinfo{author}{\bibfnamefont{K.}~\bibnamefont{Itoh}},
  \emph{\bibinfo{title}{Modern Plasma Physics}} (\bibinfo{publisher}{Cambridge
  Univ. Press}, \bibinfo{year}{2010}).

\bibitem[{\citenamefont{Nazarenko}(2011)}]{Nazarenko2011}
\bibinfo{author}{\bibfnamefont{S.}~\bibnamefont{Nazarenko}},
  \emph{\bibinfo{title}{Wave Turbulence}} (\bibinfo{publisher}{Springer},
  \bibinfo{address}{Berlin, Heidelberg}, \bibinfo{year}{2011}).

\bibitem[{\citenamefont{Schekochihin}(2025)}]{Schekochihin2025}
\bibinfo{author}{\bibfnamefont{A.~A.} \bibnamefont{Schekochihin}},
  \emph{\bibinfo{title}{{Lectures on Kinetic Theory and Magnetohydrodynamics of
  Plasmas}}},
  \bibinfo{howpublished}{\url{https://www-thphys.physics.ox.ac.uk/people/AlexanderSchekochihin/KT/2015/KTLectureNotes.pdf}}
  (\bibinfo{year}{2025}).

\bibitem[{\citenamefont{Micha and Tkachev}(2004)}]{Micha+2004}
\bibinfo{author}{\bibfnamefont{R.}~\bibnamefont{Micha}} \bibnamefont{and}
  \bibinfo{author}{\bibfnamefont{I.~I.} \bibnamefont{Tkachev}},
  \bibinfo{journal}{Phys. Rev. D} \textbf{\bibinfo{volume}{70}},
  \bibinfo{pages}{043538} (\bibinfo{year}{2004}).

\bibitem[{\citenamefont{Yokoyama and Takaoka}(2014)}]{Yokoyama+2014}
\bibinfo{author}{\bibfnamefont{N.}~\bibnamefont{Yokoyama}} \bibnamefont{and}
  \bibinfo{author}{\bibfnamefont{M.}~\bibnamefont{Takaoka}},
  \bibinfo{journal}{Phys. Rev. E} \textbf{\bibinfo{volume}{89}},
  \bibinfo{pages}{012909} (\bibinfo{year}{2014}).

\bibitem[{\citenamefont{{Tarpin} et~al.}(2019)\citenamefont{{Tarpin}, {Canet},
  {Pagani}, and {Wschebor}}}]{Tarpin+2019}
\bibinfo{author}{\bibfnamefont{M.}~\bibnamefont{{Tarpin}}},
  \bibinfo{author}{\bibfnamefont{L.}~\bibnamefont{{Canet}}},
  \bibinfo{author}{\bibfnamefont{C.}~\bibnamefont{{Pagani}}}, \bibnamefont{and}
  \bibinfo{author}{\bibfnamefont{N.}~\bibnamefont{{Wschebor}}},
  \bibinfo{journal}{J. Phys. A} \textbf{\bibinfo{volume}{52}},
  \bibinfo{eid}{085501} (\bibinfo{year}{2019}).

\bibitem[{\citenamefont{{Zhou}}(2021)}]{Zhou2021}
\bibinfo{author}{\bibfnamefont{Y.}~\bibnamefont{{Zhou}}},
  \bibinfo{journal}{Phys. Rep.} \textbf{\bibinfo{volume}{935}},
  \bibinfo{pages}{1} (\bibinfo{year}{2021}).

\bibitem[{\citenamefont{{Maier} and {Saupe}}(1958)}]{Maier+1958}
\bibinfo{author}{\bibfnamefont{W.}~\bibnamefont{{Maier}}} \bibnamefont{and}
  \bibinfo{author}{\bibfnamefont{A.}~\bibnamefont{{Saupe}}},
  \bibinfo{journal}{Z. Naturforsch. A} \textbf{\bibinfo{volume}{13}},
  \bibinfo{pages}{564} (\bibinfo{year}{1958}).

\bibitem[{\citenamefont{{Plischke} and {Bergersen}}(2006)}]{Plischke+2006}
\bibinfo{author}{\bibfnamefont{R.~K.} \bibnamefont{{Plischke}}}
  \bibnamefont{and}
  \bibinfo{author}{\bibfnamefont{B.}~\bibnamefont{{Bergersen}}},
  \emph{\bibinfo{title}{{Equilibrium Statistical Physics: Third Edition}}}
  (\bibinfo{publisher}{Singapore: World Scientific}, \bibinfo{year}{2006}).

\bibitem[{\citenamefont{{Roupas} et~al.}(2017)\citenamefont{{Roupas}, {Kocsis},
  and {Tremaine}}}]{Roupas+2017}
\bibinfo{author}{\bibfnamefont{Z.}~\bibnamefont{{Roupas}}},
  \bibinfo{author}{\bibfnamefont{B.}~\bibnamefont{{Kocsis}}}, \bibnamefont{and}
  \bibinfo{author}{\bibfnamefont{S.}~\bibnamefont{{Tremaine}}},
  \bibinfo{journal}{ApJ} \textbf{\bibinfo{volume}{842}}, \bibinfo{eid}{90}
  (\bibinfo{year}{2017}).

\bibitem[{\citenamefont{{Kraichnan}}(1958)}]{Kraichnan1958}
\bibinfo{author}{\bibfnamefont{R.~H.} \bibnamefont{{Kraichnan}}},
  \bibinfo{journal}{Phys. Rev.} \textbf{\bibinfo{volume}{109}},
  \bibinfo{pages}{1407} (\bibinfo{year}{1958}).

\bibitem[{\citenamefont{{Dupree}}(1972)}]{Dupree1972}
\bibinfo{author}{\bibfnamefont{T.~H.} \bibnamefont{{Dupree}}},
  \bibinfo{journal}{Physics of Fluids} \textbf{\bibinfo{volume}{15}},
  \bibinfo{pages}{334} (\bibinfo{year}{1972}).

\bibitem[{\citenamefont{{Boutros-Ghali} and {Dupree}}(1981)}]{Boutros+1981}
\bibinfo{author}{\bibfnamefont{T.}~\bibnamefont{{Boutros-Ghali}}}
  \bibnamefont{and} \bibinfo{author}{\bibfnamefont{T.~H.}
  \bibnamefont{{Dupree}}}, \bibinfo{journal}{Physics of Fluids}
  \textbf{\bibinfo{volume}{24}}, \bibinfo{pages}{1839} (\bibinfo{year}{1981}).

\bibitem[{\citenamefont{{Valageas}}(2004)}]{Valageas2004}
\bibinfo{author}{\bibfnamefont{P.}~\bibnamefont{{Valageas}}},
  \bibinfo{journal}{A\&A} \textbf{\bibinfo{volume}{421}}, \bibinfo{pages}{23}
  (\bibinfo{year}{2004}).

\bibitem[{\citenamefont{{Bernardeau} et~al.}(2012)\citenamefont{{Bernardeau},
  {Crocce}, and {Scoccimarro}}}]{Bernardeau+2012Corr}
\bibinfo{author}{\bibfnamefont{F.}~\bibnamefont{{Bernardeau}}},
  \bibinfo{author}{\bibfnamefont{M.}~\bibnamefont{{Crocce}}}, \bibnamefont{and}
  \bibinfo{author}{\bibfnamefont{R.}~\bibnamefont{{Scoccimarro}}},
  \bibinfo{journal}{\prd} \textbf{\bibinfo{volume}{85}}, \bibinfo{eid}{123519}
  (\bibinfo{year}{2012}).

\bibitem[{\citenamefont{{Kolmogorov}}(1941)}]{Kolmogorov1941c}
\bibinfo{author}{\bibfnamefont{A.~N.} \bibnamefont{{Kolmogorov}}},
  \bibinfo{journal}{Akademiia Nauk SSSR Doklady} \textbf{\bibinfo{volume}{32}},
  \bibinfo{pages}{16} (\bibinfo{year}{1941}).

\bibitem[{\citenamefont{{Baldauf} et~al.}(2011)\citenamefont{{Baldauf},
  {Seljak}, and {Senatore}}}]{Baldauf+2011}
\bibinfo{author}{\bibfnamefont{T.}~\bibnamefont{{Baldauf}}},
  \bibinfo{author}{\bibfnamefont{U.}~\bibnamefont{{Seljak}}}, \bibnamefont{and}
  \bibinfo{author}{\bibfnamefont{L.}~\bibnamefont{{Senatore}}},
  \bibinfo{journal}{JCAP} \textbf{\bibinfo{volume}{2011}}, \bibinfo{eid}{006}
  (\bibinfo{year}{2011}).

\bibitem[{\citenamefont{{Kraichnan}}(1959{\natexlab{a}})}]{Kraichnan1959}
\bibinfo{author}{\bibfnamefont{R.~H.} \bibnamefont{{Kraichnan}}},
  \bibinfo{journal}{J. Fluid Mech.} \textbf{\bibinfo{volume}{5}},
  \bibinfo{pages}{497} (\bibinfo{year}{1959}{\natexlab{a}}).

\bibitem[{\citenamefont{Orszag}(1970)}]{Orszag1970}
\bibinfo{author}{\bibfnamefont{S.~A.} \bibnamefont{Orszag}},
  \bibinfo{journal}{J. Fluid. Mech.} \textbf{\bibinfo{volume}{41}},
  \bibinfo{pages}{363} (\bibinfo{year}{1970}).

\bibitem[{\citenamefont{{McComb}}(1990)}]{McComb1990}
\bibinfo{author}{\bibfnamefont{W.~D.} \bibnamefont{{McComb}}},
  \emph{\bibinfo{title}{{The physics of fluid turbulence}}}
  (\bibinfo{publisher}{Clarendon Press}, \bibinfo{year}{1990}).

\bibitem[{\citenamefont{McComb}(2014)}]{McComb2014}
\bibinfo{author}{\bibfnamefont{W.~D.} \bibnamefont{McComb}},
  \emph{\bibinfo{title}{Homogeneous, Isotropic Turbulence: Phenomenology,
  Renormalization and Statistical Closures}} (\bibinfo{publisher}{Oxford Univ.
  Press}, \bibinfo{year}{2014}).

\bibitem[{\citenamefont{{Canet}}(2025)}]{Canet2025}
\bibinfo{author}{\bibfnamefont{L.}~\bibnamefont{{Canet}}}, \bibinfo{eid}{arXiv
  2509.01472} (\bibinfo{year}{2025}).

\bibitem[{\citenamefont{{Martin} et~al.}(1973)\citenamefont{{Martin}, {Siggia},
  and {Rose}}}]{MSR1973}
\bibinfo{author}{\bibfnamefont{P.~C.} \bibnamefont{{Martin}}},
  \bibinfo{author}{\bibfnamefont{E.~D.} \bibnamefont{{Siggia}}},
  \bibnamefont{and} \bibinfo{author}{\bibfnamefont{H.~A.}
  \bibnamefont{{Rose}}}, \bibinfo{journal}{Phys. Rev. A}
  \textbf{\bibinfo{volume}{8}}, \bibinfo{pages}{423} (\bibinfo{year}{1973}).

\bibitem[{\citenamefont{{Flores} and {Fouvry}}(2025)}]{Flores+2025}
\bibinfo{author}{\bibfnamefont{S.}~\bibnamefont{{Flores}}} \bibnamefont{and}
  \bibinfo{author}{\bibfnamefont{J.-B.} \bibnamefont{{Fouvry}}},
  \bibinfo{journal}{Phys. Rev. E} \textbf{\bibinfo{volume}{111}},
  \bibinfo{eid}{044111} (\bibinfo{year}{2025}).

\bibitem[{\citenamefont{{Gillessen} et~al.}(2017)}]{Gillessen+2017}
\bibinfo{author}{\bibfnamefont{S.}~\bibnamefont{{Gillessen}}}
  \bibnamefont{et~al.}, \bibinfo{journal}{ApJ} \textbf{\bibinfo{volume}{837}},
  \bibinfo{eid}{30} (\bibinfo{year}{2017}).

\bibitem[{\citenamefont{{Alexander}}(2017)}]{Alexander2017}
\bibinfo{author}{\bibfnamefont{T.}~\bibnamefont{{Alexander}}},
  \bibinfo{journal}{ARA\&A} \textbf{\bibinfo{volume}{55}}, \bibinfo{pages}{17}
  (\bibinfo{year}{2017}).

\bibitem[{\citenamefont{{Rauch} and {Tremaine}}(1996)}]{Rauch+1996}
\bibinfo{author}{\bibfnamefont{K.~P.} \bibnamefont{{Rauch}}} \bibnamefont{and}
  \bibinfo{author}{\bibfnamefont{S.}~\bibnamefont{{Tremaine}}},
  \bibinfo{journal}{New Astron.} \textbf{\bibinfo{volume}{1}},
  \bibinfo{pages}{149} (\bibinfo{year}{1996}).

\bibitem[{\citenamefont{{Kocsis} and {Tremaine}}(2011)}]{Kocsis+2011}
\bibinfo{author}{\bibfnamefont{B.}~\bibnamefont{{Kocsis}}} \bibnamefont{and}
  \bibinfo{author}{\bibfnamefont{S.}~\bibnamefont{{Tremaine}}},
  \bibinfo{journal}{MNRAS} \textbf{\bibinfo{volume}{412}}, \bibinfo{pages}{187}
  (\bibinfo{year}{2011}).

\bibitem[{\citenamefont{{Hamers} et~al.}(2018)\citenamefont{{Hamers}, {Bar-Or},
  {Petrovich}, and {Antonini}}}]{Hamers+2018}
\bibinfo{author}{\bibfnamefont{A.~S.} \bibnamefont{{Hamers}}},
  \bibinfo{author}{\bibfnamefont{B.}~\bibnamefont{{Bar-Or}}},
  \bibinfo{author}{\bibfnamefont{C.}~\bibnamefont{{Petrovich}}},
  \bibnamefont{and}
  \bibinfo{author}{\bibfnamefont{F.}~\bibnamefont{{Antonini}}},
  \bibinfo{journal}{ApJ} \textbf{\bibinfo{volume}{865}}, \bibinfo{eid}{2}
  (\bibinfo{year}{2018}).

\bibitem[{\citenamefont{{Sz{\"o}lgy{\'e}n} and {Kocsis}}(2018)}]{Szolgyen+2018}
\bibinfo{author}{\bibfnamefont{{\'A}.}~\bibnamefont{{Sz{\"o}lgy{\'e}n}}}
  \bibnamefont{and} \bibinfo{author}{\bibfnamefont{B.}~\bibnamefont{{Kocsis}}},
  \bibinfo{journal}{Phys. Rev. Lett.} \textbf{\bibinfo{volume}{121}},
  \bibinfo{eid}{101101} (\bibinfo{year}{2018}).

\bibitem[{\citenamefont{{Touma} et~al.}(2019)\citenamefont{{Touma}, {Tremaine},
  and {Kazandjian}}}]{Touma+2019}
\bibinfo{author}{\bibfnamefont{J.}~\bibnamefont{{Touma}}},
  \bibinfo{author}{\bibfnamefont{S.}~\bibnamefont{{Tremaine}}},
  \bibnamefont{and}
  \bibinfo{author}{\bibfnamefont{M.}~\bibnamefont{{Kazandjian}}},
  \bibinfo{journal}{Phys. Rev. Lett.} \textbf{\bibinfo{volume}{123}},
  \bibinfo{eid}{021103} (\bibinfo{year}{2019}).

\bibitem[{\citenamefont{{Tremaine}}(2020{\natexlab{a}})}]{Tremaine2020a}
\bibinfo{author}{\bibfnamefont{S.}~\bibnamefont{{Tremaine}}},
  \bibinfo{journal}{MNRAS} \textbf{\bibinfo{volume}{491}},
  \bibinfo{pages}{1941} (\bibinfo{year}{2020}{\natexlab{a}}).

\bibitem[{\citenamefont{{Tremaine}}(2020{\natexlab{b}})}]{Tremaine2020b}
\bibinfo{author}{\bibfnamefont{S.}~\bibnamefont{{Tremaine}}},
  \bibinfo{journal}{MNRAS} \textbf{\bibinfo{volume}{493}},
  \bibinfo{pages}{2632} (\bibinfo{year}{2020}{\natexlab{b}}).

\bibitem[{\citenamefont{{Gruzinov} et~al.}(2020)\citenamefont{{Gruzinov},
  {Levin}, and {Zhu}}}]{Gruzinov+2020}
\bibinfo{author}{\bibfnamefont{A.}~\bibnamefont{{Gruzinov}}},
  \bibinfo{author}{\bibfnamefont{Y.}~\bibnamefont{{Levin}}}, \bibnamefont{and}
  \bibinfo{author}{\bibfnamefont{J.}~\bibnamefont{{Zhu}}},
  \bibinfo{journal}{ApJ} \textbf{\bibinfo{volume}{905}}, \bibinfo{eid}{11}
  (\bibinfo{year}{2020}).

\bibitem[{\citenamefont{{Magnan} et~al.}(2022)}]{Magnan+2022}
\bibinfo{author}{\bibfnamefont{N.}~\bibnamefont{{Magnan}}}
  \bibnamefont{et~al.}, \bibinfo{journal}{MNRAS}
  \textbf{\bibinfo{volume}{514}}, \bibinfo{pages}{3452} (\bibinfo{year}{2022}).

\bibitem[{\citenamefont{{M{\'a}th{\'e}}
  et~al.}(2023)\citenamefont{{M{\'a}th{\'e}}, {Sz{\"o}lgy{\'e}n}, and
  {Kocsis}}}]{Mathe+2023}
\bibinfo{author}{\bibfnamefont{G.}~\bibnamefont{{M{\'a}th{\'e}}}},
  \bibinfo{author}{\bibfnamefont{{\'A}.}~\bibnamefont{{Sz{\"o}lgy{\'e}n}}},
  \bibnamefont{and} \bibinfo{author}{\bibfnamefont{B.}~\bibnamefont{{Kocsis}}},
  \bibinfo{journal}{MNRAS} \textbf{\bibinfo{volume}{520}},
  \bibinfo{pages}{2204} (\bibinfo{year}{2023}).

\bibitem[{\citenamefont{{Ginat} et~al.}(2023)\citenamefont{{Ginat},
  {Panamarev}, {Kocsis}, and {Perets}}}]{Ginat+2023}
\bibinfo{author}{\bibfnamefont{Y.~B.} \bibnamefont{{Ginat}}},
  \bibinfo{author}{\bibfnamefont{T.}~\bibnamefont{{Panamarev}}},
  \bibinfo{author}{\bibfnamefont{B.}~\bibnamefont{{Kocsis}}}, \bibnamefont{and}
  \bibinfo{author}{\bibfnamefont{H.~B.} \bibnamefont{{Perets}}},
  \bibinfo{journal}{MNRAS} \textbf{\bibinfo{volume}{525}},
  \bibinfo{pages}{4202} (\bibinfo{year}{2023}).

\bibitem[{\citenamefont{{Levin}}(2024)}]{Levin+2024}
\bibinfo{author}{\bibfnamefont{Y.}~\bibnamefont{{Levin}}},
  \bibinfo{journal}{ApJ} \textbf{\bibinfo{volume}{975}}, \bibinfo{eid}{278}
  (\bibinfo{year}{2024}).

\bibitem[{\citenamefont{{Panamarev} and {Kocsis}}(2022)}]{Panamarev+2022}
\bibinfo{author}{\bibfnamefont{T.}~\bibnamefont{{Panamarev}}} \bibnamefont{and}
  \bibinfo{author}{\bibfnamefont{B.}~\bibnamefont{{Kocsis}}},
  \bibinfo{journal}{MNRAS} \textbf{\bibinfo{volume}{517}},
  \bibinfo{pages}{6205} (\bibinfo{year}{2022}).

\bibitem[{\citenamefont{{Panamarev} et~al.}(2025)\citenamefont{{Panamarev},
  {Ginat}, and {Kocsis}}}]{Panamarev+2025}
\bibinfo{author}{\bibfnamefont{T.}~\bibnamefont{{Panamarev}}},
  \bibinfo{author}{\bibfnamefont{Y.~B.} \bibnamefont{{Ginat}}},
  \bibnamefont{and} \bibinfo{author}{\bibfnamefont{B.}~\bibnamefont{{Kocsis}}},
  \bibinfo{eid}{arXiv 2507.10551} (\bibinfo{year}{2025}).

\bibitem[{\citenamefont{{Fouvry} et~al.}(2023)\citenamefont{{Fouvry},
  {Bustamante-Rosell}, and {Zimmerman}}}]{Fouvry+2023}
\bibinfo{author}{\bibfnamefont{J.-B.} \bibnamefont{{Fouvry}}},
  \bibinfo{author}{\bibfnamefont{M.~J.} \bibnamefont{{Bustamante-Rosell}}},
  \bibnamefont{and}
  \bibinfo{author}{\bibfnamefont{A.}~\bibnamefont{{Zimmerman}}},
  \bibinfo{journal}{MNRAS} \textbf{\bibinfo{volume}{526}},
  \bibinfo{pages}{1471} (\bibinfo{year}{2023}).

\bibitem[{\citenamefont{{Ginat} and {Kocsis}}(2025)}]{Ginat+2025Axion}
\bibinfo{author}{\bibfnamefont{Y.~B.} \bibnamefont{{Ginat}}} \bibnamefont{and}
  \bibinfo{author}{\bibfnamefont{B.}~\bibnamefont{{Kocsis}}},
  \bibinfo{eid}{arXiv 2502.08709} (\bibinfo{year}{2025}).

\bibitem[{\citenamefont{{Fouvry} et~al.}(2019)\citenamefont{{Fouvry}, {Bar-Or},
  and {Chavanis}}}]{Fouvry+2019}
\bibinfo{author}{\bibfnamefont{J.-B.} \bibnamefont{{Fouvry}}},
  \bibinfo{author}{\bibfnamefont{B.}~\bibnamefont{{Bar-Or}}}, \bibnamefont{and}
  \bibinfo{author}{\bibfnamefont{P.-H.} \bibnamefont{{Chavanis}}},
  \bibinfo{journal}{ApJ} \textbf{\bibinfo{volume}{883}}, \bibinfo{eid}{161}
  (\bibinfo{year}{2019}).

\bibitem[{\citenamefont{{Nastac} et~al.}(2025)\citenamefont{{Nastac}, {Ewart},
  {Juno}, {Barnes}, and {Schekochihin}}}]{Nastac+2025}
\bibinfo{author}{\bibfnamefont{M.~L.} \bibnamefont{{Nastac}}},
  \bibinfo{author}{\bibfnamefont{R.~J.} \bibnamefont{{Ewart}}},
  \bibinfo{author}{\bibfnamefont{J.}~\bibnamefont{{Juno}}},
  \bibinfo{author}{\bibfnamefont{M.}~\bibnamefont{{Barnes}}}, \bibnamefont{and}
  \bibinfo{author}{\bibfnamefont{A.~A.} \bibnamefont{{Schekochihin}}},
  \bibinfo{journal}{arXiv} \bibinfo{eid}{2503.17278} (\bibinfo{year}{2025}).

\bibitem[{\citenamefont{{Kraichnan}}(1959{\natexlab{b}})}]{Kraichnan1959a}
\bibinfo{author}{\bibfnamefont{R.~H.} \bibnamefont{{Kraichnan}}},
  \bibinfo{journal}{Phys. Rev.} \textbf{\bibinfo{volume}{113}},
  \bibinfo{pages}{1181} (\bibinfo{year}{1959}{\natexlab{b}}).

\bibitem[{\citenamefont{{Bellon} and {Russo}}(2021)}]{Bellon+2021}
\bibinfo{author}{\bibfnamefont{M.~P.} \bibnamefont{{Bellon}}} \bibnamefont{and}
  \bibinfo{author}{\bibfnamefont{E.~I.} \bibnamefont{{Russo}}},
  \bibinfo{journal}{Lett. Math. Phys.} \textbf{\bibinfo{volume}{111}},
  \bibinfo{eid}{42} (\bibinfo{year}{2021}).

\bibitem[{\citenamefont{Peskin and Schroeder}(1995)}]{Peskin+1995}
\bibinfo{author}{\bibfnamefont{M.~E.} \bibnamefont{Peskin}} \bibnamefont{and}
  \bibinfo{author}{\bibfnamefont{D.~V.} \bibnamefont{Schroeder}},
  \emph{\bibinfo{title}{An Introduction to Quantum Field Theory}}
  (\bibinfo{publisher}{Addison-Wesley}, \bibinfo{year}{1995}).

\bibitem[{\citenamefont{Lancaster and Blundell}(2014)}]{Lancaster+2014}
\bibinfo{author}{\bibfnamefont{T.}~\bibnamefont{Lancaster}} \bibnamefont{and}
  \bibinfo{author}{\bibfnamefont{S.}~\bibnamefont{Blundell}},
  \emph{\bibinfo{title}{Quantum Field Theory for the Gifted Amateur}}
  (\bibinfo{publisher}{Oxford Univ. Press}, \bibinfo{year}{2014}).

\bibitem[{git()}]{github_MSR}
\bibinfo{howpublished}{\url{https://github.com/sfloresmo/VRR_MSR}}.

\bibitem[{\citenamefont{{Rose}}(1985)}]{Rose1985}
\bibinfo{author}{\bibfnamefont{H.~A.} \bibnamefont{{Rose}}},
  \bibinfo{journal}{Physica D Nonlinear Phenomena}
  \textbf{\bibinfo{volume}{14}}, \bibinfo{pages}{216} (\bibinfo{year}{1985}).

\bibitem[{\citenamefont{{Canet} et~al.}(2011)\citenamefont{{Canet}, {Chat\'e},
  and {Delamotte}}}]{Canet+2011}
\bibinfo{author}{\bibfnamefont{L.}~\bibnamefont{{Canet}}},
  \bibinfo{author}{\bibfnamefont{H.}~\bibnamefont{{Chat\'e}}},
  \bibnamefont{and}
  \bibinfo{author}{\bibfnamefont{B.}~\bibnamefont{{Delamotte}}},
  \bibinfo{journal}{J. Phys. A} \textbf{\bibinfo{volume}{44}}
  (\bibinfo{year}{2011}).

\bibitem[{\citenamefont{{Delamotte}}(2012)}]{Delamotte2012}
\bibinfo{author}{\bibfnamefont{B.}~\bibnamefont{{Delamotte}}}, in
  \emph{\bibinfo{booktitle}{Lecture Notes in Physics}}
  (\bibinfo{publisher}{Springer}, \bibinfo{year}{2012}), vol.
  \bibinfo{volume}{852}, p.~\bibinfo{pages}{49}.

\bibitem[{\citenamefont{{Dupuis} et~al.}(2021)\citenamefont{{Dupuis}, {Canet},
  {Eichhorn}, {Metzner}, {Pawlowski}, {Tissier}, and {Wschebor}}}]{Dupuis+2021}
\bibinfo{author}{\bibfnamefont{N.}~\bibnamefont{{Dupuis}}},
  \bibinfo{author}{\bibfnamefont{L.}~\bibnamefont{{Canet}}},
  \bibinfo{author}{\bibfnamefont{A.}~\bibnamefont{{Eichhorn}}},
  \bibinfo{author}{\bibfnamefont{W.}~\bibnamefont{{Metzner}}},
  \bibinfo{author}{\bibfnamefont{J.~M.} \bibnamefont{{Pawlowski}}},
  \bibinfo{author}{\bibfnamefont{M.}~\bibnamefont{{Tissier}}},
  \bibnamefont{and}
  \bibinfo{author}{\bibfnamefont{N.}~\bibnamefont{{Wschebor}}},
  \bibinfo{journal}{Phys. Rep.} \textbf{\bibinfo{volume}{910}},
  \bibinfo{pages}{1} (\bibinfo{year}{2021}).

\bibitem[{\citenamefont{{Canet}}(2022)}]{Canet2022}
\bibinfo{author}{\bibfnamefont{L.}~\bibnamefont{{Canet}}}, \bibinfo{journal}{J.
  Fluid. Mech.} \textbf{\bibinfo{volume}{950}}, \bibinfo{eid}{P1}
  (\bibinfo{year}{2022}).

\bibitem[{\citenamefont{{Fontaine} et~al.}(2023)\citenamefont{{Fontaine},
  {Tarpin}, {Bouchet}, and {Canet}}}]{Fontaine+2023}
\bibinfo{author}{\bibfnamefont{C.}~\bibnamefont{{Fontaine}}},
  \bibinfo{author}{\bibfnamefont{M.}~\bibnamefont{{Tarpin}}},
  \bibinfo{author}{\bibfnamefont{F.}~\bibnamefont{{Bouchet}}},
  \bibnamefont{and} \bibinfo{author}{\bibfnamefont{L.}~\bibnamefont{{Canet}}},
  \bibinfo{journal}{SciPost Phys.} \textbf{\bibinfo{volume}{15}},
  \bibinfo{eid}{212} (\bibinfo{year}{2023}).

\bibitem[{\citenamefont{{Goldreich} and {Sridhar}}(1995)}]{Goldreich+1995}
\bibinfo{author}{\bibfnamefont{P.}~\bibnamefont{{Goldreich}}} \bibnamefont{and}
  \bibinfo{author}{\bibfnamefont{S.}~\bibnamefont{{Sridhar}}},
  \bibinfo{journal}{ApJ} \textbf{\bibinfo{volume}{438}}, \bibinfo{pages}{763}
  (\bibinfo{year}{1995}).

\bibitem[{\citenamefont{{Ginat} et~al.}(2025)\citenamefont{{Ginat}, {Nastac},
  {Ewart}, {Konrad}, {Bartelmann}, and {Schekochihin}}}]{Ginat+2025}
\bibinfo{author}{\bibfnamefont{Y.~B.} \bibnamefont{{Ginat}}},
  \bibinfo{author}{\bibfnamefont{M.~L.} \bibnamefont{{Nastac}}},
  \bibinfo{author}{\bibfnamefont{R.~J.} \bibnamefont{{Ewart}}},
  \bibinfo{author}{\bibfnamefont{S.}~\bibnamefont{{Konrad}}},
  \bibinfo{author}{\bibfnamefont{M.}~\bibnamefont{{Bartelmann}}},
  \bibnamefont{and} \bibinfo{author}{\bibfnamefont{A.~A.}
  \bibnamefont{{Schekochihin}}}, \bibinfo{journal}{Phys. Rev. D}
  \textbf{\bibinfo{volume}{112}}, \bibinfo{eid}{063501} (\bibinfo{year}{2025}).

\bibitem[{\citenamefont{{Wang} and {Kocsis}}(2023)}]{Wang+2023}
\bibinfo{author}{\bibfnamefont{H.}~\bibnamefont{{Wang}}} \bibnamefont{and}
  \bibinfo{author}{\bibfnamefont{B.}~\bibnamefont{{Kocsis}}},
  \bibinfo{journal}{Phys. Rev. D} \textbf{\bibinfo{volume}{108}},
  \bibinfo{eid}{103004} (\bibinfo{year}{2023}).

\bibitem[{\citenamefont{{Fragione} and {Loeb}}(2022)}]{Fragione+2022}
\bibinfo{author}{\bibfnamefont{G.}~\bibnamefont{{Fragione}}} \bibnamefont{and}
  \bibinfo{author}{\bibfnamefont{A.}~\bibnamefont{{Loeb}}},
  \bibinfo{journal}{ApJL} \textbf{\bibinfo{volume}{932}}, \bibinfo{eid}{L17}
  (\bibinfo{year}{2022}).

\bibitem[{\citenamefont{{Paumard} et~al.}(2006)}]{Paumard+2006}
\bibinfo{author}{\bibfnamefont{T.}~\bibnamefont{{Paumard}}}
  \bibnamefont{et~al.}, \bibinfo{journal}{ApJ} \textbf{\bibinfo{volume}{643}},
  \bibinfo{pages}{1011} (\bibinfo{year}{2006}).

\bibitem[{\citenamefont{{Bartko} et~al.}(2009)}]{Bartko+2009}
\bibinfo{author}{\bibfnamefont{H.}~\bibnamefont{{Bartko}}}
  \bibnamefont{et~al.}, \bibinfo{journal}{ApJ} \textbf{\bibinfo{volume}{697}},
  \bibinfo{pages}{1741} (\bibinfo{year}{2009}).

\bibitem[{\citenamefont{{Lu} et~al.}(2009)}]{Lu+2009}
\bibinfo{author}{\bibfnamefont{J.~R.} \bibnamefont{{Lu}}} \bibnamefont{et~al.},
  \bibinfo{journal}{ApJ} \textbf{\bibinfo{volume}{690}}, \bibinfo{pages}{1463}
  (\bibinfo{year}{2009}).

\bibitem[{\citenamefont{{Yelda} et~al.}(2014)}]{Yelda+2014}
\bibinfo{author}{\bibfnamefont{S.}~\bibnamefont{{Yelda}}} \bibnamefont{et~al.},
  \bibinfo{journal}{ApJ} \textbf{\bibinfo{volume}{783}}, \bibinfo{eid}{131}
  (\bibinfo{year}{2014}).

\bibitem[{\citenamefont{{von Fellenberg} et~al.}(2022)}]{vonFellenberg+2022}
\bibinfo{author}{\bibfnamefont{S.~D.} \bibnamefont{{von Fellenberg}}}
  \bibnamefont{et~al.}, \bibinfo{journal}{ApJL} \textbf{\bibinfo{volume}{932}},
  \bibinfo{eid}{L6} (\bibinfo{year}{2022}).

\bibitem[{\citenamefont{{Read} et~al.}(2006)\citenamefont{{Read}, {Goerdt},
  {Moore}, {Pontzen}, {Stadel}, and {Lake}}}]{Read+2006}
\bibinfo{author}{\bibfnamefont{J.~I.} \bibnamefont{{Read}}},
  \bibinfo{author}{\bibfnamefont{T.}~\bibnamefont{{Goerdt}}},
  \bibinfo{author}{\bibfnamefont{B.}~\bibnamefont{{Moore}}},
  \bibinfo{author}{\bibfnamefont{A.~P.} \bibnamefont{{Pontzen}}},
  \bibinfo{author}{\bibfnamefont{J.}~\bibnamefont{{Stadel}}}, \bibnamefont{and}
  \bibinfo{author}{\bibfnamefont{G.}~\bibnamefont{{Lake}}},
  \bibinfo{journal}{MNRAS} \textbf{\bibinfo{volume}{373}},
  \bibinfo{pages}{1451} (\bibinfo{year}{2006}).

\bibitem[{\citenamefont{{Zelnikov} and {Kuskov}}(2016)}]{Zelnikov+2016}
\bibinfo{author}{\bibfnamefont{M.~I.} \bibnamefont{{Zelnikov}}}
  \bibnamefont{and} \bibinfo{author}{\bibfnamefont{D.~S.}
  \bibnamefont{{Kuskov}}}, \bibinfo{journal}{MNRAS}
  \textbf{\bibinfo{volume}{455}}, \bibinfo{pages}{3597} (\bibinfo{year}{2016}).

\bibitem[{\citenamefont{{Banik} and {van den Bosch}}(2022)}]{Banik+2022}
\bibinfo{author}{\bibfnamefont{U.}~\bibnamefont{{Banik}}} \bibnamefont{and}
  \bibinfo{author}{\bibfnamefont{F.~C.} \bibnamefont{{van den Bosch}}},
  \bibinfo{journal}{ApJ} \textbf{\bibinfo{volume}{926}}, \bibinfo{eid}{215}
  (\bibinfo{year}{2022}).

\bibitem[{\citenamefont{{Kaur} and {Stone}}(2022)}]{Kaur+2022}
\bibinfo{author}{\bibfnamefont{K.}~\bibnamefont{{Kaur}}} \bibnamefont{and}
  \bibinfo{author}{\bibfnamefont{N.~C.} \bibnamefont{{Stone}}},
  \bibinfo{journal}{MNRAS} \textbf{\bibinfo{volume}{515}}, \bibinfo{pages}{407}
  (\bibinfo{year}{2022}).

\bibitem[{\citenamefont{{Barab{\'a}si} and {Stanley}}(1995)}]{Barabasi+1995}
\bibinfo{author}{\bibfnamefont{A.-L.} \bibnamefont{{Barab{\'a}si}}}
  \bibnamefont{and} \bibinfo{author}{\bibfnamefont{H.~E.}
  \bibnamefont{{Stanley}}}, \emph{\bibinfo{title}{{Fractal Concepts in Surface
  Growth}}} (\bibinfo{publisher}{Cambridge Univ. Press}, \bibinfo{year}{1995}).

\bibitem[{\citenamefont{{Pr{\"a}hofer} and {Spohn}}(2004)}]{Prahofer+2004}
\bibinfo{author}{\bibfnamefont{M.}~\bibnamefont{{Pr{\"a}hofer}}}
  \bibnamefont{and} \bibinfo{author}{\bibfnamefont{H.}~\bibnamefont{{Spohn}}},
  \bibinfo{journal}{J. Stat. Phys.} \textbf{\bibinfo{volume}{115}},
  \bibinfo{pages}{255} (\bibinfo{year}{2004}).

\bibitem[{\citenamefont{{Varshalovich} et~al.}(1988)}]{Varshalovich1988}
\bibinfo{author}{\bibfnamefont{D.~A.} \bibnamefont{{Varshalovich}}}
  \bibnamefont{et~al.}, \emph{\bibinfo{title}{Quantum Theory of Angular
  Momentum}} (\bibinfo{publisher}{World Scientific}, \bibinfo{year}{1988}).

\bibitem[{\citenamefont{Commenges and Monsion}(1984)}]{Commenges+1984}
\bibinfo{author}{\bibfnamefont{D.}~\bibnamefont{Commenges}} \bibnamefont{and}
  \bibinfo{author}{\bibfnamefont{M.}~\bibnamefont{Monsion}},
  \bibinfo{journal}{IEEE Transactions on automatic control}
  \textbf{\bibinfo{volume}{29}}, \bibinfo{pages}{250} (\bibinfo{year}{1984}).

\bibitem[{\citenamefont{Munthe-Kaas}(1999)}]{MuntheKaas1999}
\bibinfo{author}{\bibfnamefont{H.}~\bibnamefont{Munthe-Kaas}},
  \bibinfo{journal}{Appl. Numer. Math.} \textbf{\bibinfo{volume}{29}},
  \bibinfo{pages}{115} (\bibinfo{year}{1999}).

\bibitem[{\citenamefont{{Fouvry} et~al.}(2022)\citenamefont{{Fouvry}, {Dehnen},
  {Tremaine}, and {Bar-Or}}}]{Fouvry+2022}
\bibinfo{author}{\bibfnamefont{J.-B.} \bibnamefont{{Fouvry}}},
  \bibinfo{author}{\bibfnamefont{W.}~\bibnamefont{{Dehnen}}},
  \bibinfo{author}{\bibfnamefont{S.}~\bibnamefont{{Tremaine}}},
  \bibnamefont{and} \bibinfo{author}{\bibfnamefont{B.}~\bibnamefont{{Bar-Or}}},
  \bibinfo{journal}{ApJ} \textbf{\bibinfo{volume}{931}}, \bibinfo{eid}{8}
  (\bibinfo{year}{2022}).

\bibitem[{\citenamefont{{Tak{\'a}cs} and {Kocsis}}(2018)}]{TakacsKocsis2018}
\bibinfo{author}{\bibfnamefont{{\'A}.}~\bibnamefont{{Tak{\'a}cs}}}
  \bibnamefont{and} \bibinfo{author}{\bibfnamefont{B.}~\bibnamefont{{Kocsis}}},
  \bibinfo{journal}{ApJ} \textbf{\bibinfo{volume}{856}}, \bibinfo{eid}{113}
  (\bibinfo{year}{2018}).

\end{thebibliography}
\end{document}